\documentclass[aps,pra,reprint,amsmath,superscriptaddress,amssymb,showpacs,nofootinbib, twocolumn]{revtex4-2}
\usepackage{amsmath,braket,amssymb,amssymb,amsthm,bbm,bm}
\usepackage[final]{graphicx}
\usepackage{subfigure}
\usepackage[english]{babel}
\usepackage[utf8]{inputenc}
\usepackage{mathtools}
\usepackage{empheq}
\usepackage{tensor}
\usepackage{cancel}
\usepackage{enumitem}
\usepackage{simplewick}
\usepackage{tikz}
\usepackage{tikz-network}
\usetikzlibrary{patterns,decorations.pathreplacing}

\usepackage{scalerel}
\usepackage{wasysym}
\usepackage{comment}

\usepackage{enumitem}
\usepackage{slashed}
\usepackage{xcolor}
\definecolor{myviolet}{RGB}{210,145,178}
\definecolor{myvioletc}{RGB}{45,130,60}
\definecolor{myorange}{RGB}{255,165,0}
\definecolor{OliveGreen}{RGB}{85,107,47}
\definecolor{NavyBlue}{RGB}{0,0,128}

\definecolor{myred}{HTML}{e41a1c}
\definecolor{myblue}{HTML}{377eb8}
\definecolor{mygreen}{HTML}{4daf4a}

\colorlet{Green}{black!30!green}

\definecolor{THc}{rgb}{0.9,0.3,0.2}

\usepackage{tikz}
\usetikzlibrary{calc,fadings,decorations.pathreplacing,shapes,shapes.multipart,arrows,shapes.misc,intersections,positioning,patterns}
\tikzset{arrow data/.style 2 args={%
		decoration={%
			markings,
			mark=at position #1 with \arrow{#2}},
		postaction=decorate}
}
\usepackage{bm}
\usepackage[colorlinks=true,citecolor=blue,linkcolor=blue,urlcolor=blue]{hyperref}
\usepackage{dsfont}
\usepackage{changepage}
\usepackage{array}
\usepackage{soul}
\usepackage[capitalise]{cleveref}
\crefname{section}{Sec.}{Secs.}
\Crefname{section}{Sec.}{Secs.}
\usepackage{xifthen}
\usepackage{xargs}
\usepackage{dynkin-diagrams}

\usepackage{amsthm}
\theoremstyle{definition}

\theoremstyle{plain}

\usepackage{bm}

\graphicspath{{./}{./figures/}}

\newcommand{\bit}{\begin{itemize}}
	\newcommand{\eit}{\end{itemize}}

\newcommand{\ba}{\begin{align}}
	\newcommand{\ea}{\end{align}}
\newcommand{\be}{\begin{equation}}
	\newcommand{\ee}{\end{equation}}
\newcommand{\bi}{\begin{itemize}}
	\newcommand{\ei}{\end{itemize}}

\newcommand{\Tr}{\operatorname{Tr}}

\DeclareMathAlphabet{\mymathbb}{U}{BOONDOX-ds}{m}{n}

\renewcommand{\log}{\ln}

\usepackage{physics}

\usepackage{braket}
\usepackage[normalem]{ulem}

\renewcommand{\boxed}[1]{%
  \framebox{\raisebox{0pt}[0.4\baselineskip][0.025\baselineskip]{\hbox to 0.25cm{\hss#1\hss}}}}

\begin{document}
\title{Enhancing entanglement asymmetry in fragmented quantum systems}

\author{Lorenzo Gotta }
\affiliation{Department of Quantum Matter Physics, University of Geneva, 1211 Geneva, Switzerland}

\author{Filiberto Ares }
\affiliation{SISSA and INFN, via Bonomea 265, 34136 Trieste, Italy}

\author{Sara Murciano }
\affiliation{Universit\'e Paris-Saclay, CNRS, LPTMS, 91405, Orsay, France}

\date{\today}
\begin{abstract}
Entanglement asymmetry provides a quantitative measure of symmetry
breaking in many-body quantum states. Focusing on inhomogeneous $U(1)$
charges, such as dipole and multipole moments, we show that the typical asymmetry is bounded by a universal fraction of its maximal value. Multipole charges naturally arise in systems with Hilbert-space fragmentation, where the dynamics splits into exponentially many disconnected sectors. Using the commutant algebra formalism, we generalize entanglement asymmetry to account for fragmentation. 
While the asymmetry grows logarithmically for conventional symmetries, it
can scale extensively in fragmented systems and 
distinguish classical from  quantum fragmentation. We derive general upper bounds for the asymmetry  and identify states that saturate them.    To study the typical behavior of the asymmetry, we consider the ensemble of random
matrix product states. By identifying the bond dimension with an effective
time parameter, we qualitatively reproduce recent results on 
asymmetry dynamics in random quantum circuits, suggesting a universal
behavior for the asymmetry of $U(1)$ charges in local ergodic
systems.

\end{abstract}
\maketitle

\section{Introduction}
The presence or absence of symmetries in many-body quantum systems has a large impact on the physics they generate. For instance, they can reshape the structure of the entanglement~\cite{goldstein18, xavier2018}, or they can affect the usual ergodic behavior which characterizes their late-time
dynamics~\cite{deutsch91, srednicki94, rigol08, rigol07}. An example is Hilbert space fragmentation~\cite{pai19,sala2020, khemani20, moudgalya21m}, a phenomenon in which there exist exponentially many dynamically disconnected
sectors, the Krylov subspaces, each of
which may be chaotic or non-ergodic; see Refs.~\cite{adler24, scherg21, wang25, guo21, zhao25obshilb} for experimental observations. Here, Krylov subspaces are usually not simply distinguished by conventional global symmetries, and the fragmentation is characterized by exponentially many conserved quantities~\cite{moudgalya2022commutant}. Another interesting example comes from open quantum systems, where the operator entanglement of the evolved density matrix can be highly affected by the presence of Abelian or non-Abelian symmetries~\cite{pablo1,pablo2}.

Given the relevance of generic symmetries both in and out of equilibrium, it is meaningful to ask what happens when these symmetries are broken. A revival in interest in the topic came from the study of the \textit{entanglement asymmetry} in many-body quantum systems~\cite{ares2023asymmetry}, showing that counterintuitive relaxation phenomena can occur under local symmetry restoration, such as the Mpemba effect~\cite{joshi2024observing, rylands2024microscopic, ares25, teza26}. In short, entanglement asymmetry quantifies how much a state breaks a given symmetry. Asymmetry was originally introduced as a quantum resource theory, in which operations that respect a given symmetry are distinguished from those capable of generating the resource of interest, i.e. asymmetric states in this case~\cite{bartlett07, vaccaro2008tradeoff, gour2009measuring, marvian2014extending, chitambar2019quantum, tarabunga25,aditya2025}. For finite and compact Lie groups, bounds on the number of qubits, $L$, have been derived~\cite{gour2009measuring}: for instance, for finite groups the asymmetry is bounded by $\log |G|$, where $|G|$ is the cardinality of the group, while for compact Lie groups the asymmetry is bounded by $c\log L$, where $c$ is a coefficient that depends on the group. Ref.~\cite{mazzoni2025breaking} has further put some bounds on the amount of asymmetry generated by
local operations acting on product states for $U(1)$ and $SU(2)$ symmetries generated by spatially homogeneous charges, finding that their asymmetry can be at most half its maximum allowed value. 

Due to the sparking interest in both quantum information and condensed matter regarding symmetry and symmetry breaking, we study the latter in the unconventional setting of Hilbert-space fragmentation. For the sake of generality, we begin our analysis by deriving the \textit{typical} and \textit{maximal} asymmetry of one of the charges responsible for extensive fragmentation of the Hilbert space, i.e. dipole and multipole charges. Using the arguments of Ref.~\cite{mazzoni2025breaking}, we show that, for these inhomogeneous charges, the asymmetry is typically at most a specific fraction of its maximum allowed value. We explicitly verify this result in various settings, including superpositions of frozen configurations, random matrix product states, Haar-random states, and squeezed fermionic Gaussian states. Random matrix product states (MPS) provide a particularly convenient framework to explore these questions analytically~\cite{rmps1,rmps2,rmps3,rmps4,rmps5,rmps6,rmps7,dowling2025,magni2025,Sierant2026,lami2025, magni2025doped, sauliere2026}. Building on previous observations that increasing the bond dimension of random MPS plays the same scaling role as increasing time in local chaotic circuits~\cite{rmps7,magni2025, sauliere2026}, we use inhomogeneous random MPS as an effective description of non-equilibrium dynamics. Within this identification, we show that the entanglement asymmetry displays the same qualitative behavior previously reported for Haar-random brickwork circuits~\cite{ares2025prr} and in Floquet experiments~\cite{Joshi2026}. 
This agreement suggests that the observed behavior is universal across local ergodic systems.

In the presence of Hilbert-space fragmentation, the number of possible sectors of the theory is however not exhausted by the multipole symmetry, since there is an exponentially large number of disconnected Krylov sectors~\cite{khemani20}. The origin of these sectors can be traced back to the commutant algebra, namely the set of operators that commute with every term of the system's Hamiltonian or with every jump or Kraus operator in non-unitary evolutions, resulting in an exponential number of strongly conserved quantities~\cite{moudgalya2022commutant,moudgalya2023symmetries,moudgalya2024scars,moudgalya2024hydro,moudgalya2023numerical}. This formalism has been widely used as a powerful tool to characterize such quantities~\cite{pablo1, pablo2,sahu2025entanglementcosthierarchiesquantum,marche2025exceptional,gotta2025opensystemquantummanybodyscars}. Here, we exploit it once again to introduce a notion of entanglement asymmetry for these unconventional symmetries that measures the extent to which a given state breaks them. This asymmetry allows us to determine the maximal symmetry breaking for fragmented symmetries and to compare it with that of conventional symmetries. We also provide a class of states that saturate these bounds.

It turns out that there is a correlation between the amount of asymmetry and the degree of fragmentation. For conventional symmetries, the number of distinct sectors grows only polynomially with the system size, so the corresponding asymmetry typically scales logarithmically. In contrast, the asymmetry associated with the commutant algebra of fragmented systems can scale extensively with the system size, reflecting the exponentially large number of dynamically disconnected sectors that they are endowed with. In this sense, the entanglement asymmetry can be interpreted as a quantitative probe of the effective number of distinguishable sectors associated with fragmentation. Moreover, it can distinguish between classical and quantum fragmentation.

A genuine question at this point is why it is interesting to identify symmetries for which the entanglement asymmetry takes larger values. The key reason is that, for pure states, the entanglement asymmetry associated with a charge operator $\hat Q$ provides a lower bound on the variance of that operator in the state under consideration~\cite{mazzoni2025breaking} A larger asymmetry therefore implies a larger variance, and hence a larger quantum Fisher information (QFI), since for pure states the QFI is directly proportional to the charge variance~\cite{pezze2009,toth2012,Hyllus2012}. In Appendix~\ref{app:QFI}, we extend this connection to mixed states, showing that, more generally, a larger entanglement asymmetry can lead to a larger QFI (see also the recent work~\cite{ferro2026}).
In fact, the QFI has also been recently used to probe the evolution of broken symmetries following quantum quenches~\cite{ferro25, yamashika25}, and the emergence of the Mpemba effect.

The QFI quantifies how sensitively a state changes under a unitary imprinting generated by $\hat Q$, and thus sets the ultimate precision achievable in estimating the associated parameter. As a consequence, a large entanglement asymmetry identifies states that are promising resources for quantum sensing. An intuitive way to understand this link is the following. If a state lies entirely within a single charge sector, the imprinting transformation only contributes a global phase. In general, it is impossible to capture a global phase through standard observables such as correlators,  so the sensitivity of the state to the parameter to estimate is zero. However, if the state has support over many charge sectors, different components acquire different phases during the imprinting process, and this can non-trivially modify specific observables depending on the estimation parameter. The broader is the distribution over charge sectors, the larger is the spread of acquired phases, and the more strongly the state responds to the transformation. Therefore, the entanglement asymmetry quantifies exactly this kind of spread in charge sectors, and states with large asymmetry naturally emerge as powerful resources for quantum sensing. This connection provides a strong motivation to look for states with a large asymmetry content, both in and out-of-equilibrium.

\subsection{Summary of the main results}

\begin{table*}[t]
\centering
\begin{tabular}{|c|c|c|}
\hline
\textbf{Symmetry structure} & \textbf{Number of sectors} & \textbf{Maximal asymmetry scaling} \\
\hline
Homogeneous $U(1)$ charge & $\sim L$ & $\sim \log L$ \\
\hline
Multipole symmetry (order $p$) & $\sim L^{p+1}$ & $\sim (p+1)\log L$ \\
\hline
Commutant algebra & $\sim e^{cL}$ & $\sim L$ \\
\hline
\end{tabular}
\caption{Symmetry structures studied in this paper and the corresponding maximal scaling of the entanglement asymmetry for a system of size $L$. Here, the asymmetry scaling refers, for all symmetry structures, to the full-system density matrix, since the commutant-algebra asymmetry in Eq.~\eqref{eq:asymm_comm} is naturally defined without reference to a subsystem due to the possible non-locality of the elements of the commutant. See Sec.~\ref{sec:generalized} for a detailed discussion on the definition of the commutant-algebra asymmetry for generic subsystems.}
\label{tab:asymmetry_scaling}
\end{table*}

In this work, we present several results that connect different symmetry
structures to the scaling of entanglement asymmetry. To guide the reader,
we briefly outline the logic of the paper.

We begin by studying the entanglement asymmetry for \emph{inhomogeneous} $U(1)$ charges, such as dipole and higher multipole moments. These symmetries split the Hilbert space into a much larger number of charge sectors than homogeneous global symmetries, and therefore provide a natural setting where asymmetry can grow faster. Within this framework, we analyze both
general bounds and the behavior of typical states. In particular, within the ensemble of random MPS we study the typical scaling of the entanglement asymmetry. By identifying the bond dimension with an effective time evolution, we reproduce the qualitative dynamical behavior previously observed in Haar-random circuits and Floquet systems. This suggests a universal dynamical structure governing the asymmetry of $U(1)$ charges in ergodic quantum systems.

The results obtained for multipole symmetries indicate that the scaling of
the entanglement asymmetry is closely related to the structure of the Hilbert-space decomposition into symmetry sectors. Motivated by this observation, we develop a more general framework based on the commutant of a bond algebra that generates a family of Hamiltonians. This construction allows us to define a generalized notion of entanglement asymmetry that applies beyond conventional symmetry groups.

Finally, we show that this algebraic perspective naturally captures  Hilbert-space fragmentation. In particular, fragmented systems can exhibit a much larger number of sectors than those generated by ordinary symmetries, which allows the entanglement asymmetry to grow extensively with system size.

The progression from multipole symmetries to the behavior of typical states explored through random MPS, and finally to the
general commutant-algebra framework reveals a hierarchy of symmetry structures and corresponding asymmetry scalings, which we explore throughout the remainder of the paper and summarize in Table~\ref{tab:asymmetry_scaling}.

More specifically, the paper is organized as follows. In Sec.~\ref{sec:recap}, we review the definition of entanglement asymmetry
for homogeneous $U(1)$ charges. In Sec.~\ref{sec:dipole} we extend the analysis to inhomogeneous charges, such as dipole and multipole symmetries. This provides the basis for the study of entanglement asymmetry for more general symmetry structures presented in Sec.~\ref{sec:generalized}, with additional examples discussed in Sec.~\ref{sec:examples}. We conclude in
Sec.~\ref{sec:conclusion} with a summary of our main findings, while technical details are postponed to the appendices.

\section{Entanglement asymmetry for $U(1)$ groups}\label{sec:recap}

In this section, we review the definition of entanglement asymmetry for global $U(1)$ symmetries. In general, we will consider a one dimensional lattice $\Lambda=\{1,\dots, L \}$ and a bipartition of it $\Lambda=A\cup B$. We assume that the charge $\hat Q$ that generates the global $U(1)$ symmetry is local in the sense that it can be decomposed into the contributions of $A$ and $B$, i.e. $\hat Q=\hat Q_A+\hat Q_B$. The (von Neumann) entanglement asymmetry $\Delta S_{A,\hat Q}$ of a state $\hat\rho$ over the subsystem $A$ with respect to the charge $\hat Q$ is defined as~\cite{ares2023asymmetry}
\begin{equation}\label{eq:def_asymm}
    \Delta S_{A,\hat Q} = S(\hat\rho_{A,\hat Q})-S(\hat\rho_A),
\end{equation}
where $S(\hat\rho)=-{\rm Tr}[\hat\rho \log\hat\rho]$ is the von Neumann entanglement entropy, $\hat\rho_A={\rm Tr}_B [\hat\rho]$ is the reduced density matrix in $A$, and $\hat\rho_{A,\hat Q}$ its symmetrization. Precisely, $\hat \rho_{A, \hat Q}=\sum_{q\in \mathbb{Z}}\hat \Pi_q \hat\rho_A \hat\Pi_q$, $\hat\Pi_q$ being the projector onto the eigenspace of the charge $\hat Q_A$ with eigenvalue $q$. The state $\hat\rho_{A,\hat Q}$ is symmetric by construction, i.e., $[\hat\rho_{A, \hat{Q}}, \hat Q_A]=0$, as it is block-diagonal in the eigenbasis of the $U(1)$ charge $\hat Q_A$. 

Since the calculation of the von Neumann entropy is often challenging, it is convenient to define the R\'enyi-$n$ entanglement asymmetry of a state $\hat\rho$ with respect to the charge $\hat Q$ over subsystem $A$ as well through the expression~\cite{ares2023asymmetry}
\begin{equation}\label{eq:renyi_ea}
    \Delta S^{(n)}_{A,\hat Q} = S^{(n)}(\hat\rho_{A,\hat Q})-S^{(n)}(\hat\rho_A),
\end{equation}
where $S^{(n)}(\hat\rho)=\frac{1}{1-n}\log {\rm Tr}[\hat\rho^n]$ is the R\'enyi-$n$ entropy, from which one can obtain the von Neumann entropy by considering its analytic continuation as $n\rightarrow 1$. Moreover, it is worth mentioning that both von Neumann and R\'enyi-$n$ entanglement asymmetries satisfy (i) $\Delta S_{A,\hat Q}^{(n)}\geq 0$ and (ii) $\Delta S_{A,\hat Q}^{(n)}=0$ iff $\hat\rho_A = \hat\rho_{A,\hat Q}$, that is, iff the state $\hat{\rho}_A$ is symmetric with respect to $\hat Q_A$.

While evaluating the R\'enyi-$n$ entanglement asymmetry, it is often convenient to take advantage of the Fourier representation of the symmetric density matrix $\hat\rho_{A,\hat Q}$, which reads
\begin{equation}
    \hat\rho_{A,\hat Q} = \int_{-\pi}^{\pi} \frac{d\alpha}{2\pi} e^{-i\alpha\hat Q_A} \hat \rho_A e^{i\alpha \hat Q_A},
\end{equation}
and allows for rewriting the R\'enyi-$n$ entanglement entropy of $\hat\rho_{A,\hat Q}$ in the form
\begin{equation}\label{Eq:fourier_n_renyi}
    S^{(n)}(\hat\rho_{A,\hat Q})=\frac{1}{1-n}\log \int_{[-\pi,\pi]^n} \frac{d\alpha_1\dots d\alpha_n}{(2\pi)^n} Z_n(\boldsymbol{\alpha}),
\end{equation}
where $\boldsymbol{\alpha}=\{ \alpha_1,\dots,\alpha_n\}$ and
$Z_n(\boldsymbol{\alpha})$ are the charged moments,
\begin{equation}\label{Eq:part_fun}
    Z_n(\boldsymbol{\alpha})={\rm Tr}\Biggl[\prod_{j=1}^n \hat\rho_A e^{i\alpha_{j,j+1}\hat Q_A}\Biggr],
\end{equation}
with $\alpha_{j,j+1}=\alpha_j -\alpha_{j+1}$ and  $\alpha_{n+1}=\alpha_1$.

The (R\'enyi) entanglement asymmetry~\eqref{eq:def_asymm}-\eqref{eq:renyi_ea} has been recently studied across a variety of settings. They include ground states of critical theories~\cite{Chen2024, fossati2024entanglement, Kusuki2024, fossati2024, lastres2024entanglement, Benini2024}, matrix product states~\cite{capizzi2024universal, mazzoni2025breaking}, and random states~\cite{ares2023entanglement, russotto2024non, chen2024entanglement, russotto25u1}. Its non-equilibrium behavior has been investigated in closed and open free and integrable systems~\cite{ares2023lack, murciano2024entanglement, caceffo2024entangled, ares2024quantum, chalas2024multiple, bertini2024dynamics, rylands2024dynamical, rylands2024microscopic, yamashika2024entanglement, yamashika2024quenching, klobas2024asymmetry, banerjee2024asymmetry}, also under measurements~\cite{DiGiulio25}, long-range spin chains~\cite{Yu2025Tuning,yamashika25-2}, and random circuits~\cite{liu2024symmetry, turkeshi2024quantum, ares2025prr, Yu2025, summer25}.   The R\'enyi version is not only easier to calculate but it is also accessible in quantum simulators via randomized measurements~\cite{joshi2024observing, Xu2025Observation, Joshi2026}.  Various generalizations of this quantity have been introduced for non-invertible and higher-form symmetries~\cite{Benini2025, benini25cat, GattoLamas25}, domain-wall number~\cite{khor2024kink}, chiral anomalies~\cite{florio25}, gauge~\cite{Zheng25gauge}, spatial~\cite{klobas2024translation, gibbins2025, hara25} as well as strong symmetries~\cite{kusuki26strong}.

\section{Dipole and multipole entanglement asymmetry}\label{sec:dipole}

In the works cited above, the study of the entanglement asymmetry~\eqref{eq:def_asymm}-\eqref{eq:renyi_ea} has been mostly restricted to $U(1)$ symmetries generated
by local, homogeneous  (translationally invariant) charges of the form 
\begin{equation}
\hat Q_0=\sum_j \hat n_j, 
\end{equation}
where the same operator $\hat{n}_j$ acts at each site $j$ of the lattice. In this section, we will extend some results 
from the literature to the case of dipole and multipole charges, which break translational invariance.
In particular, we introduce the $p$-pole symmetry operator in a $L$-site spin-$1/2$ chain,
\begin{equation}\label{eq:p-pole_charge}
    \hat Q_p = \sum_{j=1}^L j^p \hat n_j
\end{equation}
where $\hat n_j = (1+\hat\sigma_j^z)/2$ is the $z$-magnetization, or particle number upon a Jordan-Wigner transformation, 
in the site $j$. In the case $p=0$, the 
operator~\eqref{eq:p-pole_charge} corresponds to the standard global $z$-magnetization/particle number, which is the usual choice when analyzing the asymmetry. For $p\neq 0$, we expect the corresponding asymmetry to exhibit 
different features, for example, as a function of the (sub)system size. The intuitive reason behind such expectation is 
that the number of sectors resolved according to the $p$-pole symmetry grows as $\sum_{j} j^p\sim L^{p+1}$, and the extent to which a quantum state breaks a symmetry depends on its overlap with the different charge sectors. Therefore,
the entanglement asymmetry will be typically enhanced as $p$ is increased. 

As a first example, let us compute the R\'enyi-$n$ entanglement asymmetry with respect to the dipole charge $\hat Q_1$, $p=1$, for the following product state
\begin{equation}\label{Eq:prod_state}
\ket{\psi}=\prod_{j=1}^{L/2}\left(\frac{1+\hat\sigma_{2j}^+}{\sqrt{2}} \right) \ket{\downarrow\dots\downarrow},
\end{equation}
where the system size $L$ has been assumed to be even. Such a state can be characterized as the uniform superposition of one-dimensional Krylov subspaces, i.e. the eigenstates, of a simple dipole-conserving model, defined as
\begin{equation}\label{Eq:dip_cons_ham}
    \hat H = J \sum_{j=1}^{L-3}\left(\hat\sigma_j^-\hat\sigma_{j+1}^+\hat\sigma_{j+2}^+\hat\sigma_{j+3}^- +H.c. \right),
\end{equation}
which commutes with $\hat Q_1$. 
The local terms of this Hamiltonian generate the dipole-preserving rearrangement $\ket{\uparrow\downarrow\downarrow\uparrow}\leftrightarrow \ket{\downarrow\uparrow\uparrow\downarrow}$. However, none of the $z$-basis product-state configurations that appear in the expansion of the state $\ket{\psi}$ contains either two neighbouring spins in the  state $\ket{\uparrow}$ or two spins in the state $\ket{\uparrow}$ separated by $2$ sites. Therefore, no local term in the Hamiltonian~\eqref{Eq:dip_cons_ham} can have a non-vanishing action on these configurations, which are consequently frozen one-dimensional Krylov subspaces (and zero-energy eigenstates) of the Hamiltonian.

Proceeding to the evaluation of $\Delta S^{(n)}_{A,\hat Q_1}$ for $A=\{1,\dots,L/2\}$, we notice that the reduced density matrix reads $\hat\rho_A = \ket{\psi_A}\bra{\psi_A}$, where
\begin{equation}\label{Eq:state_A}
\begin{split}
\ket{\psi_A} =\left( \otimes_{j\in\Lambda_o}\ket{\downarrow}_j\right)\bigotimes\left(\otimes_{l\in\Lambda_e}\frac{\ket{\uparrow}_l+\ket{\downarrow}_l}{\sqrt{2}} \right),
\end{split}
\end{equation}
and we divided subsystem $A$ into even and odd-site sublattices $\Lambda_e=\{2,4,\dots , L/2 \}$ and $\Lambda_o = \{1,3,\dots, L/2-1 \}$, assuming, without loss of generality, that $L/2$ is even as well, i.e., that the number of lattice sites $L$ is a multiple of $4$. One can directly see that $S^{(n)}(\hat{\rho}_A)=0$, since $\ket{\psi}$ is a product state. On the other hand, applying Eqs.~\eqref{Eq:fourier_n_renyi}-\eqref{Eq:part_fun}, it is easy to obtain that
\begin{equation}\label{Eq:part_fun_prod_state}
    Z_n (\boldsymbol{\alpha})= \prod_{j\in\Lambda_e}\left[\prod_{m=1}^n \cos\left(\frac{\alpha_m -\alpha_{m+1}}{2} j \right) \right].
\end{equation}

Let us focus on the case $n=2$ for simplicity.  Inserting Eq.~\eqref{Eq:part_fun_prod_state} in the integral~\eqref{Eq:fourier_n_renyi}, we have
\begin{equation}\label{Eq:renyi_2_prod}
 \Delta S_{A,\hat Q_1}^{(2)} = -\log \int_{-\pi}^{\pi} \frac{d\alpha}{2\pi}e^{\sum_{j\in \Lambda_e}\log \cos^2\left(\frac{j\alpha }{2} \right)}.
\end{equation}
We can obtain the large-$L$ behavior of  $\Delta S_{A,\hat Q_1}^{(2)}$ by performing a saddle-point evaluation of this integral around the points $\alpha=0, \pi$. This leads to the final result
\begin{equation}
  \Delta S_{A,\hat Q_1}^{(2)} \approx \frac{1}{2}\log \left(\frac{\pi}{4} \sum_{j\in \Lambda_e} j^2 \right).
\end{equation}
Since $\sum_{j\in \Lambda_e} j^2 \sim L^3$ when $L\to \infty$, one obtains that the R\'enyi-$2$ asymmetry of the state in Eq.~\eqref{Eq:prod_state} relative to the dipole charge behaves as
\begin{equation}\label{eq:asymm_Q_1_prod_st}
\Delta S_{A, Q_1}^{(2)}\sim\frac{3}{2}\log L.
\end{equation}
in the large $L$ limit. The same scaling is found for 
all values of the R\'enyi index $n$, including the limit 
$n\to 1$, while the dependence on $n$ appears only in the subleading terms. The reason why the leading term in Eq.~\eqref{eq:asymm_Q_1_prod_st} does not depend on $n$ is the saddle point approximation used to compute ${\rm Tr}(\rho_{A, Q}^n)$,  which factorizes the dependence on the subsystem size $\ell_A$
and on the R\'enyi index $n$ into independent contributions, see e.g. Ref.~\cite{murciano2024entanglement} for a detailed discussion in the case of the magnetization.

On the other hand, if we take as charge the usual $z$-magnetization, $\hat{Q}_0$, and repeat the same calculation, then we find that the corresponding R\'enyi-$n$ asymmetry behaves for large $L$ as
\begin{equation}\label{eq:asymm_Q_0_prod_st}
\Delta S_{A, \hat{Q}_0}^{(n)}\sim \frac{1}{2}\log L.
\end{equation}
Observe that, in both cases, the asymmetry grows logarithmically
with the subsystem size, but with a prefactor that depends on the 
exponent $p$ of the charge~\eqref{eq:p-pole_charge}. In what follows, we will generalize this result to different classes of states for any 
integer values of $p$. But first, let us derive some general bounds for the entanglement asymmetry relative to the multipole charges $\hat{Q}_p$.

\subsection{Maximal asymmetry and clustering states}

We take a generic state described by the density 
matrix $\hat \rho$, which can be either pure or mixed, and, for the moment, a generic charge $\hat Q$. 
Since the corresponding symmetrized density matrix $\hat \rho_Q$ commutes with $\hat Q$, then it takes the block diagonal form
\begin{equation}\label{eq:res_rho_Q}
\hat \rho_{\hat Q}=\bigoplus_q p_q \hat \rho_{\hat Q, q},
\end{equation}
where $p_q={\rm Tr}(\hat\Pi_q \hat \rho)$ and $\hat \rho_{\hat Q, q}=\hat\Pi_q\hat{\rho}\hat\Pi_q/p_q$. We can interpret
$p_q$ as the probability of finding $q$ as outcome in a projective measurement of $\hat{Q}$ in the state $\hat \rho$
and $\hat{\rho}_{\hat{Q}, q}$ the state after the measurement.  Inserting Eq.~\eqref{eq:res_rho_Q} in the definition~\eqref{eq:def_asymm} of 
the entanglement asymmetry, we find
\begin{equation}\label{eq:asymm_dec_num_conf}
\Delta S_{\hat Q}= H[\{p_q\}]+\sum_q p_q S(\hat{\rho}_{\hat{Q}, q})- S(\hat{\rho}), 
\end{equation}
where $H[\{p_q\}]=-\sum_q p_q\log p_q$ is the Shannon entropy of the probability distribution $p_q$, also known as number entropy. We can now apply
the fact that, on average, the entanglement entropy does not grow under projective measurements~\cite{lindblad1972entropy} and, therefore,
$\sum_q p_q S(\hat{\rho}_{\hat{Q}, q})\leq S(\hat{\rho})$. This implies that 
\begin{equation}\label{eq:shannon_bound}
\Delta S_{\hat Q}\leq H[\{p_q\}].
\end{equation}
This inequality is valid for any charge $\hat Q$~\cite{mazzoni2025breaking}.  Let us now explore its implications for the multipole charges~\eqref{eq:p-pole_charge}.

First, the result in Eq.~\eqref{eq:shannon_bound} allows us to derive a generic upper bound for their entanglement asymmetry. It is well-known that the 
Shannon entropy is maximized by the uniform probability distribution. In our case, this corresponds to the states for which 
all the charge sectors in Eq.~\eqref{eq:res_rho_Q} are equiprobable. The eigenvalues $q$ of the $p$-pole charge $\hat Q_p$ are integer numbers between
$0\leq q \leq q_{\rm max}$, where $q_{\rm max}=\sum_{j=1}^L j^p$.
At large system sizes $L$,  $q_{\rm max}\sim L^{p+1}$. Therefore, we can bound Eq.~\eqref{eq:shannon_bound} with the uniform probability distribution $p_q=1/L^{p+1}$, 
$q=0, 1, \dots, L^{p+1}$, which
leads to
\begin{equation}\label{eq:max_asymm}
\Delta S_{\hat{Q}}\leq \Delta S_{\hat Q_p}^{\rm max}=(p+1)\log(L+1).
\end{equation}
In the case $p=0$, we recover the bound found in Ref.~\cite{gour2009measuring}. 

We can obtain more refined bounds of the entanglement asymmetry for different families of states by applying the following 
result on the Shannon entropy of discrete random variables. If $p_q$ is a discrete probability 
distribution with support $X\subset \mathbb{Z}$ and variance $0<\sigma^2<\infty$, as in our case, then the 
corresponding Shannon entropy satisfies~\cite{Massey1988Information, Rioul2022Information}
\begin{equation}\label{eq:variance_bound}
H[\{p_q\}]<\frac{1}{2}\log\left[2\pi\left(\sigma^2+\frac{1}{12}\right)\right]+\frac{1}{2}.
\end{equation}
In particular, combining Eqs.~\eqref{eq:shannon_bound} and~\eqref{eq:variance_bound}, we can derive a specific bound for states that satisfy the cluster property
\begin{equation}\label{eq:clustering}
\langle \hat{O}_j \hat{O}_{j'}\rangle-\langle \hat{O}_j \rangle \langle \hat{O}_{j'}\rangle=0, \,\, \textrm{for} \,\, |j-j'|>\Lambda,
\end{equation}
where $\hat{O}_j$ is an arbitary operator acting at the site $j$, $\langle \hat{O}_j \rangle:={\rm Tr}[\hat\rho \hat{O}_j]$, and $\Lambda>0$.  This was done in Ref.~\cite{mazzoni2025breaking} for $p=0$. Let us here extend it to the
multipole charges~\eqref{eq:p-pole_charge}. The variance of the $p$-pole charge probability distribution is given by $\sigma_p^2=\langle \hat{Q}_p^2\rangle-\langle \hat{Q}_p\rangle^2$.
As shown in Appendix~\ref{app:clustering}, for states with the above cluster property, their variance is bounded by
\begin{equation}\label{eq:sigma_p_bound}
\sigma_p^2 \leq \frac{4(1+2\Lambda)}{2p+1}L^{2p+1}(1+O(L^{-1})).  
\end{equation}
Inserting this result in Eq.~\eqref{eq:variance_bound}, we find that for large $L$,
\begin{equation}\label{eq:cluster_bound}
\Delta S_{\hat{Q}_p}^{\rm cluster} \leq \frac{2p+1}{2} \log L,
\end{equation}
and, comparing this bound with Eq.~\eqref{eq:max_asymm}, we can conclude that 
\begin{equation}\label{eq:bound_cluster}
\Delta S_{\hat{Q}_p}^{\rm cluster}\leq \frac{p+1/2}{p+1}\Delta S_{\hat{Q}_p}^{\rm max}.
\end{equation}
For $p=0$, the expression agrees with the bound found in Ref.~\cite{mazzoni2025breaking}, $\Delta S_{\hat{Q}_0}^{\rm cluster}\leq 1/2\Delta S_{\hat{Q}_0}^{\rm max}$.
Observe that the state we considered in Eq.~\eqref{Eq:prod_state}, which satisfies the cluster property~\eqref{eq:clustering}, saturates the bound~\eqref{eq:cluster_bound} (up to $O(L^0)$ 
terms). A natural question at this point is whether this is a specific feature of the state~\eqref{Eq:prod_state}, or if the bound~\eqref{eq:cluster_bound} is typically tight.

\subsection{Typical states}\label{sec:typ_states}

The previous question can be addressed by studying the entanglement asymmetry in ensembles of random states that effectively 
capture different relevant physical properties. By averaging over all the states in the ensemble, one can filter out the atypical, state-specific 
properties from the typical, generic features. Here, we will investigate the multipole asymmetry in two ensembles of states:  Haar random unitary states
and inhomogeneous  Haar random MPS. In the first case, since we uniformly sample over the full Hilbert space, 
these states typically display a volume law entanglement entropy~\cite{page1993average}. On the other hand, the correlations in
random MPS decay exponentially with the distance and, consequently, their entanglement entropy satisfies an area law by construction~\cite{Haag2023correlation}. 

\begin{figure}[t]
\centering
\includegraphics[width=0.42\textwidth]{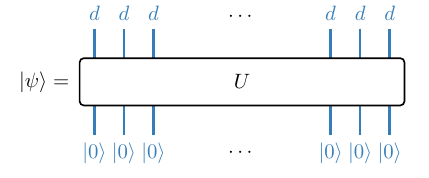}
\includegraphics[width=0.42\textwidth]{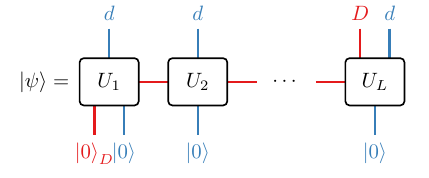}
\caption{Schematic representation of the random states that form the two classes of ensembles considered in Sec.~\ref{sec:typ_states} to study the typical multipole entanglement asymmetry: (a) Haar random states; to a set of $L$ qudits (with local Hilbert space $\mathbb{C}^d$, represented by blue lines) in the state $\ket{0}$, we apply a $d^L\times d^L$ random unitary matrix drawn from the Haar measure. (b) Inhomogeneous random MPS with bond dimension $D$; we consider again a set of $L$ qudits in the state $\ket{0}$ and we sequentially apply an independent $dD\times dD$ Haar random unitary matrix to each of them as shown in the figure. Red lines represent the bond space $\mathbb{C}^D$, which is initialized in the state $\ket{0}_D$. }
\label{fig:random_states}
\end{figure}
\textit{\textbf{Haar random states}.---} Let us start with the ensemble of Haar random states. We consider a system of $L$ qudits and choose the states ${|a\rangle_j}$, with $a=0,\dots, d-1$ and $j=1\dots, L$, as the local basis for each qudit. We take the ensemble of states $\{U|0\rangle^{\otimes L}\}$, where 
$U$ is a $d^L\times d^L$ unitary matrix drawn from the Haar random ensemble. A pictorial representation of this class of states is shown in Fig.~\ref{fig:random_states}. The analog of the $p$-pole charge~\eqref{eq:p-pole_charge} in this system is 
\begin{equation}
\hat{Q}_p=\sum_{j=1}^L \sum_{a=0}^{d-1}a j^p|a\rangle_j\langle a|_j.
\end{equation}
For a generic unitary matrix $U$, the state $U|0\rangle^{\otimes L}$  has no symmetries and, in particular, is not an eigenstate of the charge $\hat{Q}_p$. 

The asymmetry of this ensemble in the case $p=0$ was studied in Ref.~\cite{ares2023entanglement} (see also Ref.~\cite{russotto2024non}) and has recently been measured in experimental Haar random states in Ref.~\cite{Joshi2026}. 
Without loss of generality, we take as subsystem $A=\{1, \dots, \ell_A\}$, with $\ell_A\leq L$. We are interested in the average entanglement asymmetry in $A$ 
of this ensemble,
\begin{eqnarray}\label{eq:av_asymm}
\mathbb{E}[\Delta S_{A, Q_p}^{(n)}]&=&\frac{1}{1-n}\mathbb{E}\left[\log\frac{{\rm Tr}(\rho_{A, Q_p}^n)}{{\rm Tr}(\rho_{A}^n)}\right]\\&\simeq&
\frac{1}{1-n}\log\frac{\mathbb{E}[{\rm Tr}(\rho_{A, Q_p}^n)]}{\mathbb{E}[{\rm Tr}(\rho_{A}^n)]},
\end{eqnarray}
where $\mathbb{E}[\bullet]$ denotes the average over the ensemble. Notice that, in the second equality, we have introduced the 
average $\mathbb{E}[\bullet]$ inside the logarithm. For Haar random states, this a good approximation up to exponentially small subleading corrections in the system size $L$, as shown in Refs.~\cite{ares2023entanglement, russotto2024non} in the case $p=0$,
and drastically simplifies the calculations. 

The average $\mathbb{E}[{\rm Tr}(\rho_{A, Q_p}^n)]$ in Eq.~\eqref{eq:av_asymm} can be obtained from the average of the charge moments $\mathbb{E}[Z_n(\boldsymbol{\alpha})]$
using Eq.~\eqref{Eq:fourier_n_renyi}, and the latter can be computed employing the folded circuit picture~\cite{nahum17growth, nahum18operator, fisher2023random, potter2022entanglement}, as we detail in Appendix~\ref{app:haar}. For simplicity, we restrict ourselves to R\'enyi index $n=2$, although, as has been found in Ref.~\cite{ares2023entanglement} for the case $p=0$, the same qualitative behavior is expected for any positive integer $n$, including the limit
$n\to 1$.  For $\ell_A, L\to\infty$, with $\ell_A/L$ finite, we obtain that
\begin{equation}\label{eq:av_charged_mom_haar}
\mathbb{E}[Z_2(\alpha)] = \left\{\begin{array}{ll}
d^{-\ell_A}, & \ell_A<L/2,\\
d^{-L-\ell_A} \prod_{j\in A}\frac{\sin^2(d\alpha j^p)}{\sin^2(\alpha j^p)}, & \ell_A>L/2.\end{array}\right.
\end{equation}
Plugging this result in Eqs.~\eqref{Eq:fourier_n_renyi} and~\eqref{eq:av_asymm}, and performing a saddle-point approximation of the integral in $\alpha$ as in Eq.~\eqref{eq:asymm_Q_1_prod_st}, we find that
\begin{equation}\label{eq:av_asymm_haar}
\mathbb{E}[\Delta S_{A, Q_p}^{(2)}]\approx\left\{\begin{array}{ll}
\frac{2p+1}{2}\log (\ell_A), & \ell_A>L/2,\\
0, & \ell_A<L/2.
\end{array}
\right.
\end{equation}
As in the case $p=0$, the entanglement asymmetry of any $p$-pole charge presents two very different regimes 
as a function of the subsystem size $\ell_A$ in the thermodynamic limit $L\to\infty$. For subsystems $\ell_A<L/2$, the asymmetry 
is zero for any value of $p$, indicating that the corresponding reduced density matrix  is symmetric with respect to any of
these charges, even if the full system is in a state that breaks all these symmetries. At $\ell_A=L/2$, the asymmetry shows 
a jump discontinuity to a non-zero value for any $p$ and, therefore, the subsystem $A$ undergoes a sharp transition to a 
non-symmetric density matrix. In Ref.~\cite{ares2023entanglement}, the origin of the emergent 
symmetry when $\ell_A<L/2$  was explained in the case $p=0$ through the decoupling inequality~\cite{hayden2007black}
\begin{equation}\label{eq:decoupling}
\mathbb{E}\left[\bigg|\bigg| \hat\rho_A -\frac{I}{d^{\ell_A}}\bigg|\bigg|_1\right]^2\leq d^{2\ell_A-L},
\end{equation}
where $||\bullet||_1$ stands for the trace norm. We can apply their argument for a generic $p$. According to this
inequality, $\hat\rho_A$ is on average exponentially close to the identity when $\ell_A<L/2$ and $L$ is large. Since the identity commutes with any charge, the asymmetry of $\hat\rho_A$ is expected to be close to zero for any 
$\hat{Q}_p$, as we obtain in Eq.~\eqref{eq:av_asymm_haar}. Observe that the identity is invariant under any unitary transformation. Therefore, the vanishing of the asymmetry is not specific to the multipole charges considered here, but is instead a direct consequence of Page typicality. In this sense, the emergent symmetry with respect to any $U(1)$ charge 
$Q_p$ follows from the generic structure of reduced density matrices of Haar-random states, rather than from the particular choice of charge. Indeed, the same emergent symmetry of the reduced density matrix occurs for generic Lie groups~\cite{russotto2024non}.  On the other hand, when $\ell_A>L/2$, the right hand side of Eq.~\eqref{eq:decoupling}
is exponentially large and there is no bound. Notice that the decoupling inequality does not account for the jump discontinuity
of the asymmetry at $\ell_A=L/2$ and, unfortunately, we lack of a satisfactory physical explanation for it. 

Finally, if we compare 
the result in Eq.~\eqref{eq:av_asymm_haar} with Eq.~\eqref{eq:max_asymm}, we conclude that 
\begin{equation}
\mathbb{E}[\Delta S_{Q_p}]=\frac{p+1/2}{p+1}\Delta S_{\hat{Q}_p}^{\rm max}+O(L^0).
\end{equation}
That is, states uniformly sampled over the full Hilbert space typically saturate the bound we found in Eq.~\eqref{eq:bound_cluster}
for clustering states.

\textit{\textbf{Inhomogeneous Random MPS}.---} We move on now to the case of ensembles of inhomogeneous MPS. We consider a one-dimensional system 
of $L$ qudits with local physical dimension $d$. A MPS is then
any state of the form 
\begin{equation}
\ket{\psi}=\sum_{a_1, \dots, a_L=0}^{d-1}\bra{L}A_{a_1}^{(1)}\cdots A_{a_L}^{(L)}\ket{R}\ket{a_1\dots a_L},
\end{equation}
where $\ket{a_j}\in\mathbb{C}^d$ are the local basis elements of each qudit, $A_{a_j}^{(j)}$ are $D\times D$ complex matrices, $\ket{L}$, $\ket{R}\in\mathbb{C}^D$ are 
the left/right boundary conditions, and $D$ is the bond dimension of the MPS. This state can be generated by sequentially 
applying $dD\times d D$-dimensional unitary matrices $U_1, U_2, \dots, U_L$  to the qubits prepared in the product state 
$\ket{0}^{\otimes L}$ and the bond space initialized in the state $\ket{0}_D\in \mathbb{C}^D$, as we graphically show in Fig.~\ref{fig:random_states}.  In our case, to study the 
typical behavior of the entanglement asymmetry, we assume that the unitary matrices $U_j$ are $dD\times dD$ independent Haar random matrices. 

As in the Haar random states, our goal is to calculate the average entanglement asymmetry over this ensemble of MPS. Again we
will apply the approximation of Eq.~\eqref{eq:av_asymm} and introduce the average inside the logarithm.  Notice that the random MPS 
are equivalent to states resulting from a $L$-depth circuit with a 2-qudit local random unitary gates that are sequentially 
applied to an initial product state, as illustrated in Fig.~\ref{fig:random_states}. Therefore, we can follow the same strategy as in the Haar random 
states and apply the folded circuit picture to calculate the average charged moments $\mathbb{E}[Z_2(\alpha)]$. We leave the 
details of the computation to Appendix~\ref{app:random_MPS} and we report here the final result. If we consider periodic boundary conditions, then it reads
\begin{equation}\label{eq:av_charged_mom_mps_transfer_main}
\mathbb{E}[Z_2(\alpha)]=\Tr\left[\prod_{j=1}^{\ell_A} T_-(\alpha_j) T_+^{L-\ell_A}\right],
\end{equation}
with $\alpha_j=\alpha j^p$,
\begin{equation}\label{eq:T_p}
T_+=\left(\begin{array}{cc} 1 & \frac{d^2 D-D}{(d D)^2-1} \\
0 & \frac{dD^2-d}{(dD)^2-1}\end{array}\right),
\end{equation}
and
\begin{equation}\label{eq:T_m}
 T_-(\alpha)=\frac{1}{d^2 D^2-1}\begin{pmatrix}
        d D^2-\frac{\sin ^2(\alpha  d )}{d\sin^2(\alpha)} & d D-D\frac{\sin ^2(\alpha  d )}{d\sin^2(\alpha )}\\
        D\frac{\sin ^2(\alpha  d )}{\sin^2(\alpha )}-D & D^2\frac{\sin ^2(\alpha  d )}{\sin^2(\alpha )}-1
    \end{pmatrix}.
\end{equation}

\begin{figure}[t]
\centering
\includegraphics[width=0.49\textwidth]{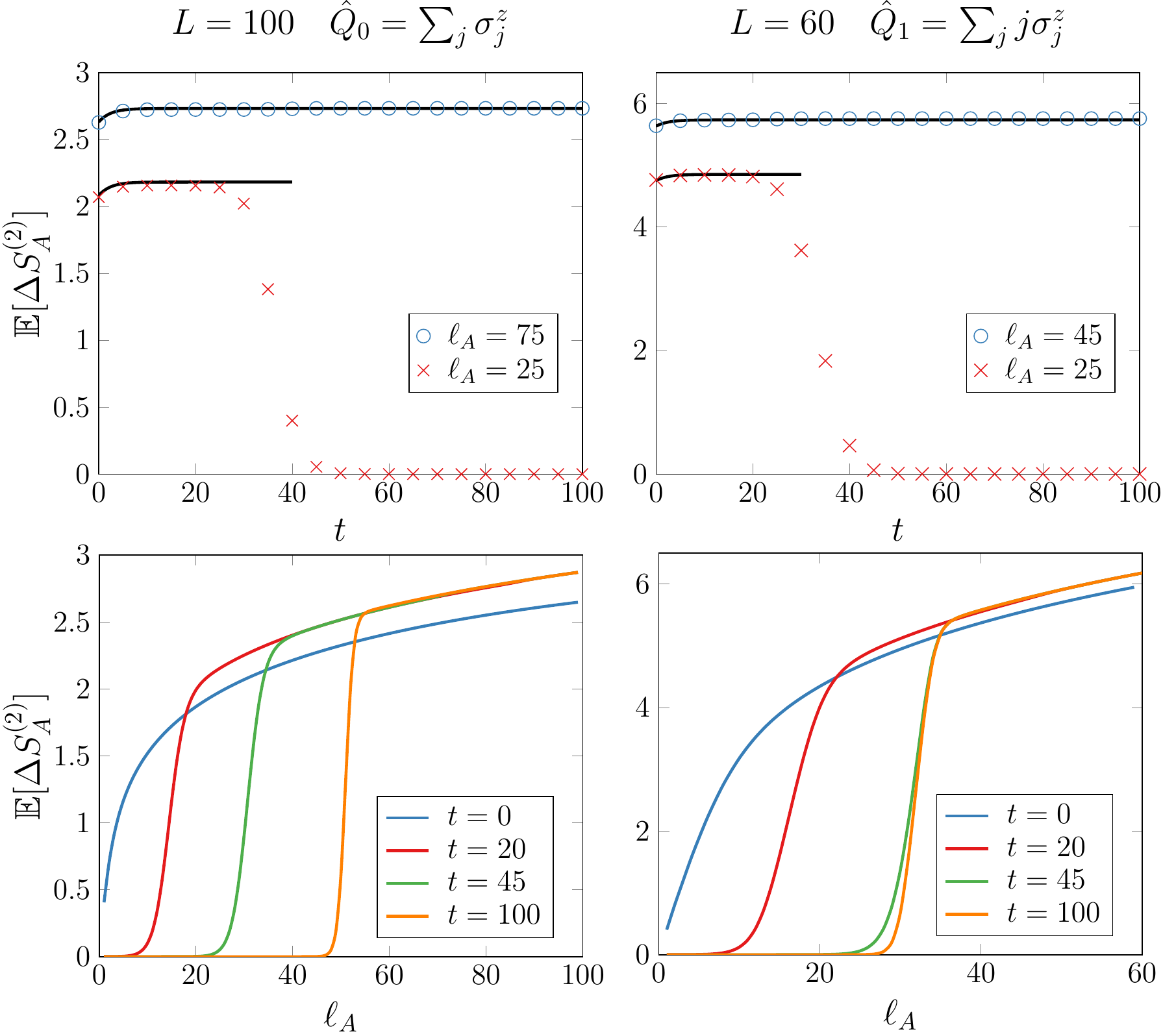}
\caption{Average Rényi-2 entanglement asymmetry in the inhomogeneous random MPS, with $d=2$, calculated using the charged moments~\eqref{eq:av_charged_mom_mps_transfer_main}. In the upper panels, we plot it as a function of the bond dimension $D$, parametrized as $D = D_0 e^{tv(d)}$, for two subsystem sizes $\ell_A$. In the lower panels, it is shown as a function of $\ell_A$ for different values of $t$. In the left panels, we consider a charge $p = 0$, while in the right panels we take $p = 1$. The black curves in the upper panels correspond to the saddle point approximation in Eq.~\eqref{eq:av_asymm_mps}. In the limit $t\to\infty$ ($D\to\infty$), we recover the results for Haar random states. }
\label{fig:rmps}
\end{figure}

Taking the Fourier transformation of the charged moments~\eqref{eq:av_charged_mom_mps_transfer_main}, we can obtain the average entanglement asymmetry in the random MPS.
In Fig.~\ref{fig:rmps}, we show the results as a function of the bond dimension $D$ and subsystem size for $p=0$ (left panels) and $p=1$ (right panels). 
Motivated by earlier works~\cite{rmps7,magni2025, sauliere2026}, which observed that increasing the bond dimension of random MPS plays the same scaling role as increasing time in local chaotic circuits, we identify $D=D_0e^{t v(d)}$, where $D_0=(d+1)/d$ and $v(d)=\log((d^2+1)/(2d))$ (see Appendix.~\ref{app:rmps_rqc} for more details). Within this identification, we can also investigate whether the entanglement asymmetry in random MPS reproduces the qualitative non-equilibrium behavior previously found for it in the case $p=0$ in Haar-random brickwork circuits ~\cite{ares2025prr} and experimentally in a Floquet system~\cite{Joshi2026}. As we show in Fig.~\ref{fig:rmps}, we find that the asymmetry exhibits the same behavior: for $\ell_A<L/2$, the plot shows an initial regime where the asymmetry is nonzero but decays rapidly as 
$D$ increases, eventually approaching zero in the large-$D$ limit. For $\ell_A>L/2$, the asymmetry is well described by the saddle-point approximation (see Appendix~\ref{app:random_MPS})
\begin{equation}\label{eq:av_asymm_mps}
\mathbb{E}[\Delta S_{A, Q_p}^{(2)}]\approx\frac{2p+1}{2}\log(\ell_A)+\frac{1}{2}\log
\frac{(d^2-1)D^2d\pi}{3(2p+1)(1+dD^2)}
\end{equation}
and saturates to a non-zero value as $D\to\infty$, which coincides with that of Haar random states, as shown in Appendix~\ref{app:random_MPS}. Eq.~\eqref{eq:av_asymm_mps}
corresponds to the black curves in the upper panels of Fig.~\ref{fig:rmps}. We observe that Eq.~\eqref{eq:av_asymm_mps} also characterizes the average asymmetry for subsystems $\ell_A < L/2$, prior to its decay to zero in the large-$D$ limit.
In Haar-random brickwork circuits and $p=0$, the relaxation time of the asymmetry, i.e., the time required to become arbitrarily close to its saturation value, is independent of the subsystem size when $\ell_A>L/2$. By contrast, for $\ell_A<L/2$, the decay to zero is governed by the scrambling time of the system. From Eq.~\eqref{eq:av_asymm_mps} and under the identification $D=D_0e^{tv(d)}$, it follows immediately that the relaxation time in random MPS is likewise independent of the subsystem size when $\ell_A>L/2$ for any $p$. For $\ell_A< L/2$, we further verify in Appendix~\ref{app:rmps_rqc} that the asymmetry behaves at large $D$ as
\begin{equation}
\mathbb{E}[\Delta S_{A, Q_p}^{(2)}]\simeq \frac{(d+1)^2}{d^2}\frac{d^{\ell_A}}{D^2}
\end{equation}
for any $p$. As a consequence, the relaxation time towards zero scales linearly with $\ell_A$, in agreement with the scrambling time of brickwork unitary circuits.


The qualitative similarity of this behavior with that found in Refs.~\cite{ares2025prr, Joshi2026} confirms that the mapping $D\sim e^t$ provides a good effective description of the dynamics. We emphasize, however, that the universality discussed here does not refer to the full time dependence of the asymmetry. Rather, 
during the growth regime the asymmetry obeys $\mathbb{E}[\Delta S_A^{(2)}]=\tfrac{1}{2}\log(\ell_A)+f(t)$,
where the logarithmic dependence on the subsystem size $\ell_A$ is universal, while the function $f(t)$ is model dependent.
The universal features also include the long-time asymptotic behavior, namely, the decay to zero for $\ell_A<L/2$ and the saturation to the Haar-random value for $\ell_A>L/2$, as well as the scaling of the relaxation time with $\ell_A$. In the random-MPS framework, the results are derived as a function of the bond dimension $D$, while their interpretation in terms of real-time dynamics relies on the identification $D\sim e^t$. This correspondence provides a simple effective description of the dynamics while requiring significantly simpler calculations than direct circuit approaches.

 Beyond typical states, we have numerically verified that the bound~\eqref{eq:cluster_bound} is also saturated for squeezed fermionic Gaussian states, and we report a detailed discussion in Appendix~\ref{app:gaussianstates}.

\section{Generalized entanglement asymmetry in the commutant-algebra framework}\label{sec:generalized}

The scaling of the entanglement asymmetry
is closely related to how the Hilbert space decomposes into sectors associated
with a given operator or algebra. In the simplest case, this decomposition is generated by a $U(1)$ charge, and in the previous section we have extended the analysis to inhomogeneous charges such as dipole and multipole moments. A natural framework to describe such decompositions in a unified way is provided by the language of commutant algebras~\cite{moudgalya2022commutant}.
In this approach, the relevant structure is encoded in the algebra of
operators commuting with the bond algebra of a Hamiltonian, which determines how the
Hilbert space splits into dynamically disconnected subspaces, usually
referred to as Krylov subspaces~\cite{khemani20}. 
Importantly, the standard $U(1)$ groups generated by the multipole moments $Q_p$ discussed in Sec.~\ref{sec:dipole} are encompassed within this broader framework. Motivated by this observation, we extend the definition of entanglement asymmetry from symmetrization with respect to a group action to symmetrization with respect to the commutant of a von Neumann algebra. While in Sec.~\ref{sec:dipole} the multipole moments $Q_p$ need not be conserved charges of a Hamiltonian, here we focus on the commutant of the bond algebra of a Hamiltonian, which is the relevant setting for fragmented systems. Nevertheless, the definition applies to arbitrary von Neumann algebras. In both constructions, the chosen group or algebra is part of the definition of the asymmetry, whereas the invariance under it is a property of the state considered. Thus, the asymmetry quantifies the extent to which the state fails to be invariant under the chosen group action or algebra, vanishing whenever the state is invariant.

Let us first quickly review the theoretical formalism introduced in Ref.~\cite{moudgalya2022commutant}. We consider the algebra generated by the identity and the set of local (generally multi-site) terms $\{\hat h_{\alpha} \}$ of a Hamiltonian, which is dubbed bond algebra and denoted as $\mathcal{A}=\langle\langle \{\hat h_{\alpha} \}   \rangle\rangle$. 
Its commutant algebra $\mathcal{C}$ can be defined as the algebra of operators that commute with each generator of the bond algebra $\mathcal{A}$. Since $\mathcal{A}$ and $\mathcal{C}$ are von Neumann algebras, the double-commutant theorem states that they are mutual centralizers. Therefore, they provide a canonical decomposition of the Hilbert space $\mathcal{H}$ on which they act
\begin{equation}\label{Eq:decomp}
    \mathcal{H}=\bigoplus_{\lambda} \left( \mathcal{H}_{\lambda}^{\mathcal{A}} \otimes \mathcal{H}_{\lambda}^{\mathcal{C}}\right),
\end{equation}
where $\mathcal{H}_{\lambda}^{\mathcal{A}}$ is a $D_{\lambda}$-dimensional irrep of $\mathcal{A}$ and $\mathcal{H}_{\lambda}^{\mathcal{C}}$ is a $d_{\lambda}$-dimensional irrep of $\mathcal{C}$. The Krylov subspaces associated with $\mathcal{A}$ correspond, for each value of $\lambda$, to the $d_{\lambda}$ copies of the irrep $\mathcal{H}^{\mathcal{A}}_{\lambda}$. Since every operator $A \in \mathcal{A}$ acts irreducibly within each of these blocks, the total number of Krylov subspaces is $N_{\mathcal{K}}=\sum_{\lambda}d_{\lambda}$. Using the decomposition in Eq.~\eqref{Eq:decomp}, one immediately obtains that $\mathrm{dim}(\mathcal{H})=\sum_{\lambda} D_{\lambda}d_{\lambda}$. Moreover, for each $\lambda$, $\mathcal{A}$ and $\mathcal{C}$
realize the full matrix algebras on $\mathcal{H}_{\lambda}^{\mathcal{A}}$ and $\mathcal{H}_{\lambda}^{\mathcal{C}}$, respectively, and this leads to $\mathrm{dim}(\mathcal{A})=\sum_{\lambda}D_{\lambda}^2$ and $\mathrm{dim}(\mathcal{C})=\sum_{\lambda}d_{\lambda}^2$.

With this unified framework in mind, we can define a generalized notion of entanglement asymmetry that quantifies the symmetry breaking for any symmetry structure captured by the commutant-algebra formalism, including those responsible for ergodicity breaking in scarred and fragmented systems.
Since the commutant algebra $\mathcal{C}$ is a finite-dimensional unital associative von Neumann algebra, we can follow the procedure outlined in Ref.~\cite{Benini2025}, originally developed in the context of generalized symmetries in quantum field theory, to construct a symmetric state $\hat{\rho}_{S}$ that satisfies
\begin{equation}
\hat O\hat{\rho}_S=\hat{\rho}_S \hat O,\quad  \forall \hat O\in \mathcal{C}.
\end{equation}
Let us take a basis $\{\hat X_a\}$ of the commutant algebra $\mathcal{C}$. 
The symmetrization of $\hat{\rho}$ is then given by~\cite{Benini2025}
\begin{equation}\label{eq:rhoS1}
\hat{\rho}_S=\sum_{a, b}\tilde{K}^{a, b} \bigoplus_{\lambda, \lambda'} \hat X_a(\lambda)\hat\rho \hat X_b(\lambda'), 
\end{equation}
where $\tilde{K}^{a, b}$ are the entries of the inverse of the matrix $K_{a, b}={\rm Tr}(\hat X_a \hat X_b)$,
and we have decomposed the basis elements of $\mathcal{C}$ into irreps as
\begin{equation}
\hat X_a=\bigoplus_\lambda \hat X_a(\lambda).
\end{equation}
It is easy to check that $\hat{\rho}_S$ satisfies several properties necessary to be a good symmetrized density matrix~\cite{Benini2025}. In particular, if the original state $\hat{\rho}$ is already invariant under the bond algebra $\mathcal{C}$, then the symmetrization leaves it unchanged, i.e., $\hat{\rho}_S=\hat{\rho}$.

Following Eq.~\eqref{eq:rhoS1}, the entanglement asymmetry of the state $\hat{\rho}$ with respect to the commutant $\mathcal{C}$ can be defined as in Ref.~\cite{Benini2025} for higher-form and non-invertible symmetries,
\begin{equation}\label{eq:asymm_comm}
\Delta S_\mathcal{C}(\hat\rho)=S(\hat\rho_S)-S(\hat\rho).
\end{equation}
This is a generalization of the entanglement asymmetry for $U(1)$ groups in Eq.~\eqref{eq:def_asymm}, and has similar properties. It can be rewritten 
as the relative entropy between $\hat\rho$ and $\hat\rho_S$ and, consequently, is non-negative. It vanishes iff $\hat\rho$ is symmetric under the commutant $\mathcal{C}$, namely $[\hat\rho, \hat O]=0$ $\forall \hat O\in\mathcal{C}$. 

Notice that we have defined the commutant asymmetry for the density matrix of the full system, rather than for a subsystem $A$ as we did for standard $U(1)$ symmetries in Sec.~\ref{sec:recap}. Unlike the multipole moments studied in Sec.~\ref{sec:dipole}, the elements of the commutant algebra do not generally satisfy the locality property $\hat{O} = \hat{O}_A + \hat{O}_B$, where $\hat{O}_A$ and $\hat{O}_B$ have support in $A$ and $B$, respectively. This property is, in principle, required to define the asymmetry in a subsystem $A$ according to Sec.~\ref{sec:recap}: one first takes the partial trace and then symmetrizes the reduced density matrix. To define a subsystem commutant asymmetry, one could adopt the approach in Ref.~\cite{klobas2024translation} for space translations: first construct the symmetrized state, as done here, and then take its partial trace. Both approaches yield the same asymmetry whenever the locality property holds. In any case, we will focus on the full system.

A more practical expression for the symmetrized density matrix in Eq.~\eqref{eq:rhoS1} can be obtained in terms of the projectors onto the irreps $\lambda$ in Eq.~\eqref{Eq:decomp}. Since the commutant $\mathcal{C}$ is a semisimple symmetry algebra, the symmetrization acts as a projector onto irreducible sectors. This is a consequence of Schur lemma. The symmetrization of $\hat{\rho}$ in Eq.~\eqref{eq:rhoS1} can be rewritten as 
\begin{equation}
\hat{\rho}_S=\bigoplus_{\lambda, \lambda'} (\hat{\rho}_S)_{\lambda, \lambda'},
\end{equation}
where
\begin{equation}
(\hat{\rho}_S)_{\lambda, \lambda'}=\sum_{a, b}\tilde{K}^{a,b}\hat{X}_a(\lambda)\hat{\rho}\hat{X}_b(\lambda').
\end{equation}
Since, by construction, 
$\hat{\rho}_S$ commutes with every element $\hat{X}_c$ of the commutant algebra, the blocks satisfy $\hat{X}_c(\lambda)(\hat{\rho}_S)_{\lambda, \lambda'}=(\hat{\rho}_S)_{\lambda, \lambda'}\hat{X}_c(\lambda')$. Then Schur lemma implies that $(\hat{\rho}_S)_{\lambda, \lambda'}=0$ when $\lambda$ and $\lambda'$
correspond to inequivalent irreducible representations, and
\begin{equation}\label{eq:sym_alg_2}
\hat{\rho}_S=\sum_{a, b}\tilde{K}^{a, b} \bigoplus_{\lambda} \hat X_a(\lambda)\hat\rho \hat X_b(\lambda). 
\end{equation}
We now introduce the projector $\hat\Pi_\lambda$ onto the irrep $\lambda$, such that we can rewrite Eq.~\eqref{eq:sym_alg_2} in the form
\begin{equation}\label{eq:sym_alg_3}
\hat{\rho}_S=\sum_\lambda\sum_{a, b}\tilde{K}^{a, b}  \hat X_a(\lambda)\hat \Pi_\lambda\hat\rho \hat\Pi_\lambda \hat X_b(\lambda). 
\end{equation}
In each irrep $\lambda$, the Hilbert space factorizes as $\mathcal{H}_\lambda^{\mathcal{A}}\otimes \mathcal{H}_\lambda^{\mathcal{C}}$, as stated in Eq.~\eqref{Eq:decomp}. Then the basis elements in each irrep can be decomposed as 
\begin{equation}
\hat X_a(\lambda)\equiv \mathbb{I}_\mathcal{A}\otimes \hat X_a(\lambda),
\end{equation}
where $\mathbb{I}_{\mathcal{A}}$ is the identity within the Hilbert space $\mathcal{H}_{\lambda}^{\mathcal{A}}$, and
\begin{equation}\label{eq:intstep}
\hat{\rho}_S=\sum_\lambda\sum_{a, b}\tilde{K}^{a, b}  (\mathbb{I}_\mathcal{A}\otimes \hat X_a(\lambda))\hat\Pi_\lambda\hat\rho \hat \Pi_\lambda (\mathbb{I}_\mathcal{A}\otimes \hat X_b(\lambda)).
\end{equation}
We assume that, within each irrep $\lambda$, the projected state $\hat \Pi_{\lambda}\hat \rho \hat \Pi_{\lambda}$ can be expanded in a basis of operators acting on $\mathcal{H}_{\lambda}^{\mathcal{A}}\otimes \mathcal{H}_{\lambda}^{\mathcal{C}}$. Since ${\rm End}(\mathcal{H}_\lambda)\cong{\rm End}(\mathcal{H}_\lambda^{\mathcal{A}})\otimes {\rm End}(\mathcal{H}_\lambda^{\mathcal{C}})$, we can choose operator bases $\{\hat A_j\}$ and $\{\hat B_j\}$ for the two factors. In terms of these bases, any projected density matrix takes the form
\begin{equation}
\hat \Pi_\lambda \hat{\rho}\hat \Pi_\lambda=\sum_{j,j'}c_{j, j'}\hat A_j\otimes \hat B_{j'}.
\end{equation}
If we plug it into Eq.~\eqref{eq:intstep},
we find
\begin{equation}
\hat{\rho}_S=\sum_\lambda\sum_{j, j'}c_{j, j'}
\left(\hat A_{j}\otimes \sum_{a, b}\tilde{K}^{a, b}\hat X_a(\lambda)\hat B_{ j'}\hat X_b(\lambda)\right).
\end{equation}
By applying the Schur lemma again~\cite{Benini2025}
\begin{equation}
\sum_{a, b}\tilde{K}^{a, b}\hat X_a(\lambda)\hat B_{ j'}\hat X_b(\lambda)={\rm Tr}(\hat B_{j'})\frac{\mathbb{I}_\mathcal{C}}{\dim \mathcal{H_\lambda^{\mathcal{C}}}},
\end{equation}
we obtain the following expression for the symmetrized density matrix
\begin{equation}\label{eq:rhoS_proj}
\hat{\rho}_S=\sum_\lambda {\rm Tr}_{\mathcal{C}}(\hat \Pi_\lambda \hat{\rho} \hat \Pi_\lambda)\otimes \frac{\mathbb{I}_\mathcal{C}}{d_\lambda}.
\end{equation}
Here, the sum over $\lambda$ can be understood as a sum of operators embedded in the full Hilbert space through the projectors $\hat\Pi_\lambda$. 
This result extends to commutant algebras the symmetrization of density matrices found in Ref.~\cite{gour2009measuring} (see also Ref.~\cite{bartlett07}) for 
standard non-Abelian compact groups. See Appendix~\ref{app:rho_S}
for an alternative derivation of Eq.~\eqref{eq:rhoS_proj}.

A particularly relevant situation is when the commutant $\mathcal{C}$ is Abelian, which corresponds to classical Hilbert space fragmentation~\cite{moudgalya2022commutant}. In this case, all irreps of $\mathcal{C}$ are one-dimensional, $d_\lambda=1$, and each irreducible subspace $\mathcal{H}_\lambda$ of the bond algebra $\mathcal{A}$ can be identified with a Krylov subspace. As follows from Eq.~\eqref{eq:rhoS_proj}, the symmetrized density matrix can then be obtained by projecting onto the Krylov subspaces, $\hat{\rho}_S=\sum_\lambda \hat\Pi_\lambda \hat{\rho} \hat\Pi_\lambda$.

\textit{\textbf{Bound for the generalized entanglement asymmetry}.---} To derive a bound on the entanglement asymmetry for pure states $\hat{\rho}=\ket{\psi}\bra{\psi}$, we use Eq.~\eqref{eq:rhoS_proj} for $\hat{\rho}_S$ and decompose its von Neumann entanglement entropy as done in Eq.~\eqref{eq:asymm_dec_num_conf} for the case of multipole symmetries,
\begin{equation}\label{eq:decom}
\begin{split}
    S(\hat{\rho}_S)&=-\sum_{\lambda} p_{\lambda}\log p_{\lambda}+\sum_{\lambda}p_{\lambda} S\left(\hat\rho_{\lambda}^{\mathcal{A}}\otimes \frac{\mathbb{I}_\mathcal{C}}{d_{\lambda}} \right),
    \end{split}
\end{equation}
where $p_{\lambda}=\mathrm{Tr}(\hat\Pi_{\lambda}\hat{\rho})$ can be interpreted as the probability of each irrep $\lambda$ under the projection $\hat{\Pi}_\lambda$, and $\rho_\lambda^\mathcal{A}$ is the 
reduced density matrix $\hat{\rho}_{\lambda}^{\mathcal{A}}=\mathrm{Tr}_{\mathcal{C}}(\hat{\rho}_{\lambda})$ of the projected state 
\begin{eqnarray}
\hat{\rho}_{\lambda}=\frac{\hat\Pi_{\lambda}\hat{\rho}\hat\Pi_{\lambda}}{p_{\lambda}}.
\end{eqnarray}
Since this state is pure on the virtual bipartition $\mathcal{H}_{\lambda}^{\mathcal{A}}\otimes \mathcal{H}_{\lambda}^{\mathcal{C}}$, the entropy of $\hat{\rho}_{\lambda}^{\mathcal{A}}$ can then be at most
$\log\left(d_{\mathrm{\min},\lambda}\right)$, with $d_{\min,\lambda}=\min \{D_{\lambda},d_{\lambda}\}$. Therefore,
\begin{equation}
    S(\hat{\rho}_S)
    \leq-\sum_{\lambda} p_{\lambda}\log(p_{\lambda})+\sum_{\lambda} p_{\lambda}\log(d_{\lambda} d_{\mathrm{\min},\lambda}).
\end{equation}
By further using the concavity of the logarithm, we obtain the final bound
\begin{equation}\label{Eq:bound_asymm_comm}
    S(\hat{\rho}_S)\leq \log\left(\sum_{\lambda} d_{\lambda} d_{\min,\lambda} \right).
\end{equation}
Since for a pure state $S(\hat\rho)=0$, the bound in Eq.~\eqref{Eq:bound_asymm_comm} is an upper bound to the commutant-algebra entanglement asymmetry~\eqref{eq:asymm_comm}. 

Observe that for Abelian commutants, since $d_\lambda=1$ for all $\lambda$, then $d_{{\rm min}, \lambda}=1$ and the bound of the asymmetry in Eq.~\eqref{Eq:bound_asymm_comm} is directly determined by the number of Krylov subspaces,
\begin{equation}
 \Delta S_{\mathcal{C}}\leq \ln N_\mathcal{K}.
\end{equation} 
When applied to conventional or unconventional symmetries, we recover the results summarized in Table~\ref{tab:asymmetry_scaling}. Indeed, if we symmetrize the state with respect to the Abelian algebra
generated by operators such as the number or multipole operators, one
retrieves the logarithmic scaling discussed in the previous sections,
since the number of corresponding sectors grows only polynomially with
system size. In contrast, resolving the full commutant algebra associated
with the bond algebra distinguishes the Krylov sectors of the dynamics.
In fragmented systems, where the number of such sectors grows
exponentially with system size, this may lead to an extensive scaling of the
entanglement asymmetry. 

Finally, pure states saturating the bound in Eq.~\eqref{Eq:bound_asymm_comm} exist, and are generically of the form:
\begin{equation} \label{Eq:max_as_st}
    \ket{\psi_{\max}}= \frac{1}{\sqrt{\sum_{\lambda} d_{\lambda}d_{\min,\lambda}}}\sum_{\lambda} \sqrt{d_{\lambda}d_{\min,\lambda}} \ket{\psi_{\max,\lambda}}.
\end{equation}
Here the maximally-entangled state $\ket{\psi_{\max,\lambda}}$ is defined as
\begin{equation}\label{eq:max}
    \ket{\psi_{\max,\lambda}}= \frac{1}{\sqrt{d_{\min,\lambda}}}\sum_{j=1}^{d_{\min,\lambda}} |\phi^{(\lambda)}_j\rangle\otimes |\varphi^{(\lambda)}_j\rangle,
\end{equation}
where the set $\{|\phi^{(\lambda)}_j\rangle \}_{j=1}^{d_{\min,\lambda}}$ is an orthonormal basis (subset) of $\mathcal{H}_{\lambda}^{\mathcal{A}}$ if $d_{\min,\lambda}=D_{\lambda}$ ($d_{\min,\lambda}=d_{\lambda}$), and, similarly, the set $\{|\varphi^{(\lambda)}_j\rangle \}_{j=1}^{d_{\min,\lambda}}$ is an ortohonormal basis (subset) of $\mathcal{H}_{\lambda}^{\mathcal{C}}$ if $d_{\min,\lambda}=d_{\lambda}$ ($d_{\min,\lambda}=D_{\lambda}$).
An intuitive way to understand why the state in Eq.~\eqref{Eq:max_as_st} maximizes the entanglement asymmetry is to note that Eq.~\eqref{eq:decom} separates the asymmetry into two contributions. The first term corresponds to the Shannon entropy associated with the probability distribution over the irreps $\lambda$, while the second term captures a quantum contribution
arising from the entanglement between $\mathcal H_\lambda^\mathcal{A}$ and $\mathcal H_\lambda^\mathcal{C}$ within each irrep. The state in Eq.~\eqref{Eq:max_as_st} maximizes both contributions simultaneously.
First, it is a uniform superposition of states belonging to each irrep
$\lambda$, so that all sectors are equiprobable. This maximizes the
Shannon entropy of the sector probabilities and therefore the first term in Eq.~\eqref{eq:decom}. Second, within each irrep we choose the maximally entangled
state between $\mathcal{H}_\lambda^{\mathcal{A}}$ and $\mathcal{H}_\lambda^{\mathcal{C}}$, which maximizes the
second term. This structure has a simple interpretation in the two limiting cases.
In the case of classical fragmentation, the second term in Eq.~\eqref{eq:decom}
vanishes, and the asymmetry is entirely determined by the classical
probability distribution over Krylov subspaces. The maximal state is
therefore the uniform superposition over all sectors. In contrast, in
the case of quantum fragmentation there is an additional quantum
contribution coming from the entanglement within each irrep, which is maximized by choosing maximally entangled states in each virtual bipartition $\mathcal{H}_\lambda^\mathcal{A}\otimes \mathcal{H}_\lambda^\mathcal{C}$ .

In the next section, we exploit the structure in Eq.~\eqref{Eq:max_as_st} to obtain states whose entanglement asymmetry grows faster than in the $U(1)$ cases explored so far.

\section{Hilbert-space fragmentation and volume-laws for the entanglement asymmetry}\label{sec:examples}

Since Hilbert-space fragmented systems are characterized by an exponentially large number of dynamically disconnected sectors, it is natural to explore a simple setting exhibiting this phenomenology, in which the entanglement asymmetry within the framework introduced in Section~\ref{sec:generalized} can be computed exactly. This will allow us to test the expectation that this quantity grows faster than logarithmic in the system size. We highlight here that fragmentation can be classified as classical or quantum, depending on whether a maximal Abelian subalgebra of the commutant admits a local product eigenbasis~\cite{pablo2}. Our bound~\eqref{Eq:bound_asymm_comm} shows that in classically fragmented systems, where fragment labels can be resolved by commuting observables, the asymmetry is bounded by the logarithm of the number of sectors. In contrast, for quantum fragmentation, no such product basis exists, and the noncommuting structure within fragments contributes through the factor $d_{\rm min,\lambda}$. Ref.~\cite{pablo1} showed that a vanishing or nonvanishing bulk autocorrelation function is not sufficient to distinguish classical from quantum fragmentation; in this sense, asymmetry provides a probe to differentiate between classical and quantum fragmentation.

We first reconsider the state~\eqref{Eq:prod_state} introduced in Sec.~\ref{sec:dipole}, where its entanglement asymmetry was evaluated with respect to the $U(1)$ groups generated by the magnetization and dipole moment. We can now compute its full-system asymmetry with respect to the commutant associated with the Hamiltonian~\eqref{Eq:dip_cons_ham}, using the definition introduced in Sec.~\ref{sec:generalized}. As we discussed in Sec.~\ref{sec:dipole}, all computational-basis configurations appearing in the expansion of the state $\ket{\psi}$ in Eq.~\eqref{Eq:prod_state} form isolated one-dimensional Krylov subspaces of the Hamiltonian~\eqref{Eq:dip_cons_ham}. As a result, the full-commutant symmetrization reduces to a projection onto these sectors, allowing us to evaluate the corresponding asymmetry analytically. Let $c$ label the computational-basis configurations appearing in the expansion of $\ket{\psi}$, namely those with all odd sites in the state $\ket{\downarrow}$ and arbitrary spin configurations on the even sites. Each of these configurations forms an isolated one-dimensional Krylov subspace of the Hamiltonian in Eq.~\eqref{Eq:dip_cons_ham}, $\mathcal{K}_c={\rm span}\{\ket{c}\}$, with corresponding projector $\hat{\Pi}_c=\ket{c}\bra{c}$. The full-commutant symmetrization of the state $\ket{\psi}$ is therefore 
\begin{equation}
\hat{\rho}_S=\sum_c\hat{\Pi}_c\ket{\psi}\bra{\psi}\hat{\Pi}_c.
\end{equation}
Thus, expanding Eq.~\eqref{Eq:prod_state} as
\begin{equation}
\ket{\psi}=2^{-L/4}\sum_{c}\ket{c},
\end{equation}
we directly find
\begin{equation}
\hat{\rho}_S=2^{-L/2}\sum_c\ket{c}\bra{c}.
\end{equation}
The symmetrized state is the maximally mixed state over the $2^{L/2}$ one-dimensional Krylov sectors $\mathcal{K}_c$, and its von Neumann entropy is $S(\hat{\rho}_S)=L/2\log 2$. Since $\ket{\psi}$ is a pure state, the full-system asymmetry is
\begin{equation}
\Delta S_\mathcal{C}(\ket{\psi})=S(\hat{\rho}_S)=L/2\log 2.
\end{equation}
This behavior differs from the $U(1)$ cases discussed in Sec.~\ref{sec:dipole}, where the number of charge sectors grows only polynomially with the subsystem size, leading to a logarithmic entanglement asymmetry, cf. Eqs.~\eqref{eq:asymm_Q_1_prod_st}-\eqref{eq:asymm_Q_0_prod_st}. Here, the full-commutant symmetrization resolves the state into $2^{L/2}$ one-dimensional Krylov sectors, resulting in an extensive asymmetry.

\subsection{$t-J_z$ model}

We now focus on a paradigmatic example of Hilbert-space fragmentation, the $t-J_z$ model~\cite{zhang97, batista00}. This model belongs to the family of spin-1 Hamiltonians generated by the bond algebra
\begin{equation}
\begin{split}
    \mathcal{A} =\Bigl\langle\Bigl\langle & \left\{ \sum_{\sigma =\pm 1}\left(\ket{0\sigma}_{j,j+1}\bra{\sigma 0}_{j,j+1}+h.c.\right)  \right\}_{j=1}^{L-1},\\
    &\left \{\hat S_j^z \hat S_{j+1}^z\right\}_{j=1}^{L-1},
   \left\{\hat S_j^z\right\}_{j=1}^{L},\left\{\left(\hat S_j^z\right)^2\right\}_{j=1}^{L}\Bigr\rangle\Bigr\rangle.
\end{split}
\end{equation}
where $\ket{0\sigma}_{j,j+1}\bra{\sigma 0}_{j,j+1}$ is the nearest-neighbor hopping operator on the bond $j,j+1$, $\hat{S}_j^z$ is the local spin-1 operator in the $z$ direction, and the double-angle bracket denotes the algebra generated by the listed operators.
This class of models describes lattice spin-$1$ systems where the only allowed transitions among the product states in the $z$-basis are the one in which the states $\ket{\pm 1}$ are free to exchange their position with a neighboring $\ket{0}$ state under a hopping process. As a result, the dynamics is constrained to disconnected Krylov subspaces of the full Hilbert space $\mathcal{H}$~\cite{rakovszky20, moudgalya2022commutant}, each of which is specified by fixing (i) the number $n$ of sites in the states $\ket{\pm 1}$, (ii) the number $0\leq n_{+}\leq n$ of sites in the state $\ket{1}$ and (iii) the number of different patterns $\vec{\sigma}(n,n_{\uparrow})=\{\sigma_j\}_{j=1}^n$, with $\sigma_j=\pm 1$ for every site $j$, under the constraint that $n_{\uparrow}$ entries of $\vec{\sigma}$ have to be equal to $+1$ and $n-n_{\uparrow}$ have instead to be equal to $-1$.

Under these constraints, the Hilbert space $\mathcal{H}$ decomposes into
\begin{equation}\label{Eq:krylov_dec}
    \mathcal{H} = \bigoplus_{n=0}^L \bigoplus_{n_{\uparrow}=0}^n \bigoplus_{\vec{\sigma}(n,n_{\uparrow})} \mathcal{K}_{n,n_{\uparrow},\vec{\sigma}(n,n_{\uparrow})},
\end{equation}
where $\mathcal{K}_{n,n_{\uparrow},\vec{\sigma}(n,n_{\uparrow})}$ denotes the Krylov subspace specified by the values of $n,n_{\uparrow}$, and the spin pattern $\vec{\sigma}(n,n_{\uparrow})$ defined above.

Let us now compute the entanglement asymmetry for three classes of pure states: a bound-saturating state, a uniform superposition of all $z$-basis product states with fixed $n$ and a uniform superposition of all $z$-basis product states with fixed $(n,n_{\uparrow})$.

\textit{\textbf{Maximal-asymmetric state.---}}
We start from a maximally asymmetric state which takes the form in Eq.~\eqref{Eq:max_as_st}. Since the commutant is Abelian, all multiplicities satisfy $d_{\lambda}=d_{\min,\lambda}=1$, and any bound-saturating state must have the form
\begin{equation}
    \ket{\psi_{\max}}= \frac{1}{\sqrt{N_{\mathcal{K}}}}\sum_{n,n_{\uparrow},\vec{\sigma}(n,n_{\uparrow})}\ket{\psi_{n,n_{\uparrow},\vec{\sigma}(n,n_{\uparrow})}},
\end{equation}
where $N_{\mathcal{K}}=2^{L+1}-1$ is the number of Krylov subspaces. The states $\ket{\psi_{n,n_{\uparrow},\vec{\sigma}(n,n_{\uparrow})}}$ are normalized and lie within the Krylov subspace labeled by the index set $\left(n,n_{\uparrow},\vec{\sigma}(n,n_{\uparrow})\right)$. The corresponding asymmetry can be read off from Eq.~\eqref{Eq:bound_asymm_comm}, giving
\begin{equation}
    \Delta S_{\mathcal{C}}(\ket{\psi_{\max}}\bra{\psi_{\max}}) = \log\left(N_{\mathcal{K}} \right)\propto L, 
\end{equation}
i.e., a volume-law behavior. 

\textit{\textbf{Fixed $n$.---}}
The second state is defined as
\begin{equation}
    \ket{\psi_n}:=\frac{1}{\sqrt{D_{\mathcal{H}_n}}}\sum_{C\in \mathcal{H}_n}\ket{C},
\end{equation}
where $\mathcal{H}_n$ denotes the sector of the Hilbert space spanned by all $z$-basis product configurations $\ket{C}$ containing exactly $n$ sites in either the
$\ket{1}$ or $\ket{-1}$ state.
Keeping in mind the decomposition in Eq.~\eqref{Eq:krylov_dec}, the state $\ket{\psi_n}$ can be rewritten by grouping the configurations according to the Krylov subspace that they belong to 
\begin{equation}\label{Eq:psi_n_exp}
\begin{split}
    &\ket{\psi_n}  = \frac{1}{\sqrt{D_{\mathcal{H}_n}}} \sum_{n_{\uparrow}=0}^n \sum_{\vec{\sigma}(n,n_{\uparrow})} \sqrt{D_{\mathcal{K}_{n,n_{\uparrow},\vec{\sigma}(n,n_{\uparrow})}}} \ket{\psi_{n,n_{\uparrow},\vec{\sigma}(n,n_{\uparrow})}},        
\end{split}
\end{equation}
where the states 
\begin{equation}\label{Eq:Krylov_gs}
  \ket{\psi_{n,n_{\uparrow},\vec{\sigma}(n,n_{\uparrow})}}=  \frac{1}{\sqrt{D_{\mathcal{K}_{n,n_{\uparrow},\vec{\sigma}(n,n_{\uparrow})}}}}\sum_{C\in \mathcal{K}_{n,n_{\uparrow},\vec{\sigma}(n,n_{\uparrow})}} \ket{C} ,
\end{equation}
form a set of orthonormal states.
Every such state is simply the equal-weight superposition of all basis configurations in the Krylov subspace $\mathcal{K}_{n,n_{\uparrow},\vec{\sigma}(n,n_{\uparrow})}$ for each allowed value of $n_{\uparrow}$ and $\vec{\sigma}(n,n_{\uparrow})$ at the given value of $n$.

The commutant-algebra asymmetry is then simply obtained from the  symmetrization of~\eqref{Eq:psi_n_exp},
\begin{equation}
\begin{split}
    &\hat{\rho}_S= \sum_{m,m_{\uparrow},\vec{\sigma}(m,m_{\uparrow})}\hat \Pi_{\mathcal{K}_{m,m_{\uparrow},\vec{\sigma}(m,m_{\uparrow})}} \ket{\psi_n}\bra{\psi_n}\hat \Pi_{\mathcal{K}_{m,m_{\uparrow},\vec{\sigma}(m,m_{\uparrow})}} \\
    &=\sum_{n_{\uparrow}=0}^n \sum_{\vec{\sigma}(n,n_{\uparrow})} \frac{D_{\mathcal{K}_{n,n_{\uparrow},\vec{\sigma}(n,n_{\uparrow})}}}{D_{\mathcal{H}_{n}}}\ket{\psi_{n,n_{\uparrow},\vec{\sigma}(n,n_{\uparrow})}}\bra{\psi_{n,n_{\uparrow},\vec{\sigma}(n,n_{\uparrow})}},     
\end{split}
\end{equation}
as the Shannon entropy of the probability distribution obtained by the square of the coefficients in the expansion of $\ket{\psi_n}$ in Eq.~\eqref{Eq:psi_n_exp}, namely
\begin{equation}
    \begin{split}
        &\Delta S_{\mathcal{C}} (\ket{\psi_n}\bra{\psi_n})=\\
        &-\sum_{n_{\uparrow}=0}^n \sum_{\vec{\sigma}(n,n_{\uparrow})} \frac{D_{\mathcal{K}_{n,n_{\uparrow},\vec{\sigma}(n,n_{\uparrow})}}}{D_{\mathcal{H}_{n}}} \log\left(\frac{D_{\mathcal{K}_{n,n_{\uparrow},\vec{\sigma}(n,n_{\uparrow})}}}{D_{\mathcal{H}_{n}}}\right).
    \end{split}
\end{equation}
Using the fact that 
\begin{eqnarray}\label{eq:space1}
 D_{\mathcal{K}_{n,n_{\uparrow},\vec{\sigma}(n,n_{\uparrow})}}= \binom{L}{n},\quad   D_{\mathcal{H}_{n}}= \binom{L}{n} 2^n, 
\end{eqnarray} 
the asymmetry reads
\begin{equation}
 \Delta S_{\mathcal{C}} (\ket{\psi_n}\bra{\psi_n})= n\log 2. 
\end{equation}
When the sector $\mathcal{H}_n$ is chosen so that $n$, i.e., the number of sites in the state $\ket{-1}$ or $\ket{+1}$, scales extensively with the system size, we obtain a volume-law for the asymmetry of the state $\ket{\psi_n}$. 

\textit{\textbf{Fixed $n$ and $n_{\uparrow}$.---}}
Finally, we choose the third state as
\begin{equation}
    \ket{\psi_{n,n_{\uparrow}}}=\frac{1}{\sqrt{D_{\mathcal{H}_{n,n_{\uparrow}}}}}\sum_{C\in \mathcal{H}_{n,n_{\uparrow}}}\ket{C},
\end{equation}
where $\mathcal{H}_{n,n_{\uparrow}}$ 
consists of all the $z$-basis product states with $n$ spins in $\ket{\pm 1}$ and exactly $n_{\uparrow}$ of them equal to $\ket{1}$.
Using the decomposition in Eq.~\eqref{Eq:krylov_dec}, we can rewrite $\ket{\psi_{n,n_{\uparrow}}}$ by assembling all configurations belonging to the same subspace
\begin{equation}\label{Eq:psi_n_n_up_exp}
\begin{split}
&\ket{\psi_{n,n_{\uparrow}}}  = \frac{1}{\sqrt{D_{\mathcal{H}_{n,n_{\uparrow}}}}}  \sum_{\vec{\sigma}(n,n_{\uparrow})} \sqrt{D_{\mathcal{K}_{n,n_{\uparrow},\vec{\sigma}(n,n_{\uparrow})}}} \ket{\psi_{n,n_{\uparrow},\vec{\sigma}(n,n_{\uparrow})}},        
\end{split}
\end{equation}
where the states $\ket{\psi_{n,n_{\uparrow},\vec{\sigma}(n,n_{\uparrow})}}$, introduced in Eq.~\eqref{Eq:Krylov_gs},
form a set of orthonormal states, as mentioned before.

Following the same reasoning as above, the commutant-algebra asymmetry is obtained by computing the Shannon entropy of the probability distribution formed by the square of the coefficients in the expansion $\ket{\psi_{n,n_{\uparrow}}}$ in Eq.\eqref{Eq:psi_n_n_up_exp}, namely
\begin{equation}
    \begin{split}
        &\Delta S_{\mathcal{C}} (\ket{\psi_{n,n_{\uparrow}}}\bra{\psi_{n,n_{\uparrow}}})=\\
        &- \sum_{\vec{\sigma}(n,n_{\uparrow})} \frac{D_{\mathcal{K}_{n,n_{\uparrow},\vec{\sigma}(n,n_{\uparrow})}}}{D_{\mathcal{H}_{n,n_{\uparrow}}}} \log\left(\frac{D_{\mathcal{K}_{n,n_{\uparrow},\vec{\sigma}(n,n_{\uparrow})}}}{D_{\mathcal{H}_{n,n_{\uparrow}}}}\right).
    \end{split}
\end{equation}
Using Eq.~\eqref{eq:space1} and
\begin{equation}
D_{\mathcal{H}_{n,n_{\uparrow}}}= \binom{L}{n} \binom{n}{n_{\uparrow}},
\end{equation} 
the asymmetry reads
\begin{equation}
    \begin{split}
        &\Delta S_{\mathcal{C}} (\ket{\psi_{n,n_{\uparrow}}}\bra{\psi_{n,n_{\uparrow}}})= \log\binom{n}{n_{\uparrow}} .
    \end{split}
\end{equation}
When $n$ and $n_{\uparrow}$ scale extensively, the asymmetry of  $\ket{\psi_{n,n_{\uparrow}}}$ follows a volume-law with logarithmic corrections. 

Beyond the technical result on the extensive scaling of the entanglement asymmetry, this finding is also notable from a resource-theoretic perspective. The symmetrized superposition of eigenstates of the $t-J_z$ model provides a clear example of a highly resourceful state within the resource theory of asymmetry.

\textit{\textbf{Typical asymmetry in a fixed-occupation sector.---}}
The examples above concern specially constructed states with controlled
weights over the Krylov sectors. It is therefore natural to ask whether
their extensive asymmetry is representative of typical states in the
fragmented Hilbert space. In general, the answer can depend on the
ensemble under consideration. If an ensemble effectively occupies
$N_{\mathrm{eff}}$ sectors, its asymmetry is expected to be of order
$\ln N_{\mathrm{eff}}$, whereas the maximum is determined by the
logarithm of the total number of available sectors. This is analogous
to the conventional $U(1)$ case, where clustering states typically
populate only $O(\sqrt{L})$ charge sectors and consequently have
$\Delta S\sim \frac{1}{2}\ln L$, while the maximal asymmetry scales as
$\ln L$. As a concrete example, we can consider Haar-random pure states within a fixed $n$-sector of the $t-J_z$ model considered in Sec.~IV. In particular, let $$\mathcal H_n=\bigoplus_{n_\uparrow=0}^{n}\bigoplus_{\vec{\sigma}(n,n_\uparrow)}
\mathcal K_{n,n_\uparrow,\vec{\sigma}(n,n_\uparrow)}$$ be the decomposition of the Hilbert space into Krylov sectors. For a Haar-random state $\ket{\Psi}\in\mathcal H_n$, the sector weights $p_\lambda$, with $\lambda=\bigl(n_\uparrow,\vec{\sigma}(n,n_\uparrow)\bigr)$, are Dirichlet distributed, with parameters given by the corresponding sector dimensions~\cite{bianchi1}. Since the commutant is Abelian, the second term in Eq.~\eqref{eq:decom} is zero and the Haar-averaged commutant asymmetry is the average Shannon entropy of these weights. Using the Haar average of this entropy derived in
Ref.~\cite{bianchi1}, we obtain
\begin{equation}
\begin{split}
    \mathbb{E}[\Delta S_{\mathcal{C}}]&=\psi(D_{{\mathcal{H}_{n,n_{\uparrow}}}}+1)\\&-\sum_{n_{\uparrow}}\sum_{\vec{\sigma}(n,n_{\uparrow})} \frac{D_{\mathcal{K}_{n,n_{\uparrow},\vec{\sigma}(n,n_{\uparrow})}}}{D_{\mathcal{H}_{n,n_{\uparrow}}}} \psi(D_{\mathcal{K}_{n,n_{\uparrow},\vec{\sigma}(n,n_{\uparrow})}}+1),
    \end{split}
\end{equation}
where $\psi$ is the digamma function. Using 
\begin{equation}
    D_{\mathcal{K}_{n,n_\uparrow,\vec{\sigma}}} =\binom{L}{n},
\qquad D_{\mathcal H_n}=2^n\binom{L}{n},
\end{equation}
we can expand the result above for $\mathbb{E}[\Delta S_{\mathcal{C}}]$ in the large $n$ limit, and we get $\mathbb{E}[\Delta S_{\mathcal{C}}]\simeq n \log 2$, up to corrections exponentially suppressed in $L$ (assuming $n$ scales extensively with the system size). Therefore, a Haar-random state in $\mathcal{ H}_n$ has the same leading commutant asymmetry, $n\log 2$, as the explicit state with exactly uniform weights over the Krylov sectors. The two results differ only by a finite-size correction arising from the fluctuations of the sector probabilities in the Haar ensemble.

\subsection{Temperley--Lieb spin chains}

We now provide a simple example in which the quantum contribution to the asymmetry can be evaluated explicitly. We consider the Temperley--Lieb models for spin-$S$ chains with open boundary conditions and local Hilbert-space dimension $N=2S+1$~\cite{pablo2}. The total Hilbert space decomposes as 
\begin{equation}
    \mathcal H=\bigoplus_{\lambda}\mathcal H^{\mathcal A}_{\lambda}\otimes\mathcal H^{\mathcal C}_{\lambda},
\end{equation}
where the dimensions of the Temperley–Lieb bond algebra and its commutant are, respectively,
\begin{equation}
    D_\lambda=\binom{L}{L/2+\lambda}-\binom{L}{L/2+\lambda+1}, \qquad d_\lambda=[2\lambda+1]_q.
\end{equation}
Here $[n]_q=(q^n-q^{-n})/(q-q^{-1})$ is the $q$-deformed integer, with $N=q+q^{-1}$. To isolate the contribution arising from the virtual bipartition
$\mathcal H^{\mathcal A}_{\lambda}\otimes
\mathcal H^{\mathcal C}_{\lambda}$,
we consider a pure state supported entirely within a single
$\lambda$ sector,
\begin{equation}
\ket{\psi_{\lambda}}=\frac{1}{\sqrt{d_{\rm min,\lambda}}}\sum_j \ket{a_j^{(\lambda)}}\otimes \ket{c_j^{(\lambda)}},
\end{equation}
where $\{|a_j^{(\lambda)}\rangle\}$ and$\{|c_j^{(\lambda)}\rangle\}$  are orthonormal basis of $\mathcal{H}_\lambda^\mathcal{A}$ and $\mathcal{H}_\lambda^\mathcal{C}$, respectively. The reduced density
matrix on $\mathcal H^{\mathcal A}_{\lambda}$ is
\begin{equation}
\hat\rho^{\mathcal{A}}_\lambda=\frac{1}{d_{\rm min,\lambda}}\sum_j\ket{a_j^{(\lambda)}}\bra{a_j^{(\lambda)}}.
\end{equation}
Since the state has support in only one sector, $p_\lambda=1$,
the Shannon contribution associated with the distribution over
different irreps in Eq.~\eqref{eq:decom} vanishes. Using the symmetrized
density matrix~\eqref{eq:rhoS_proj}, we find that the total asymmetry is
\begin{equation}
    \Delta S_{\mathcal{C}}=\log d_{\lambda}+\log d_{\rm min,\lambda}.
\end{equation}
The contribution $\log d_{\rm min,\lambda}$ comes from the entanglement between the representation spaces of the bond algebra and its commutant within a single sector, so it has an intrinsically quantum origin. For the specific Temperley–Lieb algebra, we can choose $\lambda=1$, such that $d_1=q^2+1+q^{-2}=N^2-1$ and $d_{\rm min,\lambda}=d_1$. Thus, $\Delta S_{\mathcal{C}}=2\log(N^2-1)$. 

\section{Conclusions}\label{sec:conclusion}

In this work, motivated by  the perspective of the quantum resource theory for the asymmetry, we introduced the multipole and, more generically, the commutant-algebra entanglement asymmetry with the purpose of unveiling symmetries that lead to enhanced large-size scalings of the asymmetry. We summarize here our main findings:

\textbf{Symmetry breaking of inhomogeneous charges:} 
We have studied entanglement asymmetry in the presence of multipole conserved charges, focusing on situations where a $U(1)$ symmetry is broken in a spatially inhomogeneous way, such as for dipole and higher multipole moments. In contrast to the usual homogeneous charge, these symmetries partition the Hilbert space into a much larger number of charge sectors that grows rapidly with system size. As a result, states that break such inhomogeneous symmetries can distribute their weight across many more charge sectors, which enhances the entanglement asymmetry. We show that in different settings ranging from random MPS to simple product states, the asymmetry scales logarithmically with system size, but with a larger prefactor than in the homogeneous case. We compute both the maximal asymmetry one can achieve in this case, but also the typical that one would get. We derive general upper bounds for this growth and show that they are typically saturated. 

\textbf{Universal aspects of the dynamical behavior of entanglement asymmetry for $U(1)$ charges:} As a byproduct of our analysis of random MPS, we have also investigated the dynamics of entanglement asymmetry for both homogeneous and inhomogeneous charges. By identifying the bond dimension with an effective time, we observe the same universal behavior previously found in Haar-random circuits and Floquet dynamics: for subsystems smaller than half of the system, the asymmetry is initially nonzero but decays exponentially, whereas for larger subsystems it saturates to a finite value. Although the duration of the initial transient regime depends on the specific circuit geometry, the shared qualitative behavior indicates that the underlying mechanism is universal.

\textbf{Commutant algebra framework:} To go beyond standard symmetry groups, we reformulate the entanglement asymmetry within the commutant-algebra framework, which provides a unified language to describe both conventional symmetries and the more exotic structures that appear in Hilbert-space fragmentation. Instead of characterizing symmetry sectors through the eigenvalues of a conserved charge, this approach decomposes the Hilbert space using the algebra of operators that commute with all local dynamical terms. The commutant-algebra approach naturally captures situations where the Hilbert space splits into many dynamically disconnected sectors that cannot be labeled by simple global quantum numbers. We derive general bounds for the commutant-algebra asymmetry of a pure state in terms of the internal structure of the Krylov sectors. Since commutant algebras share the algebraic structure of the higher-form and non-invertible symmetries considered in Ref.~\cite{Benini2025}, we adopt the corresponding symmetrization procedure and definition of entanglement asymmetry introduced there. This opens up the possibility of studying the breaking and restoration of unconventional symmetries across a wide range of systems and phenomena through the same lenses. 

\textbf{Extensive entanglement asymmetry:} By exploiting the underlying algebraic structure, we construct explicit states whose entanglement asymmetry grows much faster than in conventional symmetry settings. While ordinary global symmetries lead only to logarithmic growth of the asymmetry with system size, fragmented systems allow for extensive behavior, because these states coherently spread over an exponential number of dynamically disconnected sectors into which the Hilbert space decomposes. This result directly connects to our original motivation for identifying states and charges with large asymmetry. Entanglement asymmetry provides, in general, a lower bound to the quantum Fisher information, meaning that states with larger asymmetry are potentially more sensitive to symmetry-generated transformations. From this perspective, fragmented systems naturally host highly asymmetric states that may serve as promising resources for quantum technologies, such as quantum sensing, where enhanced sensitivity is a key requirement.

Our new results open up several interesting lines of investigation to be explored in the future. First of all, while the present work deals with the typical value of the dipole entanglement asymmetry, powerful techniques in the realm of free-fermion models would allow to tackle its quench dynamics under quadratic fermionic Hamiltonians from a broad class of initial states, such as those studied in Appendix~\ref{app:gaussianstates}. In particular, it is interesting to investigate whether the phenomenology of the quantum Mpemba effect~\cite{ares25} can be observed in such systems as well in the relaxation dynamics of the dipole entanglement asymmetry. Moreover, it would be interesting to understand whether the qualitative behavior observed in random MPS reflects a deeper underlying mechanism that may control other quantities like the circuit complexity~\cite{fan2025}.

Furthermore, Brownian circuits have recently proven to be a powerful tool in the study of universal properties of entanglement dynamics. Thus, it is certainly timely to adapt the findings presented in Ref.~\cite{vardhan2026entdyn} to the entanglement asymmetry. Specifically, the Brownian-circuit picture would allow to address the evaluation of the saturation value of the entanglement asymmetry under noisy symmetry-endowed evolution, as well as to analyze the entanglement asymmetry dynamics in the known case of absence of circuit symmetries, complementing previous studies in Refs.~\cite{ares2025prr, russotto25qssep} (see also Refs.~\cite{liu2024symmetry, turkeshi2024quantum, Yu2025, summer25}).

\textbf{Acknowledgments} We thank Sreemayee Aditya, Luca Capizzi, Florent Ferro, Yahui Li, Michele Mazzoni, Lorenzo Piroli, Pablo Sala, and Riccardo Senese for enlightening discussions. We especially thank Jacopo De Nardis and Xhek Turkeshi for their suggestions about random MPS. FA and SM thank Pasquale Calabrese for several collaborations on related topics. This work is supported by the Swiss National Science Foundation under Division II (Grant No. 200020-219400). FA acknowledges support from the European Research Council under the Advanced Grant no. 101199196 (MOSE).

\bibliography{bibliography}

\begin{thebibliography}{121}%
\makeatletter
\providecommand \@ifxundefined [1]{%
 \@ifx{#1\undefined}
}%
\providecommand \@ifnum [1]{%
 \ifnum #1\expandafter \@firstoftwo
 \else \expandafter \@secondoftwo
 \fi
}%
\providecommand \@ifx [1]{%
 \ifx #1\expandafter \@firstoftwo
 \else \expandafter \@secondoftwo
 \fi
}%
\providecommand \natexlab [1]{#1}%
\providecommand \enquote  [1]{``#1''}%
\providecommand \bibnamefont  [1]{#1}%
\providecommand \bibfnamefont [1]{#1}%
\providecommand \citenamefont [1]{#1}%
\providecommand \href@noop [0]{\@secondoftwo}%
\providecommand \href [0]{\begingroup \@sanitize@url \@href}%
\providecommand \@href[1]{\@@startlink{#1}\@@href}%
\providecommand \@@href[1]{\endgroup#1\@@endlink}%
\providecommand \@sanitize@url [0]{\catcode `\\12\catcode `\$12\catcode `\&12\catcode `\#12\catcode `\^12\catcode `\_12\catcode `\%12\relax}%
\providecommand \@@startlink[1]{}%
\providecommand \@@endlink[0]{}%
\providecommand \url  [0]{\begingroup\@sanitize@url \@url }%
\providecommand \@url [1]{\endgroup\@href {#1}{\urlprefix }}%
\providecommand \urlprefix  [0]{URL }%
\providecommand \Eprint [0]{\href }%
\providecommand \doibase [0]{https://doi.org/}%
\providecommand \selectlanguage [0]{\@gobble}%
\providecommand \bibinfo  [0]{\@secondoftwo}%
\providecommand \bibfield  [0]{\@secondoftwo}%
\providecommand \translation [1]{[#1]}%
\providecommand \BibitemOpen [0]{}%
\providecommand \bibitemStop [0]{}%
\providecommand \bibitemNoStop [0]{.\EOS\space}%
\providecommand \EOS [0]{\spacefactor3000\relax}%
\providecommand \BibitemShut  [1]{\csname bibitem#1\endcsname}%
\let\auto@bib@innerbib\@empty
\bibitem [{\citenamefont {Goldstein}\ and\ \citenamefont {Sela}(2018)}]{goldstein18}%
  \BibitemOpen
  \bibfield  {author} {\bibinfo {author} {\bibfnamefont {M.}~\bibnamefont {Goldstein}}\ and\ \bibinfo {author} {\bibfnamefont {E.}~\bibnamefont {Sela}},\ }\bibfield  {title} {\bibinfo {title} {{Symmetry-Resolved Entanglement in Many-Body Systems}},\ }\href {https://link.aps.org/doi/10.1103/PhysRevLett.120.200602} {\bibfield  {journal} {\bibinfo  {journal} {Phys. Rev. Lett.}\ }\textbf {\bibinfo {volume} {120}},\ \bibinfo {pages} {200602} (\bibinfo {year} {2018})}\BibitemShut {NoStop}%
\bibitem [{\citenamefont {Xavier}\ \emph {et~al.}(2018)\citenamefont {Xavier}, \citenamefont {Alcaraz},\ and\ \citenamefont {Sierra}}]{xavier2018}%
  \BibitemOpen
  \bibfield  {author} {\bibinfo {author} {\bibfnamefont {J.~C.}\ \bibnamefont {Xavier}}, \bibinfo {author} {\bibfnamefont {F.~C.}\ \bibnamefont {Alcaraz}},\ and\ \bibinfo {author} {\bibfnamefont {G.}~\bibnamefont {Sierra}},\ }\bibfield  {title} {\bibinfo {title} {{Equipartition of the Entanglement Entropy}},\ }\href {https://doi.org/10.1103/PhysRevB.98.041106} {\bibfield  {journal} {\bibinfo  {journal} {Phys. Rev. B}\ }\textbf {\bibinfo {volume} {98}},\ \bibinfo {pages} {041106} (\bibinfo {year} {2018})}\BibitemShut {NoStop}%
\bibitem [{\citenamefont {Deutsch}(1991)}]{deutsch91}%
  \BibitemOpen
  \bibfield  {author} {\bibinfo {author} {\bibfnamefont {J.~M.}\ \bibnamefont {Deutsch}},\ }\bibfield  {title} {\bibinfo {title} {Quantum statistical mechanics in a closed system},\ }\href {https://doi.org/10.1103/PhysRevA.43.2046} {\bibfield  {journal} {\bibinfo  {journal} {Phys. Rev. A}\ }\textbf {\bibinfo {volume} {43}},\ \bibinfo {pages} {2046} (\bibinfo {year} {1991})}\BibitemShut {NoStop}%
\bibitem [{\citenamefont {Srednicki}(1994)}]{srednicki94}%
  \BibitemOpen
  \bibfield  {author} {\bibinfo {author} {\bibfnamefont {M.}~\bibnamefont {Srednicki}},\ }\bibfield  {title} {\bibinfo {title} {Chaos and quantum thermalization},\ }\href {https://doi.org/10.1103/PhysRevE.50.888} {\bibfield  {journal} {\bibinfo  {journal} {Phys. Rev. E}\ }\textbf {\bibinfo {volume} {50}},\ \bibinfo {pages} {888} (\bibinfo {year} {1994})}\BibitemShut {NoStop}%
\bibitem [{\citenamefont {Rigol}\ \emph {et~al.}(2008)\citenamefont {Rigol}, \citenamefont {Dunjko},\ and\ \citenamefont {Olshanii}}]{rigol08}%
  \BibitemOpen
  \bibfield  {author} {\bibinfo {author} {\bibfnamefont {M.}~\bibnamefont {Rigol}}, \bibinfo {author} {\bibfnamefont {V.}~\bibnamefont {Dunjko}},\ and\ \bibinfo {author} {\bibfnamefont {M.}~\bibnamefont {Olshanii}},\ }\bibfield  {title} {\bibinfo {title} {Thermalization and its mechanism for generic isolated quantum systems},\ }\href {https://doi.org/10.1038/nature06838} {\bibfield  {journal} {\bibinfo  {journal} {Nature}\ }\textbf {\bibinfo {volume} {452}},\ \bibinfo {pages} {854} (\bibinfo {year} {2008})}\BibitemShut {NoStop}%
\bibitem [{\citenamefont {Rigol}\ \emph {et~al.}(2007)\citenamefont {Rigol}, \citenamefont {Dunjko}, \citenamefont {Yurovsky},\ and\ \citenamefont {Olshanii}}]{rigol07}%
  \BibitemOpen
  \bibfield  {author} {\bibinfo {author} {\bibfnamefont {M.}~\bibnamefont {Rigol}}, \bibinfo {author} {\bibfnamefont {V.}~\bibnamefont {Dunjko}}, \bibinfo {author} {\bibfnamefont {V.}~\bibnamefont {Yurovsky}},\ and\ \bibinfo {author} {\bibfnamefont {M.}~\bibnamefont {Olshanii}},\ }\bibfield  {title} {\bibinfo {title} {Relaxation in a completely integrable many-body quantum system: an ab initio study of the dynamics of the highly excited states of 1d lattice hard-core bosons},\ }\href {https://doi.org/10.1103/PhysRevLett.98.050405} {\bibfield  {journal} {\bibinfo  {journal} {Phys. Rev. Lett.}\ }\textbf {\bibinfo {volume} {98}},\ \bibinfo {pages} {050405} (\bibinfo {year} {2007})}\BibitemShut {NoStop}%
\bibitem [{\citenamefont {Pai}\ \emph {et~al.}(2019)\citenamefont {Pai}, \citenamefont {Pretko},\ and\ \citenamefont {Nandkishore}}]{pai19}%
  \BibitemOpen
  \bibfield  {author} {\bibinfo {author} {\bibfnamefont {S.}~\bibnamefont {Pai}}, \bibinfo {author} {\bibfnamefont {M.}~\bibnamefont {Pretko}},\ and\ \bibinfo {author} {\bibfnamefont {R.~M.}\ \bibnamefont {Nandkishore}},\ }\bibfield  {title} {\bibinfo {title} {Localization in fractonic random circuits},\ }\href {https://doi.org/10.1103/PhysRevX.9.021003} {\bibfield  {journal} {\bibinfo  {journal} {Phys. Rev. X}\ }\textbf {\bibinfo {volume} {9}},\ \bibinfo {pages} {021003} (\bibinfo {year} {2019})}\BibitemShut {NoStop}%
\bibitem [{\citenamefont {Sala}\ \emph {et~al.}(2020)\citenamefont {Sala}, \citenamefont {Rakovszky}, \citenamefont {Verresen}, \citenamefont {Knap},\ and\ \citenamefont {Pollmann}}]{sala2020}%
  \BibitemOpen
  \bibfield  {author} {\bibinfo {author} {\bibfnamefont {P.}~\bibnamefont {Sala}}, \bibinfo {author} {\bibfnamefont {T.}~\bibnamefont {Rakovszky}}, \bibinfo {author} {\bibfnamefont {R.}~\bibnamefont {Verresen}}, \bibinfo {author} {\bibfnamefont {M.}~\bibnamefont {Knap}},\ and\ \bibinfo {author} {\bibfnamefont {F.}~\bibnamefont {Pollmann}},\ }\bibfield  {title} {\bibinfo {title} {{Ergodicity Breaking Arising from Hilbert Space Fragmentation in Dipole-Conserving Hamiltonians}},\ }\href {https://doi.org/10.1103/PhysRevX.10.011047} {\bibfield  {journal} {\bibinfo  {journal} {Phys. Rev. X}\ }\textbf {\bibinfo {volume} {10}},\ \bibinfo {pages} {011047} (\bibinfo {year} {2020})}\BibitemShut {NoStop}%
\bibitem [{\citenamefont {Khemani}\ \emph {et~al.}(2020)\citenamefont {Khemani}, \citenamefont {Hermele},\ and\ \citenamefont {Nandkishore}}]{khemani20}%
  \BibitemOpen
  \bibfield  {author} {\bibinfo {author} {\bibfnamefont {V.}~\bibnamefont {Khemani}}, \bibinfo {author} {\bibfnamefont {M.}~\bibnamefont {Hermele}},\ and\ \bibinfo {author} {\bibfnamefont {R.}~\bibnamefont {Nandkishore}},\ }\bibfield  {title} {\bibinfo {title} {{Localization from Hilbert space shattering: From theory to physical realizations}},\ }\href {https://doi.org/10.1103/PhysRevB.101.174204} {\bibfield  {journal} {\bibinfo  {journal} {Phys. Rev. B}\ }\textbf {\bibinfo {volume} {101}},\ \bibinfo {pages} {174204} (\bibinfo {year} {2020})}\BibitemShut {NoStop}%
\bibitem [{\citenamefont {Moudgalya}\ \emph {et~al.}()\citenamefont {Moudgalya}, \citenamefont {Prem}, \citenamefont {Nandkishore}, \citenamefont {Regnault},\ and\ \citenamefont {Bernevig}}]{moudgalya21m}%
  \BibitemOpen
  \bibfield  {author} {\bibinfo {author} {\bibfnamefont {S.}~\bibnamefont {Moudgalya}}, \bibinfo {author} {\bibfnamefont {A.}~\bibnamefont {Prem}}, \bibinfo {author} {\bibfnamefont {R.}~\bibnamefont {Nandkishore}}, \bibinfo {author} {\bibfnamefont {N.}~\bibnamefont {Regnault}},\ and\ \bibinfo {author} {\bibfnamefont {B.~A.}\ \bibnamefont {Bernevig}},\ }\bibfield  {title} {\bibinfo {title} {{Thermalization and its absence within Krylov sub- spaces of a constrained Hamiltonian}},\ }\href {https://doi.org/10.1142/9789811231711_0009} {\bibinfo  {journal} {in Memorial Volume for Shoucheng Zhang (World Scientific, Singapore, 2021)}\ ,\ \bibinfo {pages} {Chap. 7, pp. 147–209}}\BibitemShut {NoStop}%
\bibitem [{\citenamefont {Adler}\ \emph {et~al.}(2024)\citenamefont {Adler}, \citenamefont {Wei}, \citenamefont {Will}, \citenamefont {Srakaew}, \citenamefont {Agrawal}, \citenamefont {Weckesser}, \citenamefont {Moessner}, \citenamefont {Pollmann}, \citenamefont {Bloch},\ and\ \citenamefont {Zeiher}}]{adler24}%
  \BibitemOpen
\bibfield  {journal} {  }\bibfield  {author} {\bibinfo {author} {\bibfnamefont {D.}~\bibnamefont {Adler}}, \bibinfo {author} {\bibfnamefont {D.}~\bibnamefont {Wei}}, \bibinfo {author} {\bibfnamefont {M.}~\bibnamefont {Will}}, \bibinfo {author} {\bibfnamefont {K.}~\bibnamefont {Srakaew}}, \bibinfo {author} {\bibfnamefont {S.}~\bibnamefont {Agrawal}}, \bibinfo {author} {\bibfnamefont {P.}~\bibnamefont {Weckesser}}, \bibinfo {author} {\bibfnamefont {R.}~\bibnamefont {Moessner}}, \bibinfo {author} {\bibfnamefont {F.}~\bibnamefont {Pollmann}}, \bibinfo {author} {\bibfnamefont {I.}~\bibnamefont {Bloch}},\ and\ \bibinfo {author} {\bibfnamefont {J.}~\bibnamefont {Zeiher}},\ }\bibfield  {title} {\bibinfo {title} {{Observation of Hilbert space fragmentation and fractonic excitations in 2D}},\ }\href {https://doi.org/10.1038/s41586-024-08188-0} {\bibfield  {journal} {\bibinfo  {journal} {Nature}\ }\textbf {\bibinfo {volume} {636}},\ \bibinfo {pages} {80} (\bibinfo {year} {2024})}\BibitemShut {NoStop}%
\bibitem [{\citenamefont {Scherg}\ \emph {et~al.}(2021)\citenamefont {Scherg}, \citenamefont {Kohlert}, \citenamefont {Sala}, \citenamefont {Pollmann}, \citenamefont {Madhusudhana}, \citenamefont {Bloch},\ and\ \citenamefont {Aidelsburger}}]{scherg21}%
  \BibitemOpen
  \bibfield  {author} {\bibinfo {author} {\bibfnamefont {S.}~\bibnamefont {Scherg}}, \bibinfo {author} {\bibfnamefont {T.}~\bibnamefont {Kohlert}}, \bibinfo {author} {\bibfnamefont {P.}~\bibnamefont {Sala}}, \bibinfo {author} {\bibfnamefont {F.}~\bibnamefont {Pollmann}}, \bibinfo {author} {\bibfnamefont {B.~H.}\ \bibnamefont {Madhusudhana}}, \bibinfo {author} {\bibfnamefont {I.}~\bibnamefont {Bloch}},\ and\ \bibinfo {author} {\bibfnamefont {M.}~\bibnamefont {Aidelsburger}},\ }\bibfield  {title} {\bibinfo {title} {{Observing non-ergodicity due to kinetic constraints in tilted Fermi-Hubbard chains}},\ }\href {https://doi.org/10.1038/s41467-021-24726-0} {\bibfield  {journal} {\bibinfo  {journal} {Nat. Commun.}\ }\textbf {\bibinfo {volume} {12}},\ \bibinfo {pages} {4490} (\bibinfo {year} {2021})}\BibitemShut {NoStop}%
\bibitem [{\citenamefont {Wang}\ \emph {et~al.}(2025)\citenamefont {Wang}, \citenamefont {Shi}, \citenamefont {Sun} \emph {et~al.}}]{wang25}%
  \BibitemOpen
  \bibfield  {author} {\bibinfo {author} {\bibfnamefont {Y.-Y.}\ \bibnamefont {Wang}}, \bibinfo {author} {\bibfnamefont {Y.-H.}\ \bibnamefont {Shi}}, \bibinfo {author} {\bibfnamefont {Z.-H.}\ \bibnamefont {Sun}}, \emph {et~al.},\ }\bibfield  {title} {\bibinfo {title} {{Exploring Hilbert-space fragmentation on a superconducting processor}},\ }\href {https://doi.org/10.1103/PRXQuantum.6.010325} {\bibfield  {journal} {\bibinfo  {journal} {PRX Quantum}\ }\textbf {\bibinfo {volume} {6}},\ \bibinfo {pages} {010325} (\bibinfo {year} {2025})}\BibitemShut {NoStop}%
\bibitem [{\citenamefont {Q.~Guo}\ \emph {et~al.}(2021)\citenamefont {Q.~Guo} \emph {et~al.}}]{guo21}%
  \BibitemOpen
  \bibfield  {author} {\bibinfo {author} {\bibfnamefont {H.~L.}\ \bibnamefont {Q.~Guo}, \bibfnamefont {C.~Cheng}} \emph {et~al.},\ }\bibfield  {title} {\bibinfo {title} {{Stark Many-Body Localization on a Superconducting Quantum Processor}},\ }\href {https://doi.org/10.1103/PhysRevLett.127.240502} {\bibfield  {journal} {\bibinfo  {journal} {Phys. Rev. Lett.}\ }\textbf {\bibinfo {volume} {127}},\ \bibinfo {pages} {240502} (\bibinfo {year} {2021})}\BibitemShut {NoStop}%
\bibitem [{\citenamefont {Zhao}\ \emph {et~al.}(2025)\citenamefont {Zhao}, \citenamefont {Datla}, \citenamefont {Tian}, \citenamefont {Aliyu},\ and\ \citenamefont {Loh}}]{zhao25obshilb}%
  \BibitemOpen
  \bibfield  {author} {\bibinfo {author} {\bibfnamefont {L.}~\bibnamefont {Zhao}}, \bibinfo {author} {\bibfnamefont {P.~R.}\ \bibnamefont {Datla}}, \bibinfo {author} {\bibfnamefont {W.}~\bibnamefont {Tian}}, \bibinfo {author} {\bibfnamefont {M.~M.}\ \bibnamefont {Aliyu}},\ and\ \bibinfo {author} {\bibfnamefont {H.}~\bibnamefont {Loh}},\ }\bibfield  {title} {\bibinfo {title} {{Observation of Quantum Thermalization Restricted to Hilbert Space Fragments and $\mathbb{Z}_{2k}$ Scars}},\ }\href {https://doi.org/10.1103/PhysRevX.15.011035} {\bibfield  {journal} {\bibinfo  {journal} {Phys. Rev. X}\ }\textbf {\bibinfo {volume} {15}},\ \bibinfo {pages} {011035} (\bibinfo {year} {2025})}\BibitemShut {NoStop}%
\bibitem [{\citenamefont {Moudgalya}\ and\ \citenamefont {Motrunich}(2022)}]{moudgalya2022commutant}%
  \BibitemOpen
  \bibfield  {author} {\bibinfo {author} {\bibfnamefont {S.}~\bibnamefont {Moudgalya}}\ and\ \bibinfo {author} {\bibfnamefont {O.~I.}\ \bibnamefont {Motrunich}},\ }\bibfield  {title} {\bibinfo {title} {{Hilbert Space Fragmentation and Commutant Algebras}},\ }\href {https://doi.org/10.1103/PhysRevX.12.011050} {\bibfield  {journal} {\bibinfo  {journal} {Phys. Rev. X}\ }\textbf {\bibinfo {volume} {12}},\ \bibinfo {pages} {011050} (\bibinfo {year} {2022})}\BibitemShut {NoStop}%
\bibitem [{\citenamefont {Li}\ \emph {et~al.}(2023)\citenamefont {Li}, \citenamefont {Sala},\ and\ \citenamefont {Pollmann}}]{pablo1}%
  \BibitemOpen
  \bibfield  {author} {\bibinfo {author} {\bibfnamefont {Y.}~\bibnamefont {Li}}, \bibinfo {author} {\bibfnamefont {P.}~\bibnamefont {Sala}},\ and\ \bibinfo {author} {\bibfnamefont {F.}~\bibnamefont {Pollmann}},\ }\bibfield  {title} {\bibinfo {title} {Hilbert space fragmentation in open quantum systems},\ }\href {https://doi.org/10.1103/PhysRevResearch.5.043239} {\bibfield  {journal} {\bibinfo  {journal} {Phys. Rev. Res.}\ }\textbf {\bibinfo {volume} {5}},\ \bibinfo {pages} {043239} (\bibinfo {year} {2023})}\BibitemShut {NoStop}%
\bibitem [{\citenamefont {Li}\ \emph {et~al.}(2025)\citenamefont {Li}, \citenamefont {Pollmann}, \citenamefont {Read},\ and\ \citenamefont {Sala}}]{pablo2}%
  \BibitemOpen
  \bibfield  {author} {\bibinfo {author} {\bibfnamefont {Y.}~\bibnamefont {Li}}, \bibinfo {author} {\bibfnamefont {F.}~\bibnamefont {Pollmann}}, \bibinfo {author} {\bibfnamefont {N.}~\bibnamefont {Read}},\ and\ \bibinfo {author} {\bibfnamefont {P.}~\bibnamefont {Sala}},\ }\bibfield  {title} {\bibinfo {title} {Highly entangled stationary states from strong symmetries},\ }\href {https://doi.org/10.1103/PhysRevX.15.011068} {\bibfield  {journal} {\bibinfo  {journal} {Phys. Rev. X}\ }\textbf {\bibinfo {volume} {15}},\ \bibinfo {pages} {011068} (\bibinfo {year} {2025})}\BibitemShut {NoStop}%
\bibitem [{\citenamefont {Ares}\ \emph {et~al.}(2023{\natexlab{a}})\citenamefont {Ares}, \citenamefont {Murciano},\ and\ \citenamefont {Calabrese}}]{ares2023asymmetry}%
  \BibitemOpen
  \bibfield  {author} {\bibinfo {author} {\bibfnamefont {F.}~\bibnamefont {Ares}}, \bibinfo {author} {\bibfnamefont {S.}~\bibnamefont {Murciano}},\ and\ \bibinfo {author} {\bibfnamefont {P.}~\bibnamefont {Calabrese}},\ }\bibfield  {title} {\bibinfo {title} {Entanglement asymmetry as a probe of symmetry breaking},\ }\href {https://doi.org/https://doi.org/10.1038/s41467-023-37747-8} {\bibfield  {journal} {\bibinfo  {journal} {Nat. Commun.}\ }\textbf {\bibinfo {volume} {14}},\ \bibinfo {pages} {2036} (\bibinfo {year} {2023}{\natexlab{a}})}\BibitemShut {NoStop}%
\bibitem [{\citenamefont {Joshi}\ \emph {et~al.}(2024)\citenamefont {Joshi}, \citenamefont {Franke}, \citenamefont {Rath}, \citenamefont {Ares}, \citenamefont {Murciano}, \citenamefont {Kranzl}, \citenamefont {Blatt}, \citenamefont {Zoller}, \citenamefont {Vermersch}, \citenamefont {Calabrese}, \citenamefont {Roos},\ and\ \citenamefont {Joshi}}]{joshi2024observing}%
  \BibitemOpen
  \bibfield  {author} {\bibinfo {author} {\bibfnamefont {L.~K.}\ \bibnamefont {Joshi}}, \bibinfo {author} {\bibfnamefont {J.}~\bibnamefont {Franke}}, \bibinfo {author} {\bibfnamefont {A.}~\bibnamefont {Rath}}, \bibinfo {author} {\bibfnamefont {F.}~\bibnamefont {Ares}}, \bibinfo {author} {\bibfnamefont {S.}~\bibnamefont {Murciano}}, \bibinfo {author} {\bibfnamefont {F.}~\bibnamefont {Kranzl}}, \bibinfo {author} {\bibfnamefont {R.}~\bibnamefont {Blatt}}, \bibinfo {author} {\bibfnamefont {P.}~\bibnamefont {Zoller}}, \bibinfo {author} {\bibfnamefont {B.}~\bibnamefont {Vermersch}}, \bibinfo {author} {\bibfnamefont {P.}~\bibnamefont {Calabrese}}, \bibinfo {author} {\bibfnamefont {C.~F.}\ \bibnamefont {Roos}},\ and\ \bibinfo {author} {\bibfnamefont {M.~K.}\ \bibnamefont {Joshi}},\ }\bibfield  {title} {\bibinfo {title} {{Observing the Quantum Mpemba Effect in Quantum Simulations}},\ }\href {https://doi.org/10.1103/PhysRevLett.133.010402} {\bibfield  {journal} {\bibinfo  {journal} {Phys. Rev. Lett.}\ }\textbf
  {\bibinfo {volume} {133}},\ \bibinfo {pages} {010402} (\bibinfo {year} {2024})}\BibitemShut {NoStop}%
\bibitem [{\citenamefont {Rylands}\ \emph {et~al.}(2024{\natexlab{a}})\citenamefont {Rylands}, \citenamefont {Klobas}, \citenamefont {Ares}, \citenamefont {Calabrese}, \citenamefont {Murciano},\ and\ \citenamefont {Bertini}}]{rylands2024microscopic}%
  \BibitemOpen
  \bibfield  {author} {\bibinfo {author} {\bibfnamefont {C.}~\bibnamefont {Rylands}}, \bibinfo {author} {\bibfnamefont {K.}~\bibnamefont {Klobas}}, \bibinfo {author} {\bibfnamefont {F.}~\bibnamefont {Ares}}, \bibinfo {author} {\bibfnamefont {P.}~\bibnamefont {Calabrese}}, \bibinfo {author} {\bibfnamefont {S.}~\bibnamefont {Murciano}},\ and\ \bibinfo {author} {\bibfnamefont {B.}~\bibnamefont {Bertini}},\ }\bibfield  {title} {\bibinfo {title} {{Microscopic Origin of the Quantum Mpemba Effect in Integrable Systems}},\ }\href {https://doi.org/10.1103/PhysRevLett.133.010401} {\bibfield  {journal} {\bibinfo  {journal} {Phys. Rev. Lett.}\ }\textbf {\bibinfo {volume} {133}},\ \bibinfo {pages} {010401} (\bibinfo {year} {2024}{\natexlab{a}})}\BibitemShut {NoStop}%
\bibitem [{\citenamefont {Ares}\ \emph {et~al.}(2025{\natexlab{a}})\citenamefont {Ares}, \citenamefont {Calabrese},\ and\ \citenamefont {Murciano}}]{ares25}%
  \BibitemOpen
  \bibfield  {author} {\bibinfo {author} {\bibfnamefont {F.}~\bibnamefont {Ares}}, \bibinfo {author} {\bibfnamefont {P.}~\bibnamefont {Calabrese}},\ and\ \bibinfo {author} {\bibfnamefont {S.}~\bibnamefont {Murciano}},\ }\bibfield  {title} {\bibinfo {title} {{The quantum Mpemba effects}},\ }\href {https://doi.org/10.1038/s42254-025-00838-0} {\bibfield  {journal} {\bibinfo  {journal} {Nat. Rev. Phys.}\ }\textbf {\bibinfo {volume} {7}},\ \bibinfo {pages} {451} (\bibinfo {year} {2025}{\natexlab{a}})}\BibitemShut {NoStop}%
\bibitem [{\citenamefont {Teza}\ \emph {et~al.}(2026)\citenamefont {Teza}, \citenamefont {Bechhoefer}, \citenamefont {Lasanta}, \citenamefont {Raz},\ and\ \citenamefont {Vucelja}}]{teza26}%
  \BibitemOpen
  \bibfield  {author} {\bibinfo {author} {\bibfnamefont {G.}~\bibnamefont {Teza}}, \bibinfo {author} {\bibfnamefont {J.}~\bibnamefont {Bechhoefer}}, \bibinfo {author} {\bibfnamefont {A.}~\bibnamefont {Lasanta}}, \bibinfo {author} {\bibfnamefont {O.}~\bibnamefont {Raz}},\ and\ \bibinfo {author} {\bibfnamefont {M.}~\bibnamefont {Vucelja}},\ }\bibfield  {title} {\bibinfo {title} {Speedups in nonequilibrium thermal relaxation: Mpemba and related effects},\ }\href {https://doi.org/10.1016/j.physrep.2025.10.009} {\bibfield  {journal} {\bibinfo  {journal} {Phys. Rep.}\ }\textbf {\bibinfo {volume} {1164}},\ \bibinfo {pages} {1} (\bibinfo {year} {2026})}\BibitemShut {NoStop}%
\bibitem [{\citenamefont {Bartlett}\ \emph {et~al.}(2007)\citenamefont {Bartlett}, \citenamefont {Rudolph},\ and\ \citenamefont {Spekkens}}]{bartlett07}%
  \BibitemOpen
  \bibfield  {author} {\bibinfo {author} {\bibfnamefont {S.~D.}\ \bibnamefont {Bartlett}}, \bibinfo {author} {\bibfnamefont {T.}~\bibnamefont {Rudolph}},\ and\ \bibinfo {author} {\bibfnamefont {R.~W.}\ \bibnamefont {Spekkens}},\ }\bibfield  {title} {\bibinfo {title} {{Reference frames, superselection rules, and quantum information}},\ }\href {https://doi.org/10.1103/RevModPhys.79.555} {\bibfield  {journal} {\bibinfo  {journal} {Rev. Mod. Phys.}\ }\textbf {\bibinfo {volume} {79}},\ \bibinfo {pages} {555} (\bibinfo {year} {2007})}\BibitemShut {NoStop}%
\bibitem [{\citenamefont {Vaccaro}\ \emph {et~al.}(2008)\citenamefont {Vaccaro}, \citenamefont {Anselmi}, \citenamefont {Wiseman},\ and\ \citenamefont {Jacobs}}]{vaccaro2008tradeoff}%
  \BibitemOpen
  \bibfield  {author} {\bibinfo {author} {\bibfnamefont {J.~A.}\ \bibnamefont {Vaccaro}}, \bibinfo {author} {\bibfnamefont {F.}~\bibnamefont {Anselmi}}, \bibinfo {author} {\bibfnamefont {H.~M.}\ \bibnamefont {Wiseman}},\ and\ \bibinfo {author} {\bibfnamefont {K.}~\bibnamefont {Jacobs}},\ }\bibfield  {title} {\bibinfo {title} {Tradeoff between extractable mechanical work, accessible entanglement, and ability to act as a reference system, under arbitrary superselection rules},\ }\href {https://doi.org/10.1103/PhysRevA.77.032114} {\bibfield  {journal} {\bibinfo  {journal} {Phys. Rev. A}\ }\textbf {\bibinfo {volume} {77}},\ \bibinfo {pages} {032114} (\bibinfo {year} {2008})}\BibitemShut {NoStop}%
\bibitem [{\citenamefont {Gour}\ \emph {et~al.}(2009)\citenamefont {Gour}, \citenamefont {Marvian},\ and\ \citenamefont {Spekkens}}]{gour2009measuring}%
  \BibitemOpen
  \bibfield  {author} {\bibinfo {author} {\bibfnamefont {G.}~\bibnamefont {Gour}}, \bibinfo {author} {\bibfnamefont {I.}~\bibnamefont {Marvian}},\ and\ \bibinfo {author} {\bibfnamefont {R.~W.}\ \bibnamefont {Spekkens}},\ }\bibfield  {title} {\bibinfo {title} {{Measuring the quality of a quantum reference frame: The relative entropy of frameness}},\ }\href {https://doi.org/10.1103/PhysRevA.80.012307} {\bibfield  {journal} {\bibinfo  {journal} {Phys. Rev. A}\ }\textbf {\bibinfo {volume} {80}},\ \bibinfo {pages} {012307} (\bibinfo {year} {2009})}\BibitemShut {NoStop}%
\bibitem [{\citenamefont {Marvian}\ and\ \citenamefont {Spekkens}(2014)}]{marvian2014extending}%
  \BibitemOpen
  \bibfield  {author} {\bibinfo {author} {\bibfnamefont {I.}~\bibnamefont {Marvian}}\ and\ \bibinfo {author} {\bibfnamefont {R.~W.}\ \bibnamefont {Spekkens}},\ }\bibfield  {title} {\bibinfo {title} {{Extending Noether’s theorem by quantifying the asymmetry of quantum states}},\ }\href {https://doi.org/10.1038/ncomms4821} {\bibfield  {journal} {\bibinfo  {journal} {Nature Comm.}\ }\textbf {\bibinfo {volume} {5}},\ \bibinfo {pages} {3821} (\bibinfo {year} {2014})}\BibitemShut {NoStop}%
\bibitem [{\citenamefont {Chitambar}\ and\ \citenamefont {Gour}(2019)}]{chitambar2019quantum}%
  \BibitemOpen
  \bibfield  {author} {\bibinfo {author} {\bibfnamefont {E.}~\bibnamefont {Chitambar}}\ and\ \bibinfo {author} {\bibfnamefont {G.}~\bibnamefont {Gour}},\ }\bibfield  {title} {\bibinfo {title} {Quantum resource theories},\ }\href {https://doi.org/10.1103/RevModPhys.91.025001} {\bibfield  {journal} {\bibinfo  {journal} {Rev. Mod. Phys.}\ }\textbf {\bibinfo {volume} {91}},\ \bibinfo {pages} {025001} (\bibinfo {year} {2019})}\BibitemShut {NoStop}%
\bibitem [{\citenamefont {Tarabunga}\ \emph {et~al.}(2025)\citenamefont {Tarabunga}, \citenamefont {Frau}, \citenamefont {Haug}, \citenamefont {Tirrito},\ and\ \citenamefont {Piroli}}]{tarabunga25}%
  \BibitemOpen
  \bibfield  {author} {\bibinfo {author} {\bibfnamefont {P.~S.}\ \bibnamefont {Tarabunga}}, \bibinfo {author} {\bibfnamefont {M.}~\bibnamefont {Frau}}, \bibinfo {author} {\bibfnamefont {T.}~\bibnamefont {Haug}}, \bibinfo {author} {\bibfnamefont {E.}~\bibnamefont {Tirrito}},\ and\ \bibinfo {author} {\bibfnamefont {L.}~\bibnamefont {Piroli}},\ }\bibfield  {title} {\bibinfo {title} {A nonstabilizerness monotone from stabilizerness asymmetry},\ }\href {https://doi.org/10.1088/2058-9565/adfd0d} {\bibfield  {journal} {\bibinfo  {journal} {Quantum Sci. Technol.}\ }\textbf {\bibinfo {volume} {10}},\ \bibinfo {pages} {045026} (\bibinfo {year} {2025})}\BibitemShut {NoStop}%
\bibitem [{\citenamefont {Aditya}\ \emph {et~al.}()\citenamefont {Aditya}, \citenamefont {Summer}, \citenamefont {Sierant},\ and\ \citenamefont {Turkeshi}}]{aditya2025}%
  \BibitemOpen
  \bibfield  {author} {\bibinfo {author} {\bibfnamefont {S.}~\bibnamefont {Aditya}}, \bibinfo {author} {\bibfnamefont {A.}~\bibnamefont {Summer}}, \bibinfo {author} {\bibfnamefont {P.}~\bibnamefont {Sierant}},\ and\ \bibinfo {author} {\bibfnamefont {X.}~\bibnamefont {Turkeshi}},\ }\bibfield  {title} {\bibinfo {title} {{Mpemba Effects in Quantum Complexity}},\ }\href {https://arxiv.org/abs/2509.22176} {\ }\Eprint {https://arxiv.org/abs/2509.22176} {arXiv:2509.22176} \BibitemShut {NoStop}%
\bibitem [{\citenamefont {Mazzoni}\ \emph {et~al.}()\citenamefont {Mazzoni}, \citenamefont {Capizzi},\ and\ \citenamefont {Piroli}}]{mazzoni2025breaking}%
  \BibitemOpen
  \bibfield  {author} {\bibinfo {author} {\bibfnamefont {M.}~\bibnamefont {Mazzoni}}, \bibinfo {author} {\bibfnamefont {L.}~\bibnamefont {Capizzi}},\ and\ \bibinfo {author} {\bibfnamefont {L.}~\bibnamefont {Piroli}},\ }\bibfield  {title} {\bibinfo {title} {Breaking global symmetries with locality-preserving operations},\ }\href {https://arxiv.org/abs/2508.15892} {\ }\Eprint {https://arxiv.org/abs/2508.15892} {arXiv:2508.15892} \BibitemShut {NoStop}%
\bibitem [{\citenamefont {Garnerone}\ \emph {et~al.}(2010{\natexlab{a}})\citenamefont {Garnerone}, \citenamefont {de~Oliveira},\ and\ \citenamefont {Zanardi}}]{rmps1}%
  \BibitemOpen
  \bibfield  {author} {\bibinfo {author} {\bibfnamefont {S.}~\bibnamefont {Garnerone}}, \bibinfo {author} {\bibfnamefont {T.~R.}\ \bibnamefont {de~Oliveira}},\ and\ \bibinfo {author} {\bibfnamefont {P.}~\bibnamefont {Zanardi}},\ }\bibfield  {title} {\bibinfo {title} {Typicality in random matrix product states},\ }\href {https://doi.org/10.1103/PhysRevA.81.032336} {\bibfield  {journal} {\bibinfo  {journal} {Phys. Rev. A}\ }\textbf {\bibinfo {volume} {81}},\ \bibinfo {pages} {032336} (\bibinfo {year} {2010}{\natexlab{a}})}\BibitemShut {NoStop}%
\bibitem [{\citenamefont {Garnerone}\ \emph {et~al.}(2010{\natexlab{b}})\citenamefont {Garnerone}, \citenamefont {de~Oliveira}, \citenamefont {Haas},\ and\ \citenamefont {Zanardi}}]{rmps2}%
  \BibitemOpen
  \bibfield  {author} {\bibinfo {author} {\bibfnamefont {S.}~\bibnamefont {Garnerone}}, \bibinfo {author} {\bibfnamefont {T.~R.}\ \bibnamefont {de~Oliveira}}, \bibinfo {author} {\bibfnamefont {S.}~\bibnamefont {Haas}},\ and\ \bibinfo {author} {\bibfnamefont {P.}~\bibnamefont {Zanardi}},\ }\bibfield  {title} {\bibinfo {title} {Statistical properties of random matrix product states},\ }\href {https://doi.org/10.1103/PhysRevA.82.052312} {\bibfield  {journal} {\bibinfo  {journal} {Phys. Rev. A}\ }\textbf {\bibinfo {volume} {82}},\ \bibinfo {pages} {052312} (\bibinfo {year} {2010}{\natexlab{b}})}\BibitemShut {NoStop}%
\bibitem [{\citenamefont {Haag}\ \emph {et~al.}(2023{\natexlab{a}})\citenamefont {Haag}, \citenamefont {Baccari},\ and\ \citenamefont {Styliaris}}]{rmps3}%
  \BibitemOpen
  \bibfield  {author} {\bibinfo {author} {\bibfnamefont {D.}~\bibnamefont {Haag}}, \bibinfo {author} {\bibfnamefont {F.}~\bibnamefont {Baccari}},\ and\ \bibinfo {author} {\bibfnamefont {G.}~\bibnamefont {Styliaris}},\ }\bibfield  {title} {\bibinfo {title} {{Typical Correlation Length of Sequentially Generated Tensor Network States}},\ }\href {https://doi.org/10.1103/PRXQuantum.4.030330} {\bibfield  {journal} {\bibinfo  {journal} {PRX Quantum}\ }\textbf {\bibinfo {volume} {4}},\ \bibinfo {pages} {030330} (\bibinfo {year} {2023}{\natexlab{a}})}\BibitemShut {NoStop}%
\bibitem [{\citenamefont {Haferkamp}\ \emph {et~al.}(2021)\citenamefont {Haferkamp}, \citenamefont {Bertoni}, \citenamefont {Roth},\ and\ \citenamefont {Eisert}}]{rmps4}%
  \BibitemOpen
  \bibfield  {author} {\bibinfo {author} {\bibfnamefont {J.}~\bibnamefont {Haferkamp}}, \bibinfo {author} {\bibfnamefont {C.}~\bibnamefont {Bertoni}}, \bibinfo {author} {\bibfnamefont {I.}~\bibnamefont {Roth}},\ and\ \bibinfo {author} {\bibfnamefont {J.}~\bibnamefont {Eisert}},\ }\bibfield  {title} {\bibinfo {title} {{Emergent Statistical Mechanics from Properties of Disordered Random Matrix Product States}},\ }\href {https://doi.org/10.1103/PRXQuantum.2.040308} {\bibfield  {journal} {\bibinfo  {journal} {PRX Quantum}\ }\textbf {\bibinfo {volume} {2}},\ \bibinfo {pages} {040308} (\bibinfo {year} {2021})}\BibitemShut {NoStop}%
\bibitem [{\citenamefont {Lami}\ \emph {et~al.}(2025{\natexlab{a}})\citenamefont {Lami}, \citenamefont {Haug},\ and\ \citenamefont {De~Nardis}}]{rmps5}%
  \BibitemOpen
  \bibfield  {author} {\bibinfo {author} {\bibfnamefont {G.}~\bibnamefont {Lami}}, \bibinfo {author} {\bibfnamefont {T.}~\bibnamefont {Haug}},\ and\ \bibinfo {author} {\bibfnamefont {J.}~\bibnamefont {De~Nardis}},\ }\bibfield  {title} {\bibinfo {title} {{Quantum State Designs with Clifford-Enhanced Matrix Product States}},\ }\href {https://doi.org/10.1103/PRXQuantum.6.010345} {\bibfield  {journal} {\bibinfo  {journal} {PRX Quantum}\ }\textbf {\bibinfo {volume} {6}},\ \bibinfo {pages} {010345} (\bibinfo {year} {2025}{\natexlab{a}})}\BibitemShut {NoStop}%
\bibitem [{\citenamefont {Lancien}\ and\ \citenamefont {Pérez-García}(2021)}]{rmps6}%
  \BibitemOpen
  \bibfield  {author} {\bibinfo {author} {\bibfnamefont {C.}~\bibnamefont {Lancien}}\ and\ \bibinfo {author} {\bibfnamefont {D.}~\bibnamefont {Pérez-García}},\ }\bibfield  {title} {\bibinfo {title} {{Correlation Length in Random MPS and PEPS}},\ }\href {https://doi.org/10.1007/s00023-021-01087-4} {\bibfield  {journal} {\bibinfo  {journal} {Annales Henri Poincaré}\ }\textbf {\bibinfo {volume} {23}},\ \bibinfo {pages} {141–222} (\bibinfo {year} {2021})}\BibitemShut {NoStop}%
\bibitem [{\citenamefont {Lami}\ \emph {et~al.}(2025{\natexlab{b}})\citenamefont {Lami}, \citenamefont {De~Nardis},\ and\ \citenamefont {Turkeshi}}]{rmps7}%
  \BibitemOpen
  \bibfield  {author} {\bibinfo {author} {\bibfnamefont {G.}~\bibnamefont {Lami}}, \bibinfo {author} {\bibfnamefont {J.}~\bibnamefont {De~Nardis}},\ and\ \bibinfo {author} {\bibfnamefont {X.}~\bibnamefont {Turkeshi}},\ }\bibfield  {title} {\bibinfo {title} {{Anticoncentration and State Design of Random Tensor Networks}},\ }\href {https://doi.org/10.1103/PhysRevLett.134.010401} {\bibfield  {journal} {\bibinfo  {journal} {Phys. Rev. Lett.}\ }\textbf {\bibinfo {volume} {134}},\ \bibinfo {pages} {010401} (\bibinfo {year} {2025}{\natexlab{b}})}\BibitemShut {NoStop}%
\bibitem [{\citenamefont {Dowling}\ \emph {et~al.}()\citenamefont {Dowling}, \citenamefont {De~Nardis}, \citenamefont {Heinrich}, \citenamefont {Turkeshi},\ and\ \citenamefont {Pappalardi}}]{dowling2025}%
  \BibitemOpen
  \bibfield  {author} {\bibinfo {author} {\bibfnamefont {N.}~\bibnamefont {Dowling}}, \bibinfo {author} {\bibfnamefont {J.}~\bibnamefont {De~Nardis}}, \bibinfo {author} {\bibfnamefont {M.}~\bibnamefont {Heinrich}}, \bibinfo {author} {\bibfnamefont {X.}~\bibnamefont {Turkeshi}},\ and\ \bibinfo {author} {\bibfnamefont {S.}~\bibnamefont {Pappalardi}},\ }\bibfield  {title} {\bibinfo {title} {Free independence and unitary design from random matrix product unitaries},\ }\href {https://arxiv.org/abs/2508.00051} {\ }\Eprint {https://arxiv.org/abs/2508.00051} {arXiv:2508.00051} \BibitemShut {NoStop}%
\bibitem [{\citenamefont {Magni}\ \emph {et~al.}(2025)\citenamefont {Magni}, \citenamefont {Christopoulos}, \citenamefont {De~Luca},\ and\ \citenamefont {Turkeshi}}]{magni2025}%
  \BibitemOpen
  \bibfield  {author} {\bibinfo {author} {\bibfnamefont {B.}~\bibnamefont {Magni}}, \bibinfo {author} {\bibfnamefont {A.}~\bibnamefont {Christopoulos}}, \bibinfo {author} {\bibfnamefont {A.}~\bibnamefont {De~Luca}},\ and\ \bibinfo {author} {\bibfnamefont {X.}~\bibnamefont {Turkeshi}},\ }\bibfield  {title} {\bibinfo {title} {{Anticoncentration in Clifford Circuits and Beyond: From Random Tensor Networks to Pseudomagic States}},\ }\href {https://doi.org/10.1103/p8dn-glcw} {\bibfield  {journal} {\bibinfo  {journal} {Phys. Rev. X}\ }\textbf {\bibinfo {volume} {15}},\ \bibinfo {pages} {031071} (\bibinfo {year} {2025})}\BibitemShut {NoStop}%
\bibitem [{\citenamefont {Sierant}\ \emph {et~al.}(2026)\citenamefont {Sierant}, \citenamefont {Stornati},\ and\ \citenamefont {Turkeshi}}]{Sierant2026}%
  \BibitemOpen
  \bibfield  {author} {\bibinfo {author} {\bibfnamefont {P.}~\bibnamefont {Sierant}}, \bibinfo {author} {\bibfnamefont {P.}~\bibnamefont {Stornati}},\ and\ \bibinfo {author} {\bibfnamefont {X.}~\bibnamefont {Turkeshi}},\ }\bibfield  {title} {\bibinfo {title} {Fermionic magic resources of quantum many-body systems},\ }\href {https://doi.org/10.1103/3yx4-1j27} {\bibfield  {journal} {\bibinfo  {journal} {PRX Quantum}\ }\textbf {\bibinfo {volume} {7}},\ \bibinfo {pages} {010302} (\bibinfo {year} {2026})}\BibitemShut {NoStop}%
\bibitem [{\citenamefont {Lami}\ \emph {et~al.}()\citenamefont {Lami}, \citenamefont {De~Luca}, \citenamefont {Turkeshi},\ and\ \citenamefont {De~Nardis}}]{lami2025}%
  \BibitemOpen
  \bibfield  {author} {\bibinfo {author} {\bibfnamefont {G.}~\bibnamefont {Lami}}, \bibinfo {author} {\bibfnamefont {A.}~\bibnamefont {De~Luca}}, \bibinfo {author} {\bibfnamefont {X.}~\bibnamefont {Turkeshi}},\ and\ \bibinfo {author} {\bibfnamefont {J.}~\bibnamefont {De~Nardis}},\ }\bibfield  {title} {\bibinfo {title} {{Quantum State Design and Emergent Confinement Mechanism in Measured Tensor Network States}},\ }\href {https://arxiv.org/abs/2504.16995} {\ }\Eprint {https://arxiv.org/abs/2504.16995} {arXiv:2504.16995} \BibitemShut {NoStop}%
\bibitem [{\citenamefont {Magni}\ \emph {et~al.}()\citenamefont {Magni}, \citenamefont {Heinrich}, \citenamefont {Leone},\ and\ \citenamefont {Turkeshi}}]{magni2025doped}%
  \BibitemOpen
  \bibfield  {author} {\bibinfo {author} {\bibfnamefont {B.}~\bibnamefont {Magni}}, \bibinfo {author} {\bibfnamefont {M.}~\bibnamefont {Heinrich}}, \bibinfo {author} {\bibfnamefont {L.}~\bibnamefont {Leone}},\ and\ \bibinfo {author} {\bibfnamefont {X.}~\bibnamefont {Turkeshi}},\ }\bibfield  {title} {\bibinfo {title} {{Anticoncentration and State Design of Doped Real Clifford Circuits and Tensor Networks}},\ }\href {https://arxiv.org/abs/2512.15880} {\ }\Eprint {https://arxiv.org/abs/2512.15880} {arXiv:2512.15880} \BibitemShut {NoStop}%
\bibitem [{\citenamefont {Sauliere}\ \emph {et~al.}()\citenamefont {Sauliere}, \citenamefont {Lami}, \citenamefont {Boyer}, \citenamefont {De~Nardis},\ and\ \citenamefont {De~Luca}}]{sauliere2026}%
  \BibitemOpen
  \bibfield  {author} {\bibinfo {author} {\bibfnamefont {A.}~\bibnamefont {Sauliere}}, \bibinfo {author} {\bibfnamefont {G.}~\bibnamefont {Lami}}, \bibinfo {author} {\bibfnamefont {C.}~\bibnamefont {Boyer}}, \bibinfo {author} {\bibfnamefont {J.}~\bibnamefont {De~Nardis}},\ and\ \bibinfo {author} {\bibfnamefont {A.}~\bibnamefont {De~Luca}},\ }\bibfield  {title} {\bibinfo {title} {{Universality in the Anticoncentration of Noisy Quantum Circuits at Finite Depths}},\ }\href {https://arxiv.org/abs/2508.14975} {\ }\Eprint {https://arxiv.org/abs/2508.14975} {arXiv:2508.14975} \BibitemShut {NoStop}%
\bibitem [{\citenamefont {Ares}\ \emph {et~al.}(2025{\natexlab{b}})\citenamefont {Ares}, \citenamefont {Murciano}, \citenamefont {Calabrese},\ and\ \citenamefont {Piroli}}]{ares2025prr}%
  \BibitemOpen
  \bibfield  {author} {\bibinfo {author} {\bibfnamefont {F.}~\bibnamefont {Ares}}, \bibinfo {author} {\bibfnamefont {S.}~\bibnamefont {Murciano}}, \bibinfo {author} {\bibfnamefont {P.}~\bibnamefont {Calabrese}},\ and\ \bibinfo {author} {\bibfnamefont {L.}~\bibnamefont {Piroli}},\ }\bibfield  {title} {\bibinfo {title} {Entanglement asymmetry dynamics in random quantum circuits},\ }\href {https://doi.org/10.1103/m3np-p5xj} {\bibfield  {journal} {\bibinfo  {journal} {Phys. Rev. Res.}\ }\textbf {\bibinfo {volume} {7}},\ \bibinfo {pages} {033135} (\bibinfo {year} {2025}{\natexlab{b}})}\BibitemShut {NoStop}%
\bibitem [{\citenamefont {Yang}\ \emph {et~al.}()\citenamefont {Yang}, \citenamefont {Joshi}, \citenamefont {Ares}, \citenamefont {Han}, \citenamefont {Zhang},\ and\ \citenamefont {Calabrese}}]{Joshi2026}%
  \BibitemOpen
  \bibfield  {author} {\bibinfo {author} {\bibfnamefont {J.-N.}\ \bibnamefont {Yang}}, \bibinfo {author} {\bibfnamefont {L.~K.}\ \bibnamefont {Joshi}}, \bibinfo {author} {\bibfnamefont {F.}~\bibnamefont {Ares}}, \bibinfo {author} {\bibfnamefont {Y.}~\bibnamefont {Han}}, \bibinfo {author} {\bibfnamefont {P.}~\bibnamefont {Zhang}},\ and\ \bibinfo {author} {\bibfnamefont {P.}~\bibnamefont {Calabrese}},\ }\bibfield  {title} {\bibinfo {title} {{Probing Entanglement and Symmetries in Random States Using a Superconducting Quantum Processor}},\ }\href {https://doi.org/10.48550/arXiv.2601.22224} {\ }\Eprint {https://arxiv.org/abs/2601.22224} {arXiv:2601.22224} \BibitemShut {NoStop}%
\bibitem [{\citenamefont {Moudgalya}\ and\ \citenamefont {Motrunich}(2023{\natexlab{a}})}]{moudgalya2023symmetries}%
  \BibitemOpen
  \bibfield  {author} {\bibinfo {author} {\bibfnamefont {S.}~\bibnamefont {Moudgalya}}\ and\ \bibinfo {author} {\bibfnamefont {O.~I.}\ \bibnamefont {Motrunich}},\ }\bibfield  {title} {\bibinfo {title} {{From symmetries to commutant algebras in standard Hamiltonians}},\ }\href {https://doi.org/10.1016/j.aop.2023.169384} {\bibfield  {journal} {\bibinfo  {journal} {Annals of Physics}\ }\textbf {\bibinfo {volume} {455}},\ \bibinfo {pages} {169384} (\bibinfo {year} {2023}{\natexlab{a}})}\BibitemShut {NoStop}%
\bibitem [{\citenamefont {Moudgalya}\ and\ \citenamefont {Motrunich}(2024{\natexlab{a}})}]{moudgalya2024scars}%
  \BibitemOpen
  \bibfield  {author} {\bibinfo {author} {\bibfnamefont {S.}~\bibnamefont {Moudgalya}}\ and\ \bibinfo {author} {\bibfnamefont {O.~I.}\ \bibnamefont {Motrunich}},\ }\bibfield  {title} {\bibinfo {title} {Exhaustive characterization of quantum many-body scars using commutant algebras},\ }\href {https://doi.org/10.1103/PhysRevX.14.041069} {\bibfield  {journal} {\bibinfo  {journal} {Phys. Rev. X}\ }\textbf {\bibinfo {volume} {14}},\ \bibinfo {pages} {041069} (\bibinfo {year} {2024}{\natexlab{a}})}\BibitemShut {NoStop}%
\bibitem [{\citenamefont {Moudgalya}\ and\ \citenamefont {Motrunich}(2024{\natexlab{b}})}]{moudgalya2024hydro}%
  \BibitemOpen
  \bibfield  {author} {\bibinfo {author} {\bibfnamefont {S.}~\bibnamefont {Moudgalya}}\ and\ \bibinfo {author} {\bibfnamefont {O.~I.}\ \bibnamefont {Motrunich}},\ }\bibfield  {title} {\bibinfo {title} {Symmetries as ground states of local superoperators: Hydrodynamic implications},\ }\href {https://doi.org/10.1103/PRXQuantum.5.040330} {\bibfield  {journal} {\bibinfo  {journal} {PRX Quantum}\ }\textbf {\bibinfo {volume} {5}},\ \bibinfo {pages} {040330} (\bibinfo {year} {2024}{\natexlab{b}})}\BibitemShut {NoStop}%
\bibitem [{\citenamefont {Moudgalya}\ and\ \citenamefont {Motrunich}(2023{\natexlab{b}})}]{moudgalya2023numerical}%
  \BibitemOpen
  \bibfield  {author} {\bibinfo {author} {\bibfnamefont {S.}~\bibnamefont {Moudgalya}}\ and\ \bibinfo {author} {\bibfnamefont {O.~I.}\ \bibnamefont {Motrunich}},\ }\bibfield  {title} {\bibinfo {title} {Numerical methods for detecting symmetries and commutant algebras},\ }\href {https://doi.org/10.1103/PhysRevB.107.224312} {\bibfield  {journal} {\bibinfo  {journal} {Phys. Rev. B}\ }\textbf {\bibinfo {volume} {107}},\ \bibinfo {pages} {224312} (\bibinfo {year} {2023}{\natexlab{b}})}\BibitemShut {NoStop}%
\bibitem [{\citenamefont {Sahu}\ \emph {et~al.}()\citenamefont {Sahu}, \citenamefont {Li},\ and\ \citenamefont {Sala}}]{sahu2025entanglementcosthierarchiesquantum}%
  \BibitemOpen
  \bibfield  {author} {\bibinfo {author} {\bibfnamefont {S.}~\bibnamefont {Sahu}}, \bibinfo {author} {\bibfnamefont {Y.}~\bibnamefont {Li}},\ and\ \bibinfo {author} {\bibfnamefont {P.}~\bibnamefont {Sala}},\ }\bibfield  {title} {\bibinfo {title} {Entanglement cost hierarchies in quantum fragmented mixed states},\ }\href {https://arxiv.org/abs/2506.04637} {\ }\Eprint {https://arxiv.org/abs/2506.04637} {arXiv:2506.04637} \BibitemShut {NoStop}%
\bibitem [{\citenamefont {March\'e}\ \emph {et~al.}(2025)\citenamefont {March\'e}, \citenamefont {Morettini}, \citenamefont {Mazza}, \citenamefont {Gotta},\ and\ \citenamefont {Capizzi}}]{marche2025exceptional}%
  \BibitemOpen
  \bibfield  {author} {\bibinfo {author} {\bibfnamefont {A.}~\bibnamefont {March\'e}}, \bibinfo {author} {\bibfnamefont {G.}~\bibnamefont {Morettini}}, \bibinfo {author} {\bibfnamefont {L.}~\bibnamefont {Mazza}}, \bibinfo {author} {\bibfnamefont {L.}~\bibnamefont {Gotta}},\ and\ \bibinfo {author} {\bibfnamefont {L.}~\bibnamefont {Capizzi}},\ }\bibfield  {title} {\bibinfo {title} {Exceptional stationary state in a dephasing many-body open quantum system},\ }\href {https://doi.org/10.1103/zn9v-k73w} {\bibfield  {journal} {\bibinfo  {journal} {Phys. Rev. Lett.}\ }\textbf {\bibinfo {volume} {135}},\ \bibinfo {pages} {020406} (\bibinfo {year} {2025})}\BibitemShut {NoStop}%
\bibitem [{\citenamefont {Gotta}()}]{gotta2025opensystemquantummanybodyscars}%
  \BibitemOpen
  \bibfield  {author} {\bibinfo {author} {\bibfnamefont {L.}~\bibnamefont {Gotta}},\ }\bibfield  {title} {\bibinfo {title} {Open-system quantum many-body scars: a theory},\ }\href {https://arxiv.org/abs/2509.18023} {\ }\Eprint {https://arxiv.org/abs/2509.18023} {arXiv:2509.18023} \BibitemShut {NoStop}%
\bibitem [{\citenamefont {Pezz\'e}\ and\ \citenamefont {Smerzi}(2009)}]{pezze2009}%
  \BibitemOpen
  \bibfield  {author} {\bibinfo {author} {\bibfnamefont {L.}~\bibnamefont {Pezz\'e}}\ and\ \bibinfo {author} {\bibfnamefont {A.}~\bibnamefont {Smerzi}},\ }\bibfield  {title} {\bibinfo {title} {{Entanglement, Nonlinear Dynamics, and the Heisenberg Limit}},\ }\href {https://doi.org/10.1103/PhysRevLett.102.100401} {\bibfield  {journal} {\bibinfo  {journal} {Phys. Rev. Lett.}\ }\textbf {\bibinfo {volume} {102}},\ \bibinfo {pages} {100401} (\bibinfo {year} {2009})}\BibitemShut {NoStop}%
\bibitem [{\citenamefont {T\'oth}(2012)}]{toth2012}%
  \BibitemOpen
  \bibfield  {author} {\bibinfo {author} {\bibfnamefont {G.}~\bibnamefont {T\'oth}},\ }\bibfield  {title} {\bibinfo {title} {Multipartite entanglement and high-precision metrology},\ }\href {https://doi.org/10.1103/PhysRevA.85.022322} {\bibfield  {journal} {\bibinfo  {journal} {Phys. Rev. A}\ }\textbf {\bibinfo {volume} {85}},\ \bibinfo {pages} {022322} (\bibinfo {year} {2012})}\BibitemShut {NoStop}%
\bibitem [{\citenamefont {Hyllus}\ \emph {et~al.}(2012)\citenamefont {Hyllus}, \citenamefont {Laskowski}, \citenamefont {Krischek}, \citenamefont {Schwemmer}, \citenamefont {Wieczorek}, \citenamefont {Weinfurter}, \citenamefont {Pezz\'e},\ and\ \citenamefont {Smerzi}}]{Hyllus2012}%
  \BibitemOpen
  \bibfield  {author} {\bibinfo {author} {\bibfnamefont {P.}~\bibnamefont {Hyllus}}, \bibinfo {author} {\bibfnamefont {W.}~\bibnamefont {Laskowski}}, \bibinfo {author} {\bibfnamefont {R.}~\bibnamefont {Krischek}}, \bibinfo {author} {\bibfnamefont {C.}~\bibnamefont {Schwemmer}}, \bibinfo {author} {\bibfnamefont {W.}~\bibnamefont {Wieczorek}}, \bibinfo {author} {\bibfnamefont {H.}~\bibnamefont {Weinfurter}}, \bibinfo {author} {\bibfnamefont {L.}~\bibnamefont {Pezz\'e}},\ and\ \bibinfo {author} {\bibfnamefont {A.}~\bibnamefont {Smerzi}},\ }\bibfield  {title} {\bibinfo {title} {Fisher information and multiparticle entanglement},\ }\href {https://doi.org/10.1103/PhysRevA.85.022321} {\bibfield  {journal} {\bibinfo  {journal} {Phys. Rev. A}\ }\textbf {\bibinfo {volume} {85}},\ \bibinfo {pages} {022321} (\bibinfo {year} {2012})}\BibitemShut {NoStop}%
\bibitem [{\citenamefont {Ferro}()}]{ferro2026}%
  \BibitemOpen
  \bibfield  {author} {\bibinfo {author} {\bibfnamefont {F.}~\bibnamefont {Ferro}},\ }\bibfield  {title} {\bibinfo {title} {Breaking of clustering and macroscopic coherence under the lens of asymmetry measures},\ }\href {https://arxiv.org/abs/2602.15969} {\ }\Eprint {https://arxiv.org/abs/2602.15969} {arXiv:2602.15969} \BibitemShut {NoStop}%
\bibitem [{\citenamefont {Ferro}\ and\ \citenamefont {Fagotti}()}]{ferro25}%
  \BibitemOpen
  \bibfield  {author} {\bibinfo {author} {\bibfnamefont {F.}~\bibnamefont {Ferro}}\ and\ \bibinfo {author} {\bibfnamefont {M.}~\bibnamefont {Fagotti}},\ }\bibfield  {title} {\bibinfo {title} {{Kicking Quantum Fisher Information out of Equilibrium}},\ }\href {https://doi.org/10.48550/arXiv.2503.21905} {\ }\Eprint {https://arxiv.org/abs/2503.21905} {arXiv:2503.21905} \BibitemShut {NoStop}%
\bibitem [{\citenamefont {Yamashika}\ \emph {et~al.}()\citenamefont {Yamashika}, \citenamefont {Endo},\ and\ \citenamefont {Tajima}}]{yamashika25}%
  \BibitemOpen
  \bibfield  {author} {\bibinfo {author} {\bibfnamefont {S.}~\bibnamefont {Yamashika}}, \bibinfo {author} {\bibfnamefont {S.}~\bibnamefont {Endo}},\ and\ \bibinfo {author} {\bibfnamefont {H.}~\bibnamefont {Tajima}},\ }\bibfield  {title} {\bibinfo {title} {{Quantum Fisher Information as a Measure of Symmetry Breaking in Quantum Many-Body Systems}},\ }\href {https://doi.org/10.48550/arXiv.2509.07468} {\ }\Eprint {https://arxiv.org/abs/2509.07468} {arXiv:2509.07468} \BibitemShut {NoStop}%
\bibitem [{\citenamefont {Chen}\ and\ \citenamefont {Chen}(2024)}]{Chen2024}%
  \BibitemOpen
  \bibfield  {author} {\bibinfo {author} {\bibfnamefont {M.}~\bibnamefont {Chen}}\ and\ \bibinfo {author} {\bibfnamefont {H.-H.}\ \bibnamefont {Chen}},\ }\bibfield  {title} {\bibinfo {title} {R\'enyi entanglement asymmetry in ($1+1$)-dimensional conformal field theories},\ }\href {https://doi.org/10.1103/PhysRevD.109.065009} {\bibfield  {journal} {\bibinfo  {journal} {Phys. Rev. D}\ }\textbf {\bibinfo {volume} {109}},\ \bibinfo {pages} {065009} (\bibinfo {year} {2024})}\BibitemShut {NoStop}%
\bibitem [{\citenamefont {Fossati}\ \emph {et~al.}(2024)\citenamefont {Fossati}, \citenamefont {Ares}, \citenamefont {Dubail},\ and\ \citenamefont {Calabrese}}]{fossati2024entanglement}%
  \BibitemOpen
  \bibfield  {author} {\bibinfo {author} {\bibfnamefont {M.}~\bibnamefont {Fossati}}, \bibinfo {author} {\bibfnamefont {F.}~\bibnamefont {Ares}}, \bibinfo {author} {\bibfnamefont {J.}~\bibnamefont {Dubail}},\ and\ \bibinfo {author} {\bibfnamefont {P.}~\bibnamefont {Calabrese}},\ }\bibfield  {title} {\bibinfo {title} {{Entanglement asymmetry in CFT and its relation to non-topological defects}},\ }\href {https://doi.org/10.1007/JHEP05(2024)059} {\bibfield  {journal} {\bibinfo  {journal} {JHEP}\ }\textbf {\bibinfo {volume} {05}},\ \bibinfo {pages} {059 (2024)}}\BibitemShut {NoStop}%
\bibitem [{\citenamefont {Kusuki}\ \emph {et~al.}(2025)\citenamefont {Kusuki}, \citenamefont {Murciano}, \citenamefont {Ooguri},\ and\ \citenamefont {Pal}}]{Kusuki2024}%
  \BibitemOpen
  \bibfield  {author} {\bibinfo {author} {\bibfnamefont {Y.}~\bibnamefont {Kusuki}}, \bibinfo {author} {\bibfnamefont {S.}~\bibnamefont {Murciano}}, \bibinfo {author} {\bibfnamefont {H.}~\bibnamefont {Ooguri}},\ and\ \bibinfo {author} {\bibfnamefont {S.}~\bibnamefont {Pal}},\ }\bibfield  {title} {\bibinfo {title} {{Entanglement asymmetry and symmetry defects in boundary conformal field theory}},\ }\href {https://doi.org/10.1007/JHEP01(2025)057} {\bibfield  {journal} {\bibinfo  {journal} {JHEP}\ }\textbf {\bibinfo {volume} {01}},\ \bibinfo {pages} {057 (2025)}}\BibitemShut {NoStop}%
\bibitem [{\citenamefont {Fossati}\ \emph {et~al.}(2025)\citenamefont {Fossati}, \citenamefont {Rylands},\ and\ \citenamefont {Calabrese}}]{fossati2024}%
  \BibitemOpen
  \bibfield  {author} {\bibinfo {author} {\bibfnamefont {M.}~\bibnamefont {Fossati}}, \bibinfo {author} {\bibfnamefont {C.}~\bibnamefont {Rylands}},\ and\ \bibinfo {author} {\bibfnamefont {P.}~\bibnamefont {Calabrese}},\ }\bibfield  {title} {\bibinfo {title} {{Entanglement asymmetry in CFT with boundary symmetry breaking}},\ }\href {https://doi.org/10.1007/JHEP06(2025)089} {\bibfield  {journal} {\bibinfo  {journal} {JHEP}\ }\textbf {\bibinfo {volume} {06}},\ \bibinfo {pages} {089 (2025)}}\BibitemShut {NoStop}%
\bibitem [{\citenamefont {Lastres}\ \emph {et~al.}(2025)\citenamefont {Lastres}, \citenamefont {Murciano}, \citenamefont {Ares},\ and\ \citenamefont {Calabrese}}]{lastres2024entanglement}%
  \BibitemOpen
  \bibfield  {author} {\bibinfo {author} {\bibfnamefont {M.}~\bibnamefont {Lastres}}, \bibinfo {author} {\bibfnamefont {S.}~\bibnamefont {Murciano}}, \bibinfo {author} {\bibfnamefont {F.}~\bibnamefont {Ares}},\ and\ \bibinfo {author} {\bibfnamefont {P.}~\bibnamefont {Calabrese}},\ }\bibfield  {title} {\bibinfo {title} {{Entanglement asymmetry in the critical XXZ spin chain}},\ }\href {https://doi.org/10.1088/1742-5468/ada497} {\bibfield  {journal} {\bibinfo  {journal} {J. Stat. Mech.}\ ,\ \bibinfo {pages} {013107}} (\bibinfo {year} {2025})}\BibitemShut {NoStop}%
\bibitem [{\citenamefont {Benini}\ \emph {et~al.}(2025{\natexlab{a}})\citenamefont {Benini}, \citenamefont {Godet},\ and\ \citenamefont {Singh}}]{Benini2024}%
  \BibitemOpen
  \bibfield  {author} {\bibinfo {author} {\bibfnamefont {F.}~\bibnamefont {Benini}}, \bibinfo {author} {\bibfnamefont {V.}~\bibnamefont {Godet}},\ and\ \bibinfo {author} {\bibfnamefont {A.~H.}\ \bibnamefont {Singh}},\ }\bibfield  {title} {\bibinfo {title} {Entanglement asymmetry in conformal field theory and holography},\ }\href {https://doi.org/10.1093/ptep/ptaf080} {\bibfield  {journal} {\bibinfo  {journal} {Prog. Theor. Exp. Phys.}\ }\textbf {\bibinfo {volume} {6}},\ \bibinfo {pages} {063B05} (\bibinfo {year} {2025}{\natexlab{a}})}\BibitemShut {NoStop}%
\bibitem [{\citenamefont {Capizzi}\ and\ \citenamefont {Vitale}(2024)}]{capizzi2024universal}%
  \BibitemOpen
  \bibfield  {author} {\bibinfo {author} {\bibfnamefont {L.}~\bibnamefont {Capizzi}}\ and\ \bibinfo {author} {\bibfnamefont {V.}~\bibnamefont {Vitale}},\ }\bibfield  {title} {\bibinfo {title} {A universal formula for the entanglement asymmetry of matrix product states},\ }\href {https://doi.org/10.1088/1751-8121/ad8796} {\bibfield  {journal} {\bibinfo  {journal} {J. Phys. A: Math. Theor.}\ }\textbf {\bibinfo {volume} {57}},\ \bibinfo {pages} {45LT01} (\bibinfo {year} {2024})}\BibitemShut {NoStop}%
\bibitem [{\citenamefont {Ares}\ \emph {et~al.}(2024)\citenamefont {Ares}, \citenamefont {Murciano}, \citenamefont {Piroli},\ and\ \citenamefont {Calabrese}}]{ares2023entanglement}%
  \BibitemOpen
  \bibfield  {author} {\bibinfo {author} {\bibfnamefont {F.}~\bibnamefont {Ares}}, \bibinfo {author} {\bibfnamefont {S.}~\bibnamefont {Murciano}}, \bibinfo {author} {\bibfnamefont {L.}~\bibnamefont {Piroli}},\ and\ \bibinfo {author} {\bibfnamefont {P.}~\bibnamefont {Calabrese}},\ }\bibfield  {title} {\bibinfo {title} {Entanglement asymmetry study of black hole radiation},\ }\href {https://doi.org/10.1103/PhysRevD.110.L061901} {\bibfield  {journal} {\bibinfo  {journal} {Phys. Rev. D}\ }\textbf {\bibinfo {volume} {110}},\ \bibinfo {pages} {L061901} (\bibinfo {year} {2024})}\BibitemShut {NoStop}%
\bibitem [{\citenamefont {Russotto}\ \emph {et~al.}(2025{\natexlab{a}})\citenamefont {Russotto}, \citenamefont {Ares},\ and\ \citenamefont {Calabrese}}]{russotto2024non}%
  \BibitemOpen
  \bibfield  {author} {\bibinfo {author} {\bibfnamefont {A.}~\bibnamefont {Russotto}}, \bibinfo {author} {\bibfnamefont {F.}~\bibnamefont {Ares}},\ and\ \bibinfo {author} {\bibfnamefont {P.}~\bibnamefont {Calabrese}},\ }\bibfield  {title} {\bibinfo {title} {{Non-Abelian entanglement asymmetry in random states}},\ }\href {https://doi.org/10.1007/JHEP06%282025%29149} {\bibfield  {journal} {\bibinfo  {journal} {JHEP}\ }\textbf {\bibinfo {volume} {06}},\ \bibinfo {pages} {149 (2025)}}\BibitemShut {NoStop}%
\bibitem [{\citenamefont {Chen}\ and\ \citenamefont {Tang}(2025)}]{chen2024entanglement}%
  \BibitemOpen
  \bibfield  {author} {\bibinfo {author} {\bibfnamefont {H.-H.}\ \bibnamefont {Chen}}\ and\ \bibinfo {author} {\bibfnamefont {Z.-J.}\ \bibnamefont {Tang}},\ }\bibfield  {title} {\bibinfo {title} {{Entanglement asymmetry in the Hayden-Preskill protocol}},\ }\href {https://doi.org/10.1103/PhysRevD.111.066003} {\bibfield  {journal} {\bibinfo  {journal} {Phys. Rev. D}\ }\textbf {\bibinfo {volume} {111}},\ \bibinfo {pages} {066003} (\bibinfo {year} {2025})}\BibitemShut {NoStop}%
\bibitem [{\citenamefont {Russotto}\ \emph {et~al.}(2025{\natexlab{b}})\citenamefont {Russotto}, \citenamefont {Ares},\ and\ \citenamefont {Calabrese}}]{russotto25u1}%
  \BibitemOpen
  \bibfield  {author} {\bibinfo {author} {\bibfnamefont {A.}~\bibnamefont {Russotto}}, \bibinfo {author} {\bibfnamefont {F.}~\bibnamefont {Ares}},\ and\ \bibinfo {author} {\bibfnamefont {P.}~\bibnamefont {Calabrese}},\ }\bibfield  {title} {\bibinfo {title} {Symmetry breaking in chaotic many-body quantum systems at finite temperature},\ }\href {https://doi.org/10.1103/kppn-3272} {\bibfield  {journal} {\bibinfo  {journal} {Phys. Rev. E}\ }\textbf {\bibinfo {volume} {112}},\ \bibinfo {pages} {L032101} (\bibinfo {year} {2025}{\natexlab{b}})}\BibitemShut {NoStop}%
\bibitem [{\citenamefont {Ares}\ \emph {et~al.}(2023{\natexlab{b}})\citenamefont {Ares}, \citenamefont {Murciano}, \citenamefont {Vernier},\ and\ \citenamefont {Calabrese}}]{ares2023lack}%
  \BibitemOpen
  \bibfield  {author} {\bibinfo {author} {\bibfnamefont {F.}~\bibnamefont {Ares}}, \bibinfo {author} {\bibfnamefont {S.}~\bibnamefont {Murciano}}, \bibinfo {author} {\bibfnamefont {E.}~\bibnamefont {Vernier}},\ and\ \bibinfo {author} {\bibfnamefont {P.}~\bibnamefont {Calabrese}},\ }\bibfield  {title} {\bibinfo {title} {Lack of symmetry restoration after a quantum quench: An entanglement asymmetry study},\ }\href {https://doi.org/10.21468/SciPostPhys.15.3.089} {\bibfield  {journal} {\bibinfo  {journal} {SciPost Phys.}\ }\textbf {\bibinfo {volume} {15}},\ \bibinfo {pages} {089} (\bibinfo {year} {2023}{\natexlab{b}})}\BibitemShut {NoStop}%
\bibitem [{\citenamefont {Murciano}\ \emph {et~al.}(2024)\citenamefont {Murciano}, \citenamefont {Ares}, \citenamefont {Klich},\ and\ \citenamefont {Calabrese}}]{murciano2024entanglement}%
  \BibitemOpen
  \bibfield  {author} {\bibinfo {author} {\bibfnamefont {S.}~\bibnamefont {Murciano}}, \bibinfo {author} {\bibfnamefont {F.}~\bibnamefont {Ares}}, \bibinfo {author} {\bibfnamefont {I.}~\bibnamefont {Klich}},\ and\ \bibinfo {author} {\bibfnamefont {P.}~\bibnamefont {Calabrese}},\ }\bibfield  {title} {\bibinfo {title} {{Entanglement asymmetry and quantum Mpemba effect in the XY spin chain}},\ }\href {https://doi.org/10.1088/1742-5468/ad17b4} {\bibfield  {journal} {\bibinfo  {journal} {J. Stat. Mech.}\ ,\ \bibinfo {pages} {013103}} (\bibinfo {year} {2024})}\BibitemShut {NoStop}%
\bibitem [{\citenamefont {Caceffo}\ \emph {et~al.}(2024)\citenamefont {Caceffo}, \citenamefont {Murciano},\ and\ \citenamefont {Alba}}]{caceffo2024entangled}%
  \BibitemOpen
  \bibfield  {author} {\bibinfo {author} {\bibfnamefont {F.}~\bibnamefont {Caceffo}}, \bibinfo {author} {\bibfnamefont {S.}~\bibnamefont {Murciano}},\ and\ \bibinfo {author} {\bibfnamefont {V.}~\bibnamefont {Alba}},\ }\bibfield  {title} {\bibinfo {title} {{Entangled multiplets, asymmetry, and quantum Mpemba effect in dissipative systems}},\ }\href {https://doi.org/10.1088/1742-5468/ad4537} {\bibfield  {journal} {\bibinfo  {journal} {J. Stat. Mech.}\ ,\ \bibinfo {pages} {063103}} (\bibinfo {year} {2024})}\BibitemShut {NoStop}%
\bibitem [{\citenamefont {Ares}\ \emph {et~al.}(2025{\natexlab{c}})\citenamefont {Ares}, \citenamefont {Vitale},\ and\ \citenamefont {Murciano}}]{ares2024quantum}%
  \BibitemOpen
  \bibfield  {author} {\bibinfo {author} {\bibfnamefont {F.}~\bibnamefont {Ares}}, \bibinfo {author} {\bibfnamefont {V.}~\bibnamefont {Vitale}},\ and\ \bibinfo {author} {\bibfnamefont {S.}~\bibnamefont {Murciano}},\ }\bibfield  {title} {\bibinfo {title} {{The quantum Mpemba effect in free-fermionic mixed states}},\ }\href {https://doi.org/10.1103/PhysRevB.111.104312} {\bibfield  {journal} {\bibinfo  {journal} {Phys. Rev. B}\ }\textbf {\bibinfo {volume} {111}},\ \bibinfo {pages} {104312} (\bibinfo {year} {2025}{\natexlab{c}})}\BibitemShut {NoStop}%
\bibitem [{\citenamefont {Chalas}\ \emph {et~al.}(2024)\citenamefont {Chalas}, \citenamefont {Ares}, \citenamefont {Rylands},\ and\ \citenamefont {Calabrese}}]{chalas2024multiple}%
  \BibitemOpen
  \bibfield  {author} {\bibinfo {author} {\bibfnamefont {K.}~\bibnamefont {Chalas}}, \bibinfo {author} {\bibfnamefont {F.}~\bibnamefont {Ares}}, \bibinfo {author} {\bibfnamefont {C.}~\bibnamefont {Rylands}},\ and\ \bibinfo {author} {\bibfnamefont {P.}~\bibnamefont {Calabrese}},\ }\bibfield  {title} {\bibinfo {title} {{Multiple crossing during dynamical symmetry restoration and implications for the quantum Mpemba effect}},\ }\href {https://doi.org/10.1088/1742-5468/ad769c} {\bibfield  {journal} {\bibinfo  {journal} {J. Stat. Mech.}\ ,\ \bibinfo {pages} {103101}} (\bibinfo {year} {2024})}\BibitemShut {NoStop}%
\bibitem [{\citenamefont {Bertini}\ \emph {et~al.}(2024)\citenamefont {Bertini}, \citenamefont {Klobas}, \citenamefont {Collura}, \citenamefont {Calabrese},\ and\ \citenamefont {Rylands}}]{bertini2024dynamics}%
  \BibitemOpen
  \bibfield  {author} {\bibinfo {author} {\bibfnamefont {B.}~\bibnamefont {Bertini}}, \bibinfo {author} {\bibfnamefont {K.}~\bibnamefont {Klobas}}, \bibinfo {author} {\bibfnamefont {M.}~\bibnamefont {Collura}}, \bibinfo {author} {\bibfnamefont {P.}~\bibnamefont {Calabrese}},\ and\ \bibinfo {author} {\bibfnamefont {C.}~\bibnamefont {Rylands}},\ }\bibfield  {title} {\bibinfo {title} {Dynamics of charge fluctuations from asymmetric initial states},\ }\href {https://doi.org/10.1103/PhysRevB.109.184312} {\bibfield  {journal} {\bibinfo  {journal} {Phys. Rev. B}\ }\textbf {\bibinfo {volume} {109}},\ \bibinfo {pages} {184312} (\bibinfo {year} {2024})}\BibitemShut {NoStop}%
\bibitem [{\citenamefont {Rylands}\ \emph {et~al.}(2024{\natexlab{b}})\citenamefont {Rylands}, \citenamefont {Vernier},\ and\ \citenamefont {Calabrese}}]{rylands2024dynamical}%
  \BibitemOpen
  \bibfield  {author} {\bibinfo {author} {\bibfnamefont {C.}~\bibnamefont {Rylands}}, \bibinfo {author} {\bibfnamefont {E.}~\bibnamefont {Vernier}},\ and\ \bibinfo {author} {\bibfnamefont {P.}~\bibnamefont {Calabrese}},\ }\bibfield  {title} {\bibinfo {title} {{Dynamical symmetry restoration in the Heisenberg spin chain}},\ }\href {https://iopscience.iop.org/article/10.1088/1742-5468/ad97b3} {\bibfield  {journal} {\bibinfo  {journal} {J. Stat. Mech.}\ ,\ \bibinfo {pages} {123102}} (\bibinfo {year} {2024}{\natexlab{b}})}\BibitemShut {NoStop}%
\bibitem [{\citenamefont {Yamashika}\ \emph {et~al.}(2024)\citenamefont {Yamashika}, \citenamefont {Ares},\ and\ \citenamefont {Calabrese}}]{yamashika2024entanglement}%
  \BibitemOpen
  \bibfield  {author} {\bibinfo {author} {\bibfnamefont {S.}~\bibnamefont {Yamashika}}, \bibinfo {author} {\bibfnamefont {F.}~\bibnamefont {Ares}},\ and\ \bibinfo {author} {\bibfnamefont {P.}~\bibnamefont {Calabrese}},\ }\bibfield  {title} {\bibinfo {title} {{Entanglement asymmetry and quantum Mpemba effect in two-dimensional free-fermion systems}},\ }\href {https://doi.org/10.1103/PhysRevB.110.085126} {\bibfield  {journal} {\bibinfo  {journal} {Phys. Rev. B}\ }\textbf {\bibinfo {volume} {110}},\ \bibinfo {pages} {085126} (\bibinfo {year} {2024})}\BibitemShut {NoStop}%
\bibitem [{\citenamefont {Yamashika}\ \emph {et~al.}(2025)\citenamefont {Yamashika}, \citenamefont {Calabrese},\ and\ \citenamefont {Ares}}]{yamashika2024quenching}%
  \BibitemOpen
  \bibfield  {author} {\bibinfo {author} {\bibfnamefont {S.}~\bibnamefont {Yamashika}}, \bibinfo {author} {\bibfnamefont {P.}~\bibnamefont {Calabrese}},\ and\ \bibinfo {author} {\bibfnamefont {F.}~\bibnamefont {Ares}},\ }\bibfield  {title} {\bibinfo {title} {{Quenching from superfluid to free bosons in two dimensions: entanglement, symmetries, and quantum Mpemba effect}},\ }\href {https://doi.org/10.1103/PhysRevA.111.043304} {\bibfield  {journal} {\bibinfo  {journal} {Phys. Rev. A}\ }\textbf {\bibinfo {volume} {111}},\ \bibinfo {pages} {043304} (\bibinfo {year} {2025})}\BibitemShut {NoStop}%
\bibitem [{\citenamefont {Klobas}(2024)}]{klobas2024asymmetry}%
  \BibitemOpen
  \bibfield  {author} {\bibinfo {author} {\bibfnamefont {K.}~\bibnamefont {Klobas}},\ }\bibfield  {title} {\bibinfo {title} {{Non-equilibrium dynamics of symmetry-resolved entanglement and entanglement asymmetry: exact asymptotics in Rule 54}},\ }\href {https://iopscience.iop.org/article/10.1088/1751-8121/ad91fd/meta} {\bibfield  {journal} {\bibinfo  {journal} {J. Phys A: Math Theor}\ }\textbf {\bibinfo {volume} {57}},\ \bibinfo {pages} {505001} (\bibinfo {year} {2024})}\BibitemShut {NoStop}%
\bibitem [{\citenamefont {Banerjee}\ \emph {et~al.}(2025)\citenamefont {Banerjee}, \citenamefont {Das},\ and\ \citenamefont {Sengupta}}]{banerjee2024asymmetry}%
  \BibitemOpen
  \bibfield  {author} {\bibinfo {author} {\bibfnamefont {T.}~\bibnamefont {Banerjee}}, \bibinfo {author} {\bibfnamefont {S.}~\bibnamefont {Das}},\ and\ \bibinfo {author} {\bibfnamefont {K.}~\bibnamefont {Sengupta}},\ }\bibfield  {title} {\bibinfo {title} {{Entanglement asymmetry in periodically driven quantum systems}},\ }\href {https://scipost.org/SciPostPhys.19.2.051} {\bibfield  {journal} {\bibinfo  {journal} {SciPost Phys.}\ }\textbf {\bibinfo {volume} {19}},\ \bibinfo {pages} {051} (\bibinfo {year} {2025})}\BibitemShut {NoStop}%
\bibitem [{\citenamefont {Giulio}\ \emph {et~al.}(2025)\citenamefont {Giulio}, \citenamefont {Turkeshi},\ and\ \citenamefont {Murciano}}]{DiGiulio25}%
  \BibitemOpen
  \bibfield  {author} {\bibinfo {author} {\bibfnamefont {G.~D.}\ \bibnamefont {Giulio}}, \bibinfo {author} {\bibfnamefont {X.}~\bibnamefont {Turkeshi}},\ and\ \bibinfo {author} {\bibfnamefont {S.}~\bibnamefont {Murciano}},\ }\bibfield  {title} {\bibinfo {title} {Measurement-induced symmetry restoration and quantum mpemba effect},\ }\href {https://doi.org/10.3390/e27040407} {\bibfield  {journal} {\bibinfo  {journal} {Entropy}\ }\textbf {\bibinfo {volume} {27}},\ \bibinfo {pages} {407} (\bibinfo {year} {2025})}\BibitemShut {NoStop}%
\bibitem [{\citenamefont {Yu}\ \emph {et~al.}(2025{\natexlab{a}})\citenamefont {Yu}, \citenamefont {Jin}, \citenamefont {Zhang}, \citenamefont {Xu},\ and\ \citenamefont {Fan}}]{Yu2025Tuning}%
  \BibitemOpen
  \bibfield  {author} {\bibinfo {author} {\bibfnamefont {Y.}~\bibnamefont {Yu}}, \bibinfo {author} {\bibfnamefont {T.}~\bibnamefont {Jin}}, \bibinfo {author} {\bibfnamefont {L.}~\bibnamefont {Zhang}}, \bibinfo {author} {\bibfnamefont {K.}~\bibnamefont {Xu}},\ and\ \bibinfo {author} {\bibfnamefont {H.}~\bibnamefont {Fan}},\ }\bibfield  {title} {\bibinfo {title} {{Tuning the quantum Mpemba effect in an isolated system by initial-state engineering}},\ }\href {https://doi.org/10.1103/yzjd-pk8h} {\bibfield  {journal} {\bibinfo  {journal} {Phys. Rev. B}\ }\textbf {\bibinfo {volume} {112}},\ \bibinfo {pages} {094315} (\bibinfo {year} {2025}{\natexlab{a}})}\BibitemShut {NoStop}%
\bibitem [{\citenamefont {Yamashika}\ and\ \citenamefont {Ares}()}]{yamashika25-2}%
  \BibitemOpen
  \bibfield  {author} {\bibinfo {author} {\bibfnamefont {S.}~\bibnamefont {Yamashika}}\ and\ \bibinfo {author} {\bibfnamefont {F.}~\bibnamefont {Ares}},\ }\bibfield  {title} {\bibinfo {title} {{The quantum Mpemba effect in long-range spin systems}},\ }\href {https://doi.org/10.48550/arXiv.2507.06636} {\ }\Eprint {https://arxiv.org/abs/2507.06636} {arXiv:2507.06636} \BibitemShut {NoStop}%
\bibitem [{\citenamefont {Liu}\ \emph {et~al.}(2024)\citenamefont {Liu}, \citenamefont {Zhang}, \citenamefont {Yin},\ and\ \citenamefont {Zhang}}]{liu2024symmetry}%
  \BibitemOpen
  \bibfield  {author} {\bibinfo {author} {\bibfnamefont {S.}~\bibnamefont {Liu}}, \bibinfo {author} {\bibfnamefont {H.-K.}\ \bibnamefont {Zhang}}, \bibinfo {author} {\bibfnamefont {S.}~\bibnamefont {Yin}},\ and\ \bibinfo {author} {\bibfnamefont {S.-X.}\ \bibnamefont {Zhang}},\ }\bibfield  {title} {\bibinfo {title} {{Symmetry Restoration and Quantum Mpemba Effect in Symmetric Random Circuits}},\ }\href {https://doi.org/10.1103/PhysRevLett.133.140405} {\bibfield  {journal} {\bibinfo  {journal} {Phys. Rev. Lett.}\ }\textbf {\bibinfo {volume} {133}},\ \bibinfo {pages} {140405} (\bibinfo {year} {2024})}\BibitemShut {NoStop}%
\bibitem [{\citenamefont {Turkeshi}\ \emph {et~al.}(2025)\citenamefont {Turkeshi}, \citenamefont {Calabrese},\ and\ \citenamefont {De~Luca}}]{turkeshi2024quantum}%
  \BibitemOpen
  \bibfield  {author} {\bibinfo {author} {\bibfnamefont {X.}~\bibnamefont {Turkeshi}}, \bibinfo {author} {\bibfnamefont {P.}~\bibnamefont {Calabrese}},\ and\ \bibinfo {author} {\bibfnamefont {A.}~\bibnamefont {De~Luca}},\ }\bibfield  {title} {\bibinfo {title} {{Quantum Mpemba Effect in Random Circuits}},\ }\href {https://doi.org/10.1103/5d6p-8d1b} {\bibfield  {journal} {\bibinfo  {journal} {Phys. Rev. Lett.}\ }\textbf {\bibinfo {volume} {135}},\ \bibinfo {pages} {040403} (\bibinfo {year} {2025})}\BibitemShut {NoStop}%
\bibitem [{\citenamefont {Yu}\ \emph {et~al.}(2025{\natexlab{b}})\citenamefont {Yu}, \citenamefont {Li},\ and\ \citenamefont {Zhang}}]{Yu2025}%
  \BibitemOpen
  \bibfield  {author} {\bibinfo {author} {\bibfnamefont {H.}~\bibnamefont {Yu}}, \bibinfo {author} {\bibfnamefont {Z.-X.}\ \bibnamefont {Li}},\ and\ \bibinfo {author} {\bibfnamefont {S.-X.}\ \bibnamefont {Zhang}},\ }\bibfield  {title} {\bibinfo {title} {Symmetry breaking dynamics in quantum many-body systems},\ }\href {https://doi.org/10.1088/0256-307X/42/11/110602} {\bibfield  {journal} {\bibinfo  {journal} {Chinese Phys. Lett.}\ }\textbf {\bibinfo {volume} {42}},\ \bibinfo {pages} {110602} (\bibinfo {year} {2025}{\natexlab{b}})}\BibitemShut {NoStop}%
\bibitem [{\citenamefont {Summer}\ \emph {et~al.}()\citenamefont {Summer}, \citenamefont {Moroder}, \citenamefont {Bettmann}, \citenamefont {Turkeshi}, \citenamefont {Marvian},\ and\ \citenamefont {Goold}}]{summer25}%
  \BibitemOpen
  \bibfield  {author} {\bibinfo {author} {\bibfnamefont {A.}~\bibnamefont {Summer}}, \bibinfo {author} {\bibfnamefont {M.}~\bibnamefont {Moroder}}, \bibinfo {author} {\bibfnamefont {L.~P.}\ \bibnamefont {Bettmann}}, \bibinfo {author} {\bibfnamefont {X.}~\bibnamefont {Turkeshi}}, \bibinfo {author} {\bibfnamefont {I.}~\bibnamefont {Marvian}},\ and\ \bibinfo {author} {\bibfnamefont {J.}~\bibnamefont {Goold}},\ }\bibfield  {title} {\bibinfo {title} {{A resource theoretical unification of Mpemba effects: classical and quantum}},\ }\href {https://doi.org/10.48550/arXiv.2507.16976} {\ }\Eprint {https://arxiv.org/abs/2507.16976} {arXiv:2507.16976} \BibitemShut {NoStop}%
\bibitem [{\citenamefont {Xu}\ \emph {et~al.}()\citenamefont {Xu}, \citenamefont {Fang}, \citenamefont {Chen} \emph {et~al.}}]{Xu2025Observation}%
  \BibitemOpen
  \bibfield  {author} {\bibinfo {author} {\bibfnamefont {Y.}~\bibnamefont {Xu}}, \bibinfo {author} {\bibfnamefont {C.-P.}\ \bibnamefont {Fang}}, \bibinfo {author} {\bibfnamefont {B.-J.}\ \bibnamefont {Chen}}, \emph {et~al.},\ }\bibfield  {title} {\bibinfo {title} {{Observation and Modulation of the Quantum Mpemba Effect on a Superconducting Quantum Processor}},\ }\href {https://doi.org/10.48550/arXiv.2508.07707} {\ }\Eprint {https://arxiv.org/abs/2508.07707} {arXiv:2508.07707} \BibitemShut {NoStop}%
\bibitem [{\citenamefont {Benini}\ \emph {et~al.}(2025{\natexlab{b}})\citenamefont {Benini}, \citenamefont {Calabrese}, \citenamefont {Fossati}, \citenamefont {Singh},\ and\ \citenamefont {Venuti}}]{Benini2025}%
  \BibitemOpen
  \bibfield  {author} {\bibinfo {author} {\bibfnamefont {F.}~\bibnamefont {Benini}}, \bibinfo {author} {\bibfnamefont {P.}~\bibnamefont {Calabrese}}, \bibinfo {author} {\bibfnamefont {M.}~\bibnamefont {Fossati}}, \bibinfo {author} {\bibfnamefont {A.~H.}\ \bibnamefont {Singh}},\ and\ \bibinfo {author} {\bibfnamefont {M.}~\bibnamefont {Venuti}},\ }\bibfield  {title} {\bibinfo {title} {Entanglement asymmetry for higher and noninvertible symmetries},\ }\href {https://arxiv.org/abs/2509.16311} {\  (\bibinfo {year} {2025}{\natexlab{b}})},\ \Eprint {https://arxiv.org/abs/2509.16311} {arXiv:2509.16311} \BibitemShut {NoStop}%
\bibitem [{\citenamefont {Benini}\ \emph {et~al.}(2025{\natexlab{c}})\citenamefont {Benini}, \citenamefont {Garcia-Valdecasas},\ and\ \citenamefont {Vitouladitis}}]{benini25cat}%
  \BibitemOpen
  \bibfield  {author} {\bibinfo {author} {\bibfnamefont {F.}~\bibnamefont {Benini}}, \bibinfo {author} {\bibfnamefont {E.}~\bibnamefont {Garcia-Valdecasas}},\ and\ \bibinfo {author} {\bibfnamefont {S.}~\bibnamefont {Vitouladitis}},\ }\bibfield  {title} {\bibinfo {title} {{Higher-form entanglement asymmetry. Part I. The limits of symmetry breaking}},\ }\href {https://doi.org/10.48550/arXiv.2512.15898} {\  (\bibinfo {year} {2025}{\natexlab{c}})},\ \Eprint {https://arxiv.org/abs/2512.15898} {arXiv:2512.15898} \BibitemShut {NoStop}%
\bibitem [{\citenamefont {Lamas}\ \emph {et~al.}()\citenamefont {Lamas}, \citenamefont {Gliozzi},\ and\ \citenamefont {Hughes}}]{GattoLamas25}%
  \BibitemOpen
  \bibfield  {author} {\bibinfo {author} {\bibfnamefont {A.~G.}\ \bibnamefont {Lamas}}, \bibinfo {author} {\bibfnamefont {J.}~\bibnamefont {Gliozzi}},\ and\ \bibinfo {author} {\bibfnamefont {T.~L.}\ \bibnamefont {Hughes}},\ }\bibfield  {title} {\bibinfo {title} {Higher-form entanglement asymmetry and topological order},\ }\href {https://doi.org/10.48550/arXiv.2510.03967} {\ }\Eprint {https://arxiv.org/abs/2510.03967} {arXiv:2510.03967} \BibitemShut {NoStop}%
\bibitem [{\citenamefont {Khor}\ \emph {et~al.}(2024)\citenamefont {Khor}, \citenamefont {Kürkçüoglu}, \citenamefont {Hobbs}, \citenamefont {Perdue},\ and\ \citenamefont {Klich}}]{khor2024kink}%
  \BibitemOpen
  \bibfield  {author} {\bibinfo {author} {\bibfnamefont {B.~J.~J.}\ \bibnamefont {Khor}}, \bibinfo {author} {\bibfnamefont {D.~M.}\ \bibnamefont {Kürkçüoglu}}, \bibinfo {author} {\bibfnamefont {T.~J.}\ \bibnamefont {Hobbs}}, \bibinfo {author} {\bibfnamefont {G.~N.}\ \bibnamefont {Perdue}},\ and\ \bibinfo {author} {\bibfnamefont {I.}~\bibnamefont {Klich}},\ }\bibfield  {title} {\bibinfo {title} {{Confinement and Kink Entanglement Asymmetry on a Quantum Ising Chain}},\ }\href {http://dx.doi.org/10.22331/q-2024-09-06-1462} {\bibfield  {journal} {\bibinfo  {journal} {Quantum}\ }\textbf {\bibinfo {volume} {8}},\ \bibinfo {pages} {1462} (\bibinfo {year} {2024})}\BibitemShut {NoStop}%
\bibitem [{\citenamefont {Florio}\ and\ \citenamefont {Murciano}()}]{florio25}%
  \BibitemOpen
  \bibfield  {author} {\bibinfo {author} {\bibfnamefont {A.}~\bibnamefont {Florio}}\ and\ \bibinfo {author} {\bibfnamefont {S.}~\bibnamefont {Murciano}},\ }\bibfield  {title} {\bibinfo {title} {{Entanglement asymmetry in gauge theories: chiral anomaly in the finite temperature massless Schwinger model}},\ }\href {https://doi.org/10.48550/arXiv.2511.01966} {\ }\Eprint {https://arxiv.org/abs/2511.01966} {arXiv:2511.01966} \BibitemShut {NoStop}%
\bibitem [{\citenamefont {Qi}\ and\ \citenamefont {Zheng}()}]{Zheng25gauge}%
  \BibitemOpen
  \bibfield  {author} {\bibinfo {author} {\bibfnamefont {H.-Y.}\ \bibnamefont {Qi}}\ and\ \bibinfo {author} {\bibfnamefont {W.}~\bibnamefont {Zheng}},\ }\bibfield  {title} {\bibinfo {title} {{Quantum Mpemba effect in Local Gauge Symmetry Restoration}},\ }\href {https://doi.org/10.48550/arXiv.2512.15223} {\ }\Eprint {https://arxiv.org/abs/2512.15223} {arXiv:2512.15223} \BibitemShut {NoStop}%
\bibitem [{\citenamefont {Klobas}\ \emph {et~al.}(2025)\citenamefont {Klobas}, \citenamefont {Rylands},\ and\ \citenamefont {Bertini}}]{klobas2024translation}%
  \BibitemOpen
  \bibfield  {author} {\bibinfo {author} {\bibfnamefont {K.}~\bibnamefont {Klobas}}, \bibinfo {author} {\bibfnamefont {C.}~\bibnamefont {Rylands}},\ and\ \bibinfo {author} {\bibfnamefont {B.}~\bibnamefont {Bertini}},\ }\bibfield  {title} {\bibinfo {title} {{Translation symmetry restoration under random unitary dynamics}},\ }\href {https://doi.org/10.1103/PhysRevB.111.L140304} {\bibfield  {journal} {\bibinfo  {journal} {Phys. Rev. B}\ }\textbf {\bibinfo {volume} {111}},\ \bibinfo {pages} {L140304} (\bibinfo {year} {2025})}\BibitemShut {NoStop}%
\bibitem [{\citenamefont {Gibbins}\ \emph {et~al.}(2025)\citenamefont {Gibbins}, \citenamefont {Gammon-Smith},\ and\ \citenamefont {Bertini}}]{gibbins2025}%
  \BibitemOpen
  \bibfield  {author} {\bibinfo {author} {\bibfnamefont {M.}~\bibnamefont {Gibbins}}, \bibinfo {author} {\bibfnamefont {A.}~\bibnamefont {Gammon-Smith}},\ and\ \bibinfo {author} {\bibfnamefont {B.}~\bibnamefont {Bertini}},\ }\bibfield  {title} {\bibinfo {title} {Translation symmetry restoration in integrable systems: the noninteracting case},\ }\href {https://doi.org/10.1103/mxhd-fswm} {\bibfield  {journal} {\bibinfo  {journal} {Phys. Rev. B}\ }\textbf {\bibinfo {volume} {112}},\ \bibinfo {pages} {L180307} (\bibinfo {year} {2025})}\BibitemShut {NoStop}%
\bibitem [{\citenamefont {Hara}\ \emph {et~al.}()\citenamefont {Hara}, \citenamefont {Endo},\ and\ \citenamefont {Yamashika}}]{hara25}%
  \BibitemOpen
  \bibfield  {author} {\bibinfo {author} {\bibfnamefont {R.}~\bibnamefont {Hara}}, \bibinfo {author} {\bibfnamefont {S.}~\bibnamefont {Endo}},\ and\ \bibinfo {author} {\bibfnamefont {S.}~\bibnamefont {Yamashika}},\ }\bibfield  {title} {\bibinfo {title} {Dynamics of entanglement asymmetry for space-inversion symmetry of free fermions on honeycomb lattices},\ }\href {https://doi.org/10.48550/arXiv.2511.14114} {\ }\Eprint {https://arxiv.org/abs/2511.14114} {arXiv:2511.14114} \BibitemShut {NoStop}%
\bibitem [{\citenamefont {Kusuki}\ \emph {et~al.}()\citenamefont {Kusuki}, \citenamefont {Pal},\ and\ \citenamefont {Tajima}}]{kusuki26strong}%
  \BibitemOpen
  \bibfield  {author} {\bibinfo {author} {\bibfnamefont {Y.}~\bibnamefont {Kusuki}}, \bibinfo {author} {\bibfnamefont {S.}~\bibnamefont {Pal}},\ and\ \bibinfo {author} {\bibfnamefont {H.}~\bibnamefont {Tajima}},\ }\bibfield  {title} {\bibinfo {title} {{Resource-Theoretic Quantifiers of Weak and Strong Symmetry Breaking: Strong Entanglement Asymmetry and Beyond}},\ }\href {https://doi.org/10.48550/arXiv.2601.20924} {\ }\Eprint {https://arxiv.org/abs/2601.20924} {arXiv:2601.20924} \BibitemShut {NoStop}%
\bibitem [{\citenamefont {Lindblad}(1972)}]{lindblad1972entropy}%
  \BibitemOpen
  \bibfield  {author} {\bibinfo {author} {\bibfnamefont {G.}~\bibnamefont {Lindblad}},\ }\bibfield  {title} {\bibinfo {title} {An entropy inequality for quantum measurements},\ }\href {https://projecteuclid.org/journals/communications-in-mathematical-physics/volume-28/issue-3/An-entropy-inequality-for-quantum-measurements/cmp/1103858406.full} {\bibfield  {journal} {\bibinfo  {journal} {Comm. Math. Phys.}\ }\textbf {\bibinfo {volume} {28}},\ \bibinfo {pages} {245} (\bibinfo {year} {1972})}\BibitemShut {NoStop}%
\bibitem [{\citenamefont {Massey}(1988)}]{Massey1988Information}%
  \BibitemOpen
  \bibfield  {author} {\bibinfo {author} {\bibfnamefont {J.~L.}\ \bibnamefont {Massey}},\ }\bibfield  {title} {\bibinfo {title} {{in Information Theory: Proceedings of 1988 Beijing International Workshop on Information Theory (BIWIT’88)}},\ }\href@noop {} {\bibfield  {journal} {\bibinfo  {journal} {July 4-7}\ } (\bibinfo {year} {1988})}\BibitemShut {NoStop}%
\bibitem [{\citenamefont {Rioul}(2022)}]{Rioul2022Information}%
  \BibitemOpen
  \bibfield  {author} {\bibinfo {author} {\bibfnamefont {O.}~\bibnamefont {Rioul}},\ }\bibfield  {title} {\bibinfo {title} {{in IEEE Information Theory and Applications Workshop (ITA 2022)}},\ }\href@noop {} {\  (\bibinfo {year} {2022})}\BibitemShut {NoStop}%
\bibitem [{\citenamefont {Page}(1993)}]{page1993average}%
  \BibitemOpen
  \bibfield  {author} {\bibinfo {author} {\bibfnamefont {D.~N.}\ \bibnamefont {Page}},\ }\bibfield  {title} {\bibinfo {title} {Average entropy of a subsystem},\ }\href {https://doi.org/10.1103/PhysRevLett.71.1291} {\bibfield  {journal} {\bibinfo  {journal} {Phys. Rev. Lett}\ }\textbf {\bibinfo {volume} {71}},\ \bibinfo {pages} {1291} (\bibinfo {year} {1993})}\BibitemShut {NoStop}%
\bibitem [{\citenamefont {Haag}\ \emph {et~al.}(2023{\natexlab{b}})\citenamefont {Haag}, \citenamefont {Baccari},\ and\ \citenamefont {Styliaris}}]{Haag2023correlation}%
  \BibitemOpen
  \bibfield  {author} {\bibinfo {author} {\bibfnamefont {D.}~\bibnamefont {Haag}}, \bibinfo {author} {\bibfnamefont {F.}~\bibnamefont {Baccari}},\ and\ \bibinfo {author} {\bibfnamefont {G.}~\bibnamefont {Styliaris}},\ }\bibfield  {title} {\bibinfo {title} {{Typical Correlation Length of Sequentially Generated Tensor Network States}},\ }\href {https://doi.org/10.1103/PRXQuantum.4.030330} {\bibfield  {journal} {\bibinfo  {journal} {PRX Quantum}\ }\textbf {\bibinfo {volume} {4}},\ \bibinfo {pages} {030330} (\bibinfo {year} {2023}{\natexlab{b}})}\BibitemShut {NoStop}%
\bibitem [{\citenamefont {Nahum}\ \emph {et~al.}(2017)\citenamefont {Nahum}, \citenamefont {Ruhman}, \citenamefont {Vijay},\ and\ \citenamefont {Haah}}]{nahum17growth}%
  \BibitemOpen
  \bibfield  {author} {\bibinfo {author} {\bibfnamefont {A.}~\bibnamefont {Nahum}}, \bibinfo {author} {\bibfnamefont {J.}~\bibnamefont {Ruhman}}, \bibinfo {author} {\bibfnamefont {S.}~\bibnamefont {Vijay}},\ and\ \bibinfo {author} {\bibfnamefont {J.}~\bibnamefont {Haah}},\ }\bibfield  {title} {\bibinfo {title} {Quantum entanglement growth under random unitary dynamics},\ }\href {https://doi.org/10.1103/PhysRevX.7.031016} {\bibfield  {journal} {\bibinfo  {journal} {Phys. Rev. X}\ }\textbf {\bibinfo {volume} {7}},\ \bibinfo {pages} {031016} (\bibinfo {year} {2017})}\BibitemShut {NoStop}%
\bibitem [{\citenamefont {Nahum}\ \emph {et~al.}(2018)\citenamefont {Nahum}, \citenamefont {Vijay},\ and\ \citenamefont {Haah}}]{nahum18operator}%
  \BibitemOpen
  \bibfield  {author} {\bibinfo {author} {\bibfnamefont {A.}~\bibnamefont {Nahum}}, \bibinfo {author} {\bibfnamefont {S.}~\bibnamefont {Vijay}},\ and\ \bibinfo {author} {\bibfnamefont {J.}~\bibnamefont {Haah}},\ }\bibfield  {title} {\bibinfo {title} {Operator spreading in random unitary circuits},\ }\href {https://doi.org/10.1103/PhysRevX.8.021014} {\bibfield  {journal} {\bibinfo  {journal} {Phys. Rev. X}\ }\textbf {\bibinfo {volume} {8}},\ \bibinfo {pages} {021014} (\bibinfo {year} {2018})}\BibitemShut {NoStop}%
\bibitem [{\citenamefont {Fisher}\ \emph {et~al.}(2023)\citenamefont {Fisher}, \citenamefont {Khemani}, \citenamefont {Nahum},\ and\ \citenamefont {Vijay}}]{fisher2023random}%
  \BibitemOpen
  \bibfield  {author} {\bibinfo {author} {\bibfnamefont {M.~P.}\ \bibnamefont {Fisher}}, \bibinfo {author} {\bibfnamefont {V.}~\bibnamefont {Khemani}}, \bibinfo {author} {\bibfnamefont {A.}~\bibnamefont {Nahum}},\ and\ \bibinfo {author} {\bibfnamefont {S.}~\bibnamefont {Vijay}},\ }\bibfield  {title} {\bibinfo {title} {Random quantum circuits},\ }\href {https://doi.org/10.1146/annurev-conmatphys-031720-030658} {\bibfield  {journal} {\bibinfo  {journal} {Ann. Rev. Cond. Matt. Phys.}\ }\textbf {\bibinfo {volume} {14}},\ \bibinfo {pages} {335} (\bibinfo {year} {2023})}\BibitemShut {NoStop}%
\bibitem [{\citenamefont {Potter}\ and\ \citenamefont {Vasseur}(2022)}]{potter2022entanglement}%
  \BibitemOpen
  \bibfield  {author} {\bibinfo {author} {\bibfnamefont {A.~C.}\ \bibnamefont {Potter}}\ and\ \bibinfo {author} {\bibfnamefont {R.}~\bibnamefont {Vasseur}},\ }\bibfield  {title} {\bibinfo {title} {Entanglement dynamics in hybrid quantum circuits},\ }in\ \href {https://doi.org/10.1007/978-3-031-03998-0_9} {\emph {\bibinfo {booktitle} {Entanglement in Spin Chains: From Theory to Quantum Technology Applications}}}\ (\bibinfo  {publisher} {Springer},\ \bibinfo {year} {2022})\ pp.\ \bibinfo {pages} {211--249}\BibitemShut {NoStop}%
\bibitem [{\citenamefont {Hayden}\ and\ \citenamefont {Preskill}(2007)}]{hayden2007black}%
  \BibitemOpen
  \bibfield  {author} {\bibinfo {author} {\bibfnamefont {P.}~\bibnamefont {Hayden}}\ and\ \bibinfo {author} {\bibfnamefont {J.}~\bibnamefont {Preskill}},\ }\bibfield  {title} {\bibinfo {title} {Black holes as mirrors: quantum information in random subsystems},\ }\href {https://doi.org/10.1088/1126-6708/2007/09/120} {\bibfield  {journal} {\bibinfo  {journal} {JHEP}\ }\textbf {\bibinfo {volume} {09}},\ \bibinfo {pages} {(2007) 120}}\BibitemShut {NoStop}%
\bibitem [{\citenamefont {Zhang}\ \emph {et~al.}(1997)\citenamefont {Zhang}, \citenamefont {Karbach}, \citenamefont {M\"uller},\ and\ \citenamefont {Stolze}}]{zhang97}%
  \BibitemOpen
  \bibfield  {author} {\bibinfo {author} {\bibfnamefont {S.}~\bibnamefont {Zhang}}, \bibinfo {author} {\bibfnamefont {M.}~\bibnamefont {Karbach}}, \bibinfo {author} {\bibfnamefont {G.}~\bibnamefont {M\"uller}},\ and\ \bibinfo {author} {\bibfnamefont {J.}~\bibnamefont {Stolze}},\ }\bibfield  {title} {\bibinfo {title} {{Charge and spin dynamics in the one-dimensional $t-J_z$ and $t-J$ models}},\ }\href {https://doi.org/10.1103/PhysRevB.55.6491} {\bibfield  {journal} {\bibinfo  {journal} {Phys. Rev. B}\ }\textbf {\bibinfo {volume} {55}},\ \bibinfo {pages} {6491} (\bibinfo {year} {1997})}\BibitemShut {NoStop}%
\bibitem [{\citenamefont {Batista}\ and\ \citenamefont {Ortiz}(2000)}]{batista00}%
  \BibitemOpen
  \bibfield  {author} {\bibinfo {author} {\bibfnamefont {C.~D.}\ \bibnamefont {Batista}}\ and\ \bibinfo {author} {\bibfnamefont {G.}~\bibnamefont {Ortiz}},\ }\bibfield  {title} {\bibinfo {title} {{Quantum phase diagram of the $t-J_z$ chain model}},\ }\href {https://doi.org/10.1103/PhysRevLett.85.4755} {\bibfield  {journal} {\bibinfo  {journal} {Phys. Rev. Lett.}\ }\textbf {\bibinfo {volume} {85}},\ \bibinfo {pages} {4755} (\bibinfo {year} {2000})}\BibitemShut {NoStop}%
\bibitem [{\citenamefont {Rakovszky}\ \emph {et~al.}(2020)\citenamefont {Rakovszky}, \citenamefont {Sala}, \citenamefont {Verresen}, \citenamefont {Knap},\ and\ \citenamefont {Pollmann}}]{rakovszky20}%
  \BibitemOpen
  \bibfield  {author} {\bibinfo {author} {\bibfnamefont {T.}~\bibnamefont {Rakovszky}}, \bibinfo {author} {\bibfnamefont {P.}~\bibnamefont {Sala}}, \bibinfo {author} {\bibfnamefont {R.}~\bibnamefont {Verresen}}, \bibinfo {author} {\bibfnamefont {M.}~\bibnamefont {Knap}},\ and\ \bibinfo {author} {\bibfnamefont {F.}~\bibnamefont {Pollmann}},\ }\bibfield  {title} {\bibinfo {title} {{Statistical localization: From strong fragmentation to strong edge modes}},\ }\href {https://doi.org/10.1103/PhysRevB.101.125126} {\bibfield  {journal} {\bibinfo  {journal} {Phys. Rev. B}\ }\textbf {\bibinfo {volume} {101}},\ \bibinfo {pages} {125126} (\bibinfo {year} {2020})}\BibitemShut {NoStop}%
\bibitem [{\citenamefont {Bianchi}\ and\ \citenamefont {Don\`a}(2019)}]{bianchi1}%
  \BibitemOpen
  \bibfield  {author} {\bibinfo {author} {\bibfnamefont {E.}~\bibnamefont {Bianchi}}\ and\ \bibinfo {author} {\bibfnamefont {P.}~\bibnamefont {Don\`a}},\ }\bibfield  {title} {\bibinfo {title} {Typical entanglement entropy in the presence of a center: Page curve and its variance},\ }\href {https://doi.org/10.1103/PhysRevD.100.105010} {\bibfield  {journal} {\bibinfo  {journal} {Phys. Rev. D}\ }\textbf {\bibinfo {volume} {100}},\ \bibinfo {pages} {105010} (\bibinfo {year} {2019})}\BibitemShut {NoStop}%
\bibitem [{\citenamefont {Fan}\ \emph {et~al.}()\citenamefont {Fan}, \citenamefont {Hunter-Jones}, \citenamefont {Karch},\ and\ \citenamefont {Mittal}}]{fan2025}%
  \BibitemOpen
  \bibfield  {author} {\bibinfo {author} {\bibfnamefont {Y.}~\bibnamefont {Fan}}, \bibinfo {author} {\bibfnamefont {N.}~\bibnamefont {Hunter-Jones}}, \bibinfo {author} {\bibfnamefont {A.}~\bibnamefont {Karch}},\ and\ \bibinfo {author} {\bibfnamefont {S.}~\bibnamefont {Mittal}},\ }\bibfield  {title} {\bibinfo {title} {{Sharp Transitions for Subsystem Complexity}},\ }\href {https://arxiv.org/abs/2510.18832} {\ }\Eprint {https://arxiv.org/abs/2510.18832} {arXiv:2510.18832} \BibitemShut {NoStop}%
\bibitem [{\citenamefont {Vardhan}\ and\ \citenamefont {Moudgalya}(2026)}]{vardhan2026entdyn}%
  \BibitemOpen
  \bibfield  {author} {\bibinfo {author} {\bibfnamefont {S.}~\bibnamefont {Vardhan}}\ and\ \bibinfo {author} {\bibfnamefont {S.}~\bibnamefont {Moudgalya}},\ }\bibfield  {title} {\bibinfo {title} {Entanglement dynamics from universal low-lying modes},\ }\href {https://doi.org/10.1103/prp6-y5hl} {\bibfield  {journal} {\bibinfo  {journal} {Phys. Rev. B}\ }\textbf {\bibinfo {volume} {113}},\ \bibinfo {pages} {014308} (\bibinfo {year} {2026})}\BibitemShut {NoStop}%
\bibitem [{\citenamefont {Russotto}\ \emph {et~al.}()\citenamefont {Russotto}, \citenamefont {Ares}, \citenamefont {Calabrese},\ and\ \citenamefont {Alba}}]{russotto25qssep}%
  \BibitemOpen
  \bibfield  {author} {\bibinfo {author} {\bibfnamefont {A.}~\bibnamefont {Russotto}}, \bibinfo {author} {\bibfnamefont {F.}~\bibnamefont {Ares}}, \bibinfo {author} {\bibfnamefont {P.}~\bibnamefont {Calabrese}},\ and\ \bibinfo {author} {\bibfnamefont {V.}~\bibnamefont {Alba}},\ }\bibfield  {title} {\bibinfo {title} {{Dynamics of entanglement fluctuations and quantum Mpemba effect in the $\nu=1$ QSSEP model}},\ }\href {https://doi.org/10.48550/arXiv.2510.25519} {\ }\Eprint {https://arxiv.org/abs/2510.25519} {arXiv:2510.25519} \BibitemShut {NoStop}%
\bibitem [{\citenamefont {Yu}(2013)}]{yu_2013_qfi_convex_roof}%
  \BibitemOpen
  \bibfield  {author} {\bibinfo {author} {\bibfnamefont {S.}~\bibnamefont {Yu}},\ }\bibfield  {title} {\bibinfo {title} {{Quantum Fisher Information as the Convex Roof of Variance}},\ }\href {https://arxiv.org/abs/1302.5311} {\  (\bibinfo {year} {2013})},\ \Eprint {https://arxiv.org/abs/1302.5311} {arXiv:1302.5311 [quant-ph]} \BibitemShut {NoStop}%
\bibitem [{\citenamefont {Weingarten}(1978)}]{Weingarten1978}%
  \BibitemOpen
  \bibfield  {author} {\bibinfo {author} {\bibfnamefont {D.}~\bibnamefont {Weingarten}},\ }\bibfield  {title} {\bibinfo {title} {{Asymptotic behavior of group integrals in the limit of infinite rank}},\ }\href {https://doi.org/10.1063/1.523807} {\bibfield  {journal} {\bibinfo  {journal} {J. Math. Phys.}\ }\textbf {\bibinfo {volume} {19}},\ \bibinfo {pages} {999} (\bibinfo {year} {1978})}\BibitemShut {NoStop}%
\bibitem [{\citenamefont {Collins}\ and\ \citenamefont {Sniady}(2006)}]{Collins2006}%
  \BibitemOpen
  \bibfield  {author} {\bibinfo {author} {\bibfnamefont {B.}~\bibnamefont {Collins}}\ and\ \bibinfo {author} {\bibfnamefont {P.}~\bibnamefont {Sniady}},\ }\bibfield  {title} {\bibinfo {title} {{Integration with Respect to the Haar Measure on Unitary, Orthogonal and Symplectic Group}},\ }\href {https://doi.org/10.1007/s00220-006-1554-3} {\bibfield  {journal} {\bibinfo  {journal} {Comm. Math. Phys.}\ }\textbf {\bibinfo {volume} {264}},\ \bibinfo {pages} {773} (\bibinfo {year} {2006})}\BibitemShut {NoStop}%
\bibitem [{\citenamefont {Peschel}(2003)}]{Peschel2003reduced}%
  \BibitemOpen
  \bibfield  {author} {\bibinfo {author} {\bibfnamefont {I.}~\bibnamefont {Peschel}},\ }\bibfield  {title} {\bibinfo {title} {Calculation of reduced density matrices from correlation functions},\ }\href {https://doi.org/10.1088/0305-4470/36/14/101} {\bibfield  {journal} {\bibinfo  {journal} {J. Phys. A: Math. Gen.}\ }\textbf {\bibinfo {volume} {36}},\ \bibinfo {pages} {L205} (\bibinfo {year} {2003})}\BibitemShut {NoStop}%
\bibitem [{\citenamefont {Ferro}\ \emph {et~al.}(2024)\citenamefont {Ferro}, \citenamefont {Ares},\ and\ \citenamefont {Calabrese}}]{ferro2024non}%
  \BibitemOpen
  \bibfield  {author} {\bibinfo {author} {\bibfnamefont {F.}~\bibnamefont {Ferro}}, \bibinfo {author} {\bibfnamefont {F.}~\bibnamefont {Ares}},\ and\ \bibinfo {author} {\bibfnamefont {P.}~\bibnamefont {Calabrese}},\ }\bibfield  {title} {\bibinfo {title} {{Non-equilibrium entanglement asymmetry for discrete groups: the example of the XY spin chain}},\ }\href {https://doi.org/10.1088/1742-5468/ad138f} {\bibfield  {journal} {\bibinfo  {journal} {J. Stat. Mech.}\ ,\ \bibinfo {pages} {023101}} (\bibinfo {year} {2024})}\BibitemShut {NoStop}%
\end{thebibliography}%
 
\appendix
\onecolumngrid

\section{Connection between asymmetry and quantum Fisher information}\label{app:QFI}

In this Appendix, we derive some rigorous connections between the entanglement asymmetry for a $U(1)$ symmetry and the quantum Fisher information of the charge $\hat{Q}$ that generates it, thereby showing that states with large asymmetry can be considered as promising candidates for quantum sensing applications. 

The QFI determines how sensitive is a state $\hat{\rho}$ towards a unitary imprinting $e^{-i\vartheta \hat{Q}}\hat \rho e^{i\vartheta \hat{Q}}$ generated by $\hat Q$. For a pure state $\ket{\psi}$, the QFI is directly given by the variance of $\hat Q$, $\sigma^2=\langle \hat Q^2\rangle-\langle \hat Q \rangle^2$,
\begin{equation}\label{eq:QFI_pure}
    F_{\hat Q}(\ket{\psi}) = 4 \sigma^2.
\end{equation}
On the other hand, from Eqs.~\eqref{eq:shannon_bound} and~\eqref{eq:variance_bound}, we have that the asymmetry of $\ket{\psi}$ for the charge $\hat Q$ is bounded by $\sigma^2$ according to
\begin{equation}\label{Eq:var_asymm}
\Delta S_{\hat{Q}}(\ket{\psi})<\frac{1}{2}\log\left[2\pi\left(\sigma^2+\frac{1}{12}\right)\right]+\frac{1}{2}.
\end{equation}
Combining Eqs.~\eqref{eq:QFI_pure} and~\eqref{Eq:var_asymm}, we obtain
\begin{equation}
    \frac{F_{\hat Q}(\ket{\psi})}{4} > \frac{e^{2\Delta S_{\hat Q}(\ket{\psi})}}{2\pi e} -\frac{1}{12}.
\end{equation}
In particular, for multipole charges $\hat Q_p$, the leading behavior of $\Delta S_{\hat Q_p}(\ket{\psi})\approx \frac{2p+1}{2}\log L$  for typical pure states thus leads to a lower bound on the QFI of the form:
\begin{equation}
  F_{\hat Q}(\hat\rho) > \frac{2}{\pi e} L^{2p+1} +O(L^0).  
\end{equation}
This result suggests that higher-$p$ multipole charges may serve as more effective imprinting operators in quantum sensing applications.

If we weaken our assumptions by allowing the state $\hat\rho$ to be mixed, we note that the quantum Fisher information is, in general, the convex roof of the variance~\cite{yu_2013_qfi_convex_roof}, i.e.,
\begin{equation}\label{Eq:convex_roof}
    F_{\hat Q}(\hat\rho) = 4 \inf_{\{p_k, \ket{\psi_k} \}} \sum_k p_k \sigma^2_{\ket{\psi_k}},
\end{equation}
where $\{p_k, \ket{\psi_k}\}$ represent all the pure-state ensemble decompositions of $\hat{\rho}$. The infimum is actually a minimum, meaning that there exists an ensemble decomposition $\hat\rho = \sum_k p^*_k \ket{\psi^*_k}\bra{\psi^*_k}$ such that the expression on the right-hand side of Eq.~\eqref{Eq:convex_roof} equals the QFI $F_{\hat Q}(\hat\rho)$. Consequently, we may express $F_{\hat Q}(\hat\rho)$ as
\begin{equation}
    F_{\hat Q}(\hat\rho) = 4  \sum_k p^*_k \sigma^2_{\ket{\psi_k^*}}.
\end{equation}
Since, however, the variance of the charge $\hat Q$ over the pure states $\ketbra{\psi_k^*}{\psi_k^*}$ satisfies Eq.~\eqref{Eq:var_asymm}, we may conclude that
\begin{equation}\label{SM:Eq:lower_bound_mixed}
   F_{\hat Q}(\hat\rho) > 4  \sum_k p^*_k \left(\frac{e^{2\Delta S_{\hat Q}(\ketbra{\psi^*_k}{\psi^*_k})}}{2\pi e} -\frac{1}{12}\right)=\frac{2}{\pi e}  \sum_k p^*_k \,e^{2\Delta S_{\hat Q}(\ketbra{\psi^*_k}{\psi^*_k})} -\frac{1}{3}. 
\end{equation}
Eq.~\eqref{SM:Eq:lower_bound_mixed} represents a lower bound on the quantum Fisher information of $\hat Q$ over the state $\hat\rho$. The bound is expressed as an ensemble average of an exponential function of the asymmetry, with the ensemble chosen to minimize the average charge variance. While the result depends on the quantitative form of the ensemble distribution and on the entanglement asymmetry properties of the ensemble states, Eq.~\eqref{SM:Eq:lower_bound_mixed} provides a compelling argument for searching for charges with typically enhanced large-size scaling asymmetry.

\section{Variance of multipole charges in clustering states}\label{app:clustering}

In this Appendix, we derive the bound in Eq.~\eqref{eq:sigma_p_bound} for the variance $\sigma^2=\langle \hat{Q}_p^2\rangle-\langle \hat{Q}_p\rangle^2$ of the multipole charge~\eqref{eq:p-pole_charge} in the states that satisfy the clustering property~\eqref{eq:clustering}, following the same reasoning as for the case $p=0$ in Ref.~\cite{mazzoni2025breaking}.

Let us assume here that the charge $\hat{Q}_p$ is of the form $\hat{Q}_p=\sum_{j=1}^L \hat q_j$ with $\hat q_j=j^p \hat\sigma_j^z$. In this case, using the triangle inequality, we have 
\begin{equation}
|\langle \hat q_{j_1} \hat q_{j_2}\rangle-\langle \hat q_{j_1}\rangle \langle \hat q_{j_2}\rangle|\leq
|\langle \hat q_{j_1} \hat q_{j_2}\rangle|+|\langle \hat q_{j_1}\rangle||\langle \hat q_{j_2}\rangle|.
\end{equation}
Taking the Frobenius operator norm $||\hat O||=\sqrt{{\rm Tr}(\hat O^\dagger \hat O)}$ and applying $|\langle \hat O\rangle|\leq ||\hat O||$, $||\hat O_1 \hat O_2||\leq ||\hat O_1||\cdot||\hat O_2||$
and $||\hat q_j||=\sqrt{2} j^p$, we find
\begin{equation}
|\langle \hat q_{j_1} \hat q_{j_2}\rangle-\langle \hat q_{j_1}\rangle \langle \hat q_{j_2}\rangle|\leq 2 ||\hat q_{j_1}||\cdot ||\hat q_{j_2}||=4 j_1^p j_2^p.
\end{equation}
Therefore, combining the clustering property~\eqref{eq:clustering} and the previous bound, we have
\begin{equation}
\sigma^2=\left|\sum_{j_1=1}^L\sum_{j_2\in I_{j_1}^\Lambda}\langle \hat q_{j_1} \hat q_{j_2}\rangle-\langle \hat q_{j_1}\rangle \langle \hat q_{j_2}\rangle\right|\leq 
\sum_{j_1=1}^L\sum_{j_2\in I_{j_1}^\Lambda}\left|\langle \hat q_{j_1} \hat q_{j_2}\rangle-\langle \hat q_{j_1}\rangle \langle \hat q_{j_2}\rangle\right|\leq 4 \sum_{j_1=1}^L\sum_{j_2\in I_{j_1}^\Lambda} j_1^p j_2^p,
\end{equation}
where $I_x^\Lambda=\{x'\,|\,\delta(x, x')\leq\Lambda\}$. The last sum can be 
be bounded by
\begin{equation}
\sum_{j_1=1}^L\sum_{j_2\in I_{j_1}^\Lambda} j_1^p j_2^p=\sum_{j_1=1}^L\sum_{j_2=\max(1, -\Lambda+j_1)}^{\min(L, \Lambda+j_1)}j_1^p j_2^p\leq\sum_{j_1=1}^L
\sum_{j_2=-\Lambda+j_1}^{\Lambda+j_1} j_1^p j_2^p=s(p).
\end{equation}
For large $L$, $s(p)= L^{2p+1}(1+2\Lambda)/(2p+1)+O(L^{2p})$, and, finally,
\begin{equation}
\sigma^2\leq \frac{4(1+2\Lambda)}{2p+1}L^{2p+1}(1+O(L^{-1})).
\end{equation}

\section{Calculation of the average entanglement asymmetry for Haar random states}\label{app:haar}

In this Appendix, we show the details of  the calculation of the average charged moments in the ensemble of Haar random states, reported in Eq.~\eqref{eq:av_charged_mom_haar} of the main text. To obtain them, we can apply the folded circuit picture as in the case of the homogeneous charge, $p=0$, studied in Ref.~\cite{ares2023entanglement}. Using the Choi-Jamiolkowski mapping, we can rewrite the total density matrix $\hat \rho=U\ket{0}\bra{0}U^\dagger$ as the vector $(U\otimes U^*)\ket{0}^{\otimes 2}$ 
in the doubled total Hilbert space $\mathcal{H}\otimes \mathcal{H}$, which can be graphically represented as
\begin{equation}\label{eq:folded_circuit}
\hat\rho\mapsto (U\otimes U^*)\ket{0}^{\otimes 2}=\begin{tikzpicture}[baseline={(0, -0.08cm)},
scale=1]
\def\eps{0.5};
\begin{scope}[xshift=0.1cm, yshift=0.15cm]
\draw[very thick, myblue] (-1.75,0.9) -- (-1.75,0);
\draw[very thick, myblue] (-1.5,0)-- (-0.5,0);
\draw[very thick, myblue] (-1.75,0)-- (-1.75,-0.9) ;
\draw[very thick, myblue] (-1.25,0)-- (-1.25,-0.9) ;
\draw[very thick, myblue] (2.75,0)-- (2.75,-0.9);
\draw[very thick, myblue] (2.25,0)-- (2.25,-0.9) ;
\draw[very thick, myblue] (-1.25,0.9) -- (-1.25,0);
\draw[very thick, myblue] (2.75,0.9) -- (2.75,0);
\draw[very thick, myblue] (2.25,0.9) -- (2.25,0);
\draw[thick, fill=white, rounded corners=2pt] (-2,0.4) rectangle node{$U$}(3,-0.4);
\end{scope}
\draw[very thick, myblue] (-1.75,0.9) -- (-1.75,0);
\draw[very thick, myblue] (-1.5,0)-- (-0.5,0);
\draw[very thick, myblue] (-1.75,0)-- (-1.75,-0.9) node[below]{$\ket{0}$};
\draw[very thick, myblue] (-1.25,0)-- (-1.25,-0.9) node[below]{$\ket{0}$};
\draw[very thick, myblue] (2.75,0)-- (2.75,-0.9) node[below]{$\ket{0}$};
\draw[very thick, myblue] (2.25,0)-- (2.25,-0.9) node[below]{$\ket{0}$};
\draw[very thick, myblue] (-1.25,0.9) -- (-1.25,0);
\draw[very thick, myblue] (2.75,0.9) -- (2.75,0);
\draw[very thick, myblue] (2.25,0.9) -- (2.25,0);
\draw[thick, fill=white, rounded corners=2pt] (-2,0.4) rectangle node{$U$}(3,-0.4);

\node at (0.75,-1.2) {$\cdots$};
\node at (0.75, 0.9) {$\cdots$};
\end{tikzpicture}
\end{equation}
Since the charges $\hat{Q}_p$ are diagonal in the computational basis, the charged moment $Z_2(\alpha)=\Tr(\hat\rho_A e^{i\alpha \hat{Q}_{p,A}} \hat\rho_A e^{-i\alpha \hat{Q}_{p,A}})$
can be obtained by piling up two replicas of the folded circuit~\eqref{eq:folded_circuit},
\begin{equation}
\begin{tikzpicture}[baseline={(0, 0cm)},
scale=1]
\def\eps{0.5};
\begin{scope}[xshift=-5.7cm, yshift=0.45cm]
\draw[very thick, myblue] (-1.75,0.9) -- (-1.75,0);
\draw[very thick, myblue] (-1.5,0)-- (-0.5,0);
\draw[very thick, myblue] (-1.75,0)-- (-1.75,-0.9) ;
\draw[very thick, myblue] (-1.25,0)-- (-1.25,-0.9) ;
\draw[very thick, myblue] (2.75,0)-- (2.75,-0.9);
\draw[very thick, myblue] (2.25,0)-- (2.25,-0.9) ;
\draw[very thick, myblue] (-1.25,0.9) -- (-1.25,0);
\draw[very thick, myblue] (2.75,0.9) -- (2.75,0);
\draw[very thick, myblue] (2.25,0.9) -- (2.25,0);
\draw[thick, fill=white, rounded corners=2pt] (-2,0.4) rectangle node{$U$}(3,-0.4);
\end{scope}
\begin{scope}[xshift=-5.8cm, yshift=0.3cm]
\draw[very thick, myblue] (-1.75,0.9) -- (-1.75,0);
\draw[very thick, myblue] (-1.5,0)-- (-0.5,0);
\draw[very thick, myblue] (-1.75,0)-- (-1.75,-0.9) ;
\draw[very thick, myblue] (-1.25,0)-- (-1.25,-0.9) ;
\draw[very thick, myblue] (2.75,0)-- (2.75,-0.9);
\draw[very thick, myblue] (2.25,0)-- (2.25,-0.9) ;
\draw[very thick, myblue] (-1.25,0.9) -- (-1.25,0);
\draw[very thick, myblue] (2.75,0.9) -- (2.75,0);
\draw[very thick, myblue] (2.25,0.9) -- (2.25,0);
\draw[thick, fill=white, rounded corners=2pt] (-2,0.4) rectangle node{$U$}(3,-0.4);
\end{scope}

\begin{scope}[xshift=-5.9cm, yshift=0.15cm]
\draw[very thick, myblue] (-1.75,0.9) -- (-1.75,0);
\draw[very thick, myblue] (-1.5,0)-- (-0.5,0);
\draw[very thick, myblue] (-1.75,0)-- (-1.75,-0.9) ;
\draw[very thick, myblue] (-1.25,0)-- (-1.25,-0.9) ;
\draw[very thick, myblue] (2.75,0)-- (2.75,-0.9);
\draw[very thick, myblue] (2.25,0)-- (2.25,-0.9) ;
\draw[very thick, myblue] (-1.25,0.9) -- (-1.25,0);
\draw[very thick, myblue] (2.75,0.9) -- (2.75,0);
\draw[very thick, myblue] (2.25,0.9) -- (2.25,0);
\draw[thick, fill=white, rounded corners=2pt] (-2,0.4) rectangle node{$U$}(3,-0.4);
\end{scope}
\begin{scope}[xshift=1cm, yshift=0.225cm]
\draw[ultra thick, myblue] (-1.75,0.9) -- (-1.75,0.4);
\draw[ultra thick, myblue] (-1.75,-0.4)-- (-1.75,-0.9);
\draw[ultra thick, myblue] (-1.25,-0.4)-- (-1.25,-0.9);
\draw[ultra thick, myblue] (2.75,-0.4)-- (2.75,-0.9);
\draw[ultra thick, myblue] (2.25,-0.4)-- (2.25,-0.9);
\draw[ultra thick, myblue] (-1.25,0.9) -- (-1.25,0.4);
\draw[ultra thick, myblue] (2.75,0.9) -- (2.75,0.4);
\draw[ultra thick, myblue] (2.25,0.9) -- (2.25,0.4);
\draw[thick, fill=white, rounded corners=2pt, pattern=dots] (-2,0.4) rectangle (3,-0.4);
\end{scope}

\node at (-5.5,-0.75) {$\cdots$};
\node at (-5.5, 1.1) {$\cdots$};

\begin{scope}[xshift=-6cm, yshift=0cm]
\draw[very thick, myblue] (-1.75,0.9) -- (-1.75,0);
\draw[very thick, myblue] (-1.5,0)-- (-0.5,0);
\draw[very thick, myblue] (-1.75,0)-- (-1.75,-0.9) ;
\draw[very thick, myblue] (-1.25,0)-- (-1.25,-0.9) ;
\draw[very thick, myblue] (2.75,0)-- (2.75,-0.9);
\draw[very thick, myblue] (2.25,0)-- (2.25,-0.9) ;
\draw[very thick, myblue] (-1.25,0.9) -- (-1.25,0);
\draw[very thick, myblue] (2.75,0.9) -- (2.75,0);
\draw[very thick, myblue] (2.25,0.9) -- (2.25,0);
\draw[thick, fill=white, rounded corners=2pt] (-2,0.4) rectangle node{$U$}(3,-0.4);
\end{scope}
\node at (-1.8, 0.225) {$=$};
\node at (1.6, 1) {$\cdots$};
\node at (1.6, -0.6) {$\cdots$};
\end{tikzpicture}
\end{equation}
and contracting them, 
\begin{equation}\label{eq:charged_mom_graph_haar}
Z_2(\alpha)=
\begin{tikzpicture}[baseline={(0, -0.08cm)},
scale=1]
\def\eps{0.5};
\draw[very thick, myblue] (-1.75,0.9)node[above]{\footnotesize $\bra{I_d^-(\alpha_1)}$} -- (-1.75,0.4); 
\draw[very thick, myblue] (-1.75,-0.4)-- (-1.75,-0.9)node[below]{\footnotesize $\ket{0}^{\otimes 4}$};
\draw[very thick, myblue] (-0.1,0.9)node[above]{\footnotesize $\bra{I_d^-(\alpha_{\ell_A})}$} -- (-0.1,0.4); 
\draw[very thick, myblue] (-0.1,-0.4)-- (-0.1,-0.9)node[below]{\footnotesize $\ket{0}^{\otimes 4}$};
\draw[very thick, myblue] (2.75,-0.4)-- (2.75,-0.9)node[below]{\footnotesize $\ket{0}^{\otimes 4}$};
\draw[very thick, myblue] (1.1,-0.4)-- (1.1,-0.9)node[below]{\footnotesize $\ket{0}^{\otimes 4}$};
\draw[very thick, myblue] (2.75,0.9) node[above]{\footnotesize $\bra{I_d^+}$} -- (2.75,0.4);
\draw[very thick, myblue] (1.1,0.9) node[above]{\footnotesize $\bra{I_d^+}$}-- (1.1,0.4);
\draw[thick, fill=white, rounded corners=2pt, pattern=dots] (-2,0.4) rectangle (3,-0.4);
\node at (-0.9, 0.7) {$\cdots$};
\node at (-0.9, -0.7) {$\cdots$};
\node at (2, 0.7) {$\cdots$};
\node at (2, -0.7) {$\cdots$};
\end{tikzpicture}
\end{equation} 
with the boundary states
\begin{equation}\label{eq:I_p}
\ket{I_d^+}_j=\sum_{a,b=0}^{d-1}\ket{a}_j\otimes\ket{a}_j\otimes\ket{b}_j\otimes\ket{b}_j,
\end{equation}
and 
\begin{equation}\label{eq:I_m}
\ket{I_{d}^-(\alpha)}_j=\sum_{a,b=0}^{d-1}e^{2i\alpha (a-b)}\ket{a}_j\otimes \ket{b}_j \otimes\ket{b}_j \otimes \ket{a}_j.
\end{equation}
In Eq.~\eqref{eq:charged_mom_graph_haar}, $\alpha_j$ is defined as $\alpha_j=\alpha j^p$. In formulas, Eq.~\eqref{eq:charged_mom_graph_haar} reads
\begin{equation}\label{eq:charged_mom_haar_form}
Z_2(\alpha)=\bra{+-;\alpha}(U\otimes U^*)^{\otimes 2}\ket{0}^{\otimes 4},
\end{equation}
with
\begin{equation}
\ket{+-, \alpha}=\bigotimes_{j\in B}\ket{I_d^+}_j\bigotimes_{j\in A}\ket{I_{d}^-(\alpha_j)}_j.
\end{equation}
To calculate the average charged moments over the Haar random states, we can now apply
the well-known Weingarten formula from random matrix theory~\cite{Weingarten1978, Collins2006}
\begin{equation}\label{eq:weingarten}
\mathbb{E}[(U\otimes U^*)^{\otimes 2}]=\sum_{\sigma, \tau\in\{\pm\}}w(\sigma, \tau)\ket{I_d^\sigma}\bra{I_d^\tau},
\end{equation}
where $\ket{I_d^\pm}=\otimes_{j=1}^L\ket{I_d^\pm}_j$, with $\ket{I_d^-}_j=\ket{I_d^-(0)}_j$, and 
\begin{equation}
    w(+,+)=w(-,-)=\frac{1}{d^{2L}-1},\quad w(+,-)=w(-, +)=-\frac{1}{d^{2L}-1}\frac{1}{d^{2L}}.
\end{equation}
Inserting Eq.~\eqref{eq:weingarten} in~\eqref{eq:charged_mom_haar_form}, we have
\begin{equation}\label{eq:exact_cm_haar}
\mathbb{E}[Z_2(\alpha)]=d^{-\ell_A}+d^{-L-\ell_A}\prod_{j=1}^{\ell_A}\frac{\sin^2(d\alpha_j)}{\sin^2(\alpha_j)}.
\end{equation}
In particular, in the thermodynamic limit $\ell_A,L\to\infty$ with $\ell_A/L$ finite, we obtain Eq.~\eqref{eq:av_charged_mom_haar} in the main text.

\section{Calculation of the average entanglement asymmetry in random MPS}\label{app:random_MPS}

In this Appendix, we present the calculation of the average entanglement asymmetry~\eqref{eq:av_asymm_mps} for the random MPS, defined in Fig.~\ref{fig:random_states} of the main text. We can again follow the same procedure as in the case of Haar random states, see Appendix~\ref{app:haar}. We construct the density matrix $\hat\rho=\ket{\psi}\bra{\psi}$, apply the Choi-Jamiolkowski mapping to vectorize it, and eventually take the $n=2$ replica. Graphically,
\begin{equation} 
\hat\rho^2\mapsto 
\begin{tikzpicture}[baseline={(0, -0.1cm)}, scale=1]
\def\eps{0.5};
    \draw[ultra thick, myred] (-1,0)-- (-0.5,0);
\draw[ultra thick, myblue] (0,0.9) -- (0,0.4);
\draw[ultra thick, myred] (0.5,0)-- (1,0);
\draw[ultra thick, myblue] (0,-0.4)-- (0,-0.9) node[below]{$\ket{0}^{\otimes 4}$};
\draw[thick, fill=white, rounded corners=2pt, pattern=dots] (-0.5,0.4) rectangle (0.5,-0.4);
\draw[ultra thick, myblue] (1.5,0.9)-- (1.5,0.4);
\draw[ultra thick, myred] (2,0)-- (2.5,0);
\draw[ultra thick, myblue] (1.5,-0.4)-- (1.5,-0.9) node[below]{$\ket{0}^{\otimes 4}$};
\draw[thick, fill=white, rounded corners=2pt, pattern=dots] (1,0.4) rectangle (2,-0.4);
\begin{scope}[xshift=5cm]
 \draw[ultra thick, myred] (-1,0)-- (-0.5,0);
\draw[ultra thick, myblue] (0,0.9)-- (0,0.4);
\draw[ultra thick, myred] (0.5,0)-- (1,0);
\draw[ultra thick, myblue] (0,-0.4)-- (0,-0.9) node[below]{$\ket{0}^{\otimes 4}$};
\draw[thick, fill=white, rounded corners=2pt, pattern=dots] (-0.5,0.4) rectangle (0.5,-0.4);
\draw[ultra thick, myblue] (1.5,0.9)-- (1.5,0.4);
\draw[ultra thick, myred] (2,0)-- (2.5,0);
\draw[ultra thick, myblue] (1.5,-0.4)-- (1.5,-0.9) node[below]{$\ket{0}^{\otimes 4}$};
\draw[thick, fill=white, rounded corners=2pt, pattern=dots] (1,0.4) rectangle (2,-0.4);
\node at (-1.75, 0) {$\cdots$};
\node at (3, 0) {$\cdots$};
\end{scope}
\end{tikzpicture}
\end{equation}
where
\begin{equation}
\begin{tikzpicture}[baseline={(0, 0.2cm)},
scale=1]
\def\eps{0.5};
\begin{scope}[xshift=0.75cm, yshift=0.75cm]
\draw[very thick, myblue] (0.5,0.9)-- (0.5,0);
\draw[very thick, myblue] (0.5,-0.9)-- (0.5,0);
\draw[very thick, myred] (1,0)-- (1.5,0);
\draw[very thick, myred] (-0.5,0)-- (0,0);
\draw[thick, fill=white, rounded corners=2pt] (0,0.4) rectangle (1,-0.4);
\end{scope}
\begin{scope}[xshift=0.5cm, yshift=0.5cm]
\draw[very thick, myblue] (0.5,0.9)-- (0.5,0);
\draw[very thick, myblue] (0.5,-0.9)-- (0.5,0);
\draw[very thick, myred] (1,0)-- (1.5,0);
\draw[very thick, myred] (-0.5,0)-- (0,0);
\draw[thick, fill=white, rounded corners=2pt] (0,0.4) rectangle (1,-0.4);
\end{scope}
\begin{scope}[xshift=0.25cm, yshift=0.25cm]
\draw[very thick, myblue] (0.5,0.9)-- (0.5,0);
\draw[very thick, myblue] (0.5,-0.9)-- (0.5,0);
\draw[very thick, myred] (1,0)-- (1.5,0);
\draw[very thick, myred] (-0.5,0)-- (0,0);
\draw[thick, fill=white, rounded corners=2pt] (0,0.4) rectangle (1,-0.4);
\end{scope}
\draw[very thick, myblue] (0.5,0.9)-- (0.5,0);
\draw[very thick, myblue] (0.5,-0.9) -- (0.5,0) ;
\draw[very thick, myred] (1,0)-- (1.5,0);
\draw[very thick, myred] (-0.5,0)-- (0,0);
\draw[thick, fill=white, rounded corners=2pt] (0,0.4) rectangle (1,-0.4);
\begin{scope}[xshift=-3cm, yshift=0.4cm]
\draw[ultra thick, myblue] (0.5,0.9)-- (0.5,0.4);
\draw[ultra thick, myblue] (0.5,-0.9)-- (0.5,-0.4);
\draw[ultra thick, myred] (1,0)-- (1.5,0);
\draw[ultra thick, myred] (-0.5,0)-- (0,0);
\draw[thick, fill=white, rounded corners=2pt, pattern=dots] (0,0.4) rectangle (1,-0.4);
\end{scope}
\node at (-0.9, 0.36) {$=$};
\end{tikzpicture}= (U_j\otimes U_j^*)^{\otimes 2}
\end{equation}
Since each random matrix $U_j$ is drawn independently from the rest, we can then 
average them separately. Applying the Weingarten formula,
\begin{eqnarray}
\mathbb{E}[(U_j\otimes U_j^*)^{\otimes 2}]\ket{0}^{\otimes 4}=
\begin{tikzpicture}[baseline={(0, -0.1cm)},
scale=1]
\def\eps{0.5};
\draw[ultra thick, myblue] (0.5,0.9)-- (0.5,0);
\draw[ultra thick, myblue] (0.5,-0.9) node[below]{$\ket{0}^{\otimes 4}$} -- (0.5,0) ;
\draw[ultra thick, myred] (1,0)-- (1.5,0);
\draw[ultra thick, myred] (-0.5,0)-- (0,0);
\draw[thick, fill=gray, rounded corners=2pt] (0,0.4) rectangle (1,-0.4);
\end{tikzpicture}&=&\left(\sum_{\sigma, \tau\in\{\pm\}}w(\sigma, \tau)\ket{I_d^\sigma}_j\otimes \ket{I_D^\sigma}_j\bra{I_d^\tau}_j\otimes\bra{I_D^\tau}_j\right)\ket{0}^{\otimes 4}\nonumber\\
&=& \sum_{\sigma,\tau\in\{\pm\}} w(\sigma, \tau)\ket{I_d^\sigma}_j\otimes \ket{I_D^\sigma}_j\bra{I_D^\tau}_j,
\end{eqnarray}
where 
\begin{equation}
    w(+,+)=w(-,-)=\frac{1}{(dD)^2-1},\quad w(+,-)=w(-, +)=-\frac{1}{(dD)^2-1}\frac{1}{dD},
\end{equation}
and the states $\ket{I_d^\pm}_j$ were defined in Eqs.~\eqref{eq:I_p} and~\eqref{eq:I_m} . Therefore,
\begin{equation}
\mathbb{E}[Z_2(\alpha)]=
\begin{tikzpicture}[baseline={(0, -0.07cm)}, scale=1]
\def\eps{0.5};
    \draw[ultra thick, myred] (-1,0)-- (-0.5,0);
\draw[ultra thick, myblue] (0,0.9)node[above]{$\bra{I_d^-(\alpha_1)}$}-- (0,0);
\draw[ultra thick, myred] (0.5,0)-- (1,0);
\draw[ultra thick, myblue] (0,0)-- (0,-0.9) node[below]{$\ket{0}^{\otimes 4}$};
\draw[thick, fill=gray, rounded corners=2pt] (-0.5,0.4) rectangle (0.5,-0.4);
\draw[ultra thick, myblue] (1.5,0.9)node[above]{$\bra{I_d^-(\alpha_2)}$}-- (1.5,0);
\draw[ultra thick, myred] (2,0)-- (2.5,0);
\draw[ultra thick, myblue] (1.5,0)-- (1.5,-0.9) node[below]{$\ket{0}^{\otimes 4}$};
\draw[thick, fill=gray, rounded corners=2pt] (1,0.4) rectangle (2,-0.4);
\begin{scope}[xshift=5cm]
 \draw[ultra thick, myred] (-1,0)-- (-0.5,0);
\draw[ultra thick, myblue] (0,0.9)node[above]{$\bra{I_d^+}$}-- (0,0);
\draw[ultra thick, myred] (0.5,0)-- (1,0);
\draw[ultra thick, myblue] (0,0)-- (0,-0.9) node[below]{$\ket{0}^{\otimes 4}$};
\draw[thick, fill=gray, rounded corners=2pt] (-0.5,0.4) rectangle (0.5,-0.4);
\draw[ultra thick, myblue] (1.5,0.9)node[above]{$\bra{I_d^+}$}-- (1.5,0);
\draw[ultra thick, myred] (2,0)-- (2.5,0);
\draw[ultra thick, myblue] (1.5,0)-- (1.5,-0.9) node[below]{$\ket{0}^{\otimes 4}$};
\draw[thick, fill=gray, rounded corners=2pt] (1,0.4) rectangle (2,-0.4);
\node at (-1.75, 0) {$\cdots$};
\node at (3, 0) {$\cdots$};
\end{scope}
\end{tikzpicture}
\end{equation}
where the upper blue legs are contracted with the boundary states $\bra{I_d^-(\alpha)}$ or $\bra{I_d^+}$ depending on whether the site belongs to $A$ or $B$ respectively.
We can now rewrite $\mathbb{E}[Z_2(\alpha)]$ in terms of the transfer matrices $T_-(\alpha)$ and $T_+$,
\begin{equation}\label{eq:av_charged_mom_mps_transfer_graph}
\mathbb{E}[Z_2(\alpha)]=
\begin{tikzpicture}[baseline={(0, -0.07cm)}, scale=1]
\def\eps{0.5};
    \draw[ultra thick, mygreen] (-1,0)-- (-0.5,0);
\draw[ultra thick, mygreen] (0.5,0)-- (1,0);
\draw[thick, fill=lightgray, rounded corners=2pt] (-0.6,0.5) rectangle   node{\small $T_-(\alpha_1)$}  (0.6,-0.5);
\draw[ultra thick, mygreen] (2,0)-- (2.5,0);
\draw[thick, fill=lightgray, rounded corners=2pt] (0.9,0.5) rectangle  node{\small $T_-(\alpha_2)$} (2.1,-0.5);
\begin{scope}[xshift=5cm]
 \draw[ultra thick, mygreen] (-1,0)-- (-0.5,0);
\draw[ultra thick, mygreen] (0.5,0)-- (1,0);
\draw[thick, fill=lightgray, rounded corners=2pt] (-0.6,0.5) rectangle node{\small $T_+$} (0.6,-0.5);
\draw[ultra thick, mygreen] (2,0)-- (2.5,0);
\draw[thick, fill=lightgray, rounded corners=2pt] (0.9,0.5) rectangle node{\small $T_+$} (2.1,-0.5);
\node at (-1.75, 0) {$\cdots$};
\node at (3, 0) {$\cdots$};
\end{scope}
\end{tikzpicture}
\end{equation}
defined as
\begin{equation}\label{eq:T_pm_graph}
\begin{tikzpicture}[baseline={(0, -0.07cm)},
scale=1]
\def\eps{0.5};
    \draw[ultra thick, mygreen] (-1,0) node[left] {$\sigma$}-- (-0.5,0);

\draw[ultra thick, mygreen] (0.5,0)-- (1,0) node[right] {$\tau$};

\draw[thick, fill=lightgray, rounded corners=2pt] (-0.6,0.5) rectangle node{\small $T_\pm(\alpha)$}  (0.6,-0.5);

\end{tikzpicture}=\sum_{\theta\in\{\pm\}}w(\theta, \sigma)\langle I_d^{\pm}(\alpha)|I_d^\theta\rangle_j \langle I_D^\tau|I_D^\theta\rangle_j.
\end{equation}
If we consider periodic boundary conditions, then Eq.~\eqref{eq:av_charged_mom_mps_transfer_graph} can expressed algebraically as
\begin{equation}\label{eq:av_charged_mom_mps_transfer}
\mathbb{E}[Z_2(\alpha)]=\Tr\left[\prod_{j=1}^{\ell_A} T_-(\alpha_j) T_+^{L-\ell_A}\right],
\end{equation}
which corresponds to Eq.~\eqref{eq:av_charged_mom_mps_transfer_main} in the main text. Here  $T_+$ and $T_-(\alpha)$ are the representation of the transfer matrices~\eqref{eq:T_pm_graph} in the basis $\{\ket{I_d^+}, \ket{I_d^{-}}\}$, whose explicit expressions are given in Eqs.~\eqref{eq:T_p} and~\eqref{eq:T_m}, respectively. 

Let us now derive the asymptotic behavior of $\mathbb{E}[Z_2(\alpha)]$ from this result. Although non-symmetric, both transfer matrices $T_+$ and $T_-(\alpha)$ are diagonalizable. In particular, the spectral decomposition of $T_+$ reads
\begin{equation}\label{eq:spectral_T_p}
T_+=\lambda_1^+ \ket{R_1^+}\bra{L_1^+}+\lambda_2^+\ket{R_2^+}\bra{L_2^+},
\end{equation}
where $\ket{R_j^+}$, $\ket{L_j^+}$ are the right and left eigenvectors of $T_+$ with eigenvalues
\begin{equation}
\lambda_1^+=1,\quad \lambda_2^+=\frac{d D^2-d}{(d D)^2-1}.
\end{equation}
The eigenvectors are biorthonormal, 
i.e. $\langle L_{j}^+ | R_{j'}^+\rangle=\delta_{jj'}$. Therefore, since $\lambda_2^+<1$, in the thermodynamic limit $L\to\infty$, Eq.~\eqref{eq:av_charged_mom_mps_transfer} tends to
\begin{equation}\label{eq:av_charged_mom_mps_transfer_thermo}
 \mathbb{E}[Z_2(\alpha)]=\bra{L_1^+}\prod_{j=1}^{\ell_A} T_-(\alpha_j)\ket{R_1^+}.
\end{equation}
Applying now the spectral decomposition of $T_-(\alpha)$,
\begin{equation}\label{eq:spectral_T_m}
T_-(\alpha)=\lambda_1^-(\alpha) \ket{R_1^-(\alpha)}\bra{L_1^-(\alpha)}+\lambda_2^-(\alpha)\ket{R_2^-(\alpha)}\bra{L_2^-(\alpha)},
\end{equation}
and assuming $\ell_A\gg 1$, the leading order term in Eq.~\eqref{eq:av_charged_mom_mps_transfer_thermo} is
\begin{equation}
\mathbb{E}[Z_2(\alpha)]\approx \langle L_1^+| R_1^-(\alpha_1)\rangle \prod_{j=1}^{\ell_A-1}\left[\langle L_1^-(\alpha_{j})|R_1^-(\alpha_{j+1})\rangle\right]\langle L_1^{-}(\alpha_{\ell_A})|R_1^+\rangle \prod_{j=1}^{\ell_A}\lambda_1^-(\alpha_j). 
\end{equation}
Computing the scalar products, expanding them around $\alpha = 0$, and taking the leading term in $\ell_A$, we find
\begin{equation}
 \mathbb{E}[Z_2(\alpha)]\approx  \frac{(1+d)^2 D^2}{(1+d D^2)^2}e^{-\frac{(d^2-1)D^2d}{3(2p+1)(dD^2+1)}\ell_A^{2p+1}\alpha^2}.
 \end{equation}
 The same expansion is found around the other saddle point at $\alpha=\pi$. Therefore,
using this result in Eq.~\eqref{Eq:fourier_n_renyi}, we can compute the integral in $\alpha$ by performing a saddle point approximation at $\alpha=0$ and $\alpha=\pi$, as we did in our first example in the main text, and we finally find that 
\begin{equation}
\mathbb{E}[\Delta S_{A, Q_p}^{(2)}]\approx\frac{2p+1}{2}\log(\ell_A)+\frac{1}{2}\log
\frac{(d^2-1)D^2d\pi}{3(2p+1)(1+dD^2)}.
\end{equation}
This is Eq.~\eqref{eq:av_asymm_mps} in the main text. As shown in Fig.~\ref{fig:rmps}, this approximation works well for subsystems with $\ell_A > L/2$ at any bond dimension $D$, and for $\ell_A < L/2$ when $D$ is sufficiently small. In the latter case, the limits $L \to \infty$ and $D \to \infty$ do not commute.

Finally, we can also study the limit $D\to\infty$, with $L$ finite. In that case, the transfer matrices $T_+$ and $T_-(\alpha)$ simplify as
\begin{equation}
T_+= \left(\begin{array}{cc}
 1 & 0 \\
 0 & \frac{1}{d}
 \end{array}\right),\quad
T_-(\alpha)=\left(\begin{array}{cc} 
 \frac{1}{d} & 0 \\
 0 & \frac{\sin^2(d\alpha)}{d^2\sin^2(\alpha)}\quad 
 \end{array}\right).
 \end{equation}
Inserting them in Eq.~\eqref{eq:av_charged_mom_mps_transfer}, we find
\begin{equation}\label{eq:charged_mom_RMPS_D_infty}
\mathbb{E}[Z_2(\alpha)]\overset{D\to\infty}{=}d^{-\ell_A}+d^{-L-\ell_A}\prod_{j=1}^{\ell_A}\frac{\sin^2(d\alpha_j)}{\sin^2(\alpha_j)},
\end{equation}
which coincide with the charged moments~\eqref{eq:exact_cm_haar} of the Haar random states and, therefore, have the same asymmetry. 

\section{More on the connection between random MPS and brickwork circuits}\label{app:rmps_rqc}

\begin{figure}[t]
\centering
\includegraphics[width=0.45\textwidth]{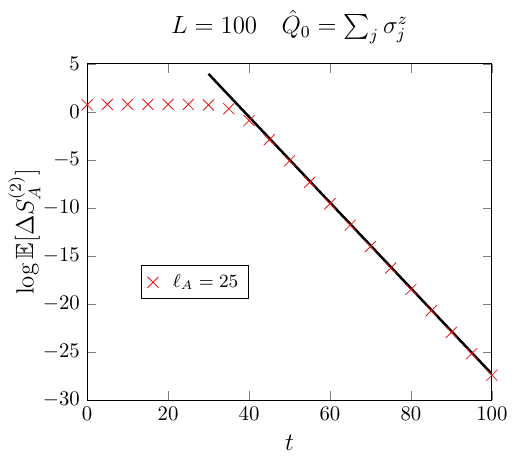}
\includegraphics[width=0.45\textwidth]{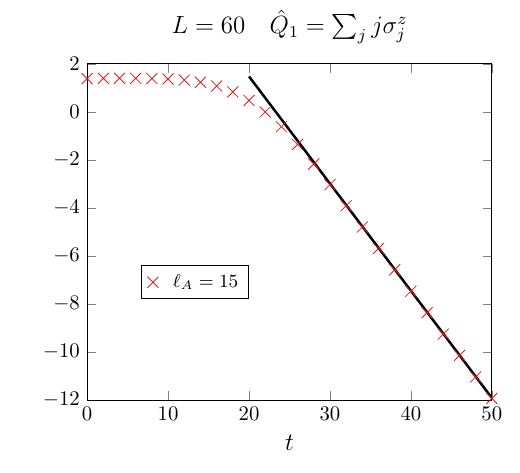}
\caption{Check of the asymptotic behavior of the asymmetry in a random MPS with $d=2$ for large bond dimension $D$ and a subsystem with $\ell_A<L/2$. We parameterize $D$ as $D=D_0e^{tv(d)}$. On the left panel we take as charge the magnetization ($p=0$) and on the right panel the dipole moment ($p=1$). Symbols are the exact asymmetry calculated using Eq.~\eqref{eq:av_charged_mom_mps_transfer}. Solid lines are the large-$D$ prediction in Eq.~\eqref{eq:relax_asymm_rmps}.   }
\label{fig:asymm_rmps_relax}
\end{figure}

In this Appendix, we discuss in more detail the relation between the results for the entanglement asymmetry found in this paper and those obtained in Ref.~\cite{ares2025prr} for random brickwork circuits, by identifying the bond dimension $D$ of the MPS with the time $t$ in the brickwork circuit via $D\sim e^t$. We can naively guess this relation as follows. A well-known result in random unitary brickwork circuits is that, in the thermodynamic limit $L\to\infty$, the average purity of an interval at times $t\ll\ell_A$ behaves as~\cite{nahum17growth}
\begin{equation}\label{eq:purity_circuit}
\mathbb{E}_{\rm circuit}[{\rm Tr}(\hat\rho_A^2)]\sim e^{-2tv(d)},
\end{equation}
where $v(d)= \log((d^2+1)/(2d))$. On the other hand, the purity in the random MPS in the thermodynamic limit is given by Eq.~\eqref{eq:av_charged_mom_mps_transfer_thermo} with $\alpha_j=0$,
\begin{equation}\label{eq:purity_rmps}
\mathbb{E}[\Tr(\hat\rho_A^2)]=\frac{(1 + d)^2 D^2}{(1 + d D^2)^2}+\left(\frac{-d + d D^2}{-1 + d^2 D^2}\right)^{\ell_A}
\frac{(-1 + D^2) (-1 + d^2 D^2)}{(1 + d D^2)^2}.
\end{equation}
Taking in this expression $\ell_A\to\infty$ and then assuming $D\gg1$, 
\begin{equation}\label{eq:purity_rmps_2}
\mathbb{E}[\Tr(\hat\rho_A^2)]\sim \frac{(d+1)^2}{d^2}\frac{1}{D^2}.
\end{equation}
Comparing Eqs.~\eqref{eq:purity_circuit} and~\eqref{eq:purity_rmps_2}, we conclude that $D=D_0e^{tv(d)}$, with $D_0=(d+1)/d$.

Let us show that, under this correspondence, the relaxation time of the asymmetry is proportional to $\ell_A$ for subsystems with $\ell_A<L/2$.
The relaxation time is defined as the time $t=t_\varepsilon^*$ from which $|\mathbb{E}[\Delta S_{A, Q_p}^{(2)}(t_\varepsilon^*)]-\mathbb{E}[\Delta S_{A, Q_p}^{(2)}(t\to\infty)]|\leq \epsilon$, for $\epsilon\ll 1$. To simplify the calculations, we take first the thermodynamic limit $L\to \infty$, in which the charged moments are given by Eq.~\eqref{eq:av_charged_mom_mps_transfer_thermo}. In this limit, $\mathbb{E}[\Delta S_{A, Q_p}^{(2)}(t\to\infty)]=0$, as follows directly from Eq.~\eqref{eq:charged_mom_RMPS_D_infty}. To determine $t_\varepsilon^*$, we study the asymptotic behavior of the two terms of the asymmetry as $D\to\infty$. Using Eq.~\eqref{eq:purity_rmps} with $\alpha_j=0$, we can easily find the expansion of the average R\'enyi-2 entropy of $\hat\rho_A$ around $D=\infty$ ($t=\infty)$,
\begin{equation}\label{eq:largeD_renyi2_mps}
\mathbb{E}[S^{(2)}(\hat\rho_A)]=\ell_A\log d-\frac{(d+1)^2}{d^2}\frac{d^{\ell_A}}{D^2}+O(D^{-4}),
\end{equation}
where we eventually assumed $\ell_A\gg 1$. 
Obtaining from Eq.~\eqref{eq:av_charged_mom_mps_transfer_thermo}, with now $\alpha_j\neq0$, the analogous expansion for the average R\'enyi-2 entropy of the symmetrized density matrix $\hat\rho_{A, \hat Q_p}$ is more difficult. Yet, in Fig.~\ref{fig:asymm_rmps_relax}, we check for $p=0$ and $p=1$ that the decay of the average asymmetry to zero in the limit $D\to\infty$ is well-described by
\begin{equation}\label{eq:relax_asymm_rmps}
\mathbb{E}[\Delta S_{A, Q_p}^{(2)}]\simeq \frac{(d+1)^2}{d^2}\frac{d^{\ell_A}}{D^2}.
\end{equation}
This means that the average R\'enyi-2 entropy of $\hat\rho_{A,\hat Q_p}$ tends to its $D\to\infty$ value much faster than that of $\hat\rho_A$, so that the relaxation of the asymmetry for $\ell_A<L/2$ is dictated by the latter, as also observed in brickwork circuits when $p=0$~\cite{ares2025prr}. From Eq.~\eqref{eq:relax_asymm_rmps}, taking $D=D_0e^{tv(d)}$, it follows directly that $t_\varepsilon^*\sim\ell_A$. 

Finally, it is interesting to compare the infinite bond-dimension expansion of the average Rényi-2 entropy~\eqref{eq:largeD_renyi2_mps} with the corresponding infinite-time expansion in brickwork circuits~\cite{ares2025prr},
\begin{equation}
\mathbb{E}_{\rm circuit}[S^{(2)}(\hat\rho_A)]\sim\ell_A\log d-\frac{\ell_A}{\sqrt{\pi t}}e^{-2tv(d)+\ell_A\log d}.
\end{equation}
In this regime, the two expansions are not exactly equal under the correspondence $D=D_0e^{tv(d)}$, although they lead to the same qualitative behavior for the relaxation time.

 \section{Squeezed fermionic Gaussian states}\label{app:gaussianstates}

In this Appendix, we consider another relevant class of states where the entanglement asymmetry has been exhaustively studied in the 
case $p=0$: the squeezed fermionic Gaussian states 
(see~\cite{murciano2024entanglement, rylands2024microscopic, chalas2024multiple}) 
\begin{equation}\label{eq:squeezed_state}
\ket{\psi}\propto  \exp\left(-\int_{0}^\pi \frac{dk}{2\pi} \mathcal{M}(k) \eta_{k}^\dagger \eta_{-k}^\dagger\right)\ket{0},
\end{equation}
where $\ket{0}$ is the vacuum of the fermionic modes $\eta_k $, i.e. $\eta_k\ket{0}$ for all $k$, and $\mathcal{M}(k)$ is an arbitrary and real odd function
of $k$. For a given $\mathcal{M}(k)$, they are the ground state of a one-dimensional, translational invariant, quadratic Hamiltonian of the form
\begin{equation}\label{eq:quadratic_Ham}
H=\sum_{l, m=1}^L(A_{|l-m|}c_l^\dagger c_m+ B_{|l-m|} c_l^\dagger c_m^\dagger+h.c.),
\end{equation}
where $c_n^\dagger$ and $c_n$ are standard creation and annihilation operators of a spinless fermion at the site $j$ of the lattice. 
The multipole charge~\eqref{eq:p-pole_charge} can be expressed in terms of them through the Jordan-Wigner transformation, and $n_j=c_j^\dagger c_j$, 
which is the particle-number operator. In the case $p=0$, it has been analytically found that, in the thermodynamic limit $L\to\infty$ and
large subsystems~\cite{rylands2024microscopic, chalas2024multiple}, 
\begin{equation}\label{eq:asymm_var_gauss}
\Delta S_{A, \hat{Q}_0}^{(n)}\approx\frac{1}{2}\log(2\pi n^{1/(1-n)}\sigma_0^2),
\end{equation}
where $\sigma_0^2$ is the variance of the charge $\hat{Q}_0$ in the state~\eqref{eq:squeezed_state}. Let us check 
whether a similar formula holds for any $p$-pole charge. 

To this end, we can use the fact that the squeezed states~\eqref{eq:squeezed_state} are Gaussian, i.e. they satisfy the Wick theorem.
In this case, the charged moments~\eqref{Eq:part_fun} can be calculated using the restriction of the two-point correlation function~\cite{Peschel2003reduced}
\begin{equation}\label{eq:two-point_corr}
\Gamma_{jj'}=2\langle  \boldsymbol{c}_j^\dagger  \boldsymbol{c}_{j'}\rangle-\delta_{jj'}, \quad \boldsymbol{c}_j=(c_j, c_j^\dagger),
\end{equation} 
to the subsystem $A$, i.e. $j, j'\in A$, and the formula~\cite{ares2023asymmetry, ares2023lack}
\begin{equation}\label{eq:charged_mom_gauss}
Z_n(\boldsymbol{\alpha})=
\sqrt{{\rm det}\left[\left(\frac{I-\Gamma}{2}\right)^n\left(I+\prod_{l=1}^n W_l\right)\right]},
\end{equation}
where $W_l=(I+\Gamma)(I-\Gamma)^{-1}e^{i\alpha_{j,j+1}n_{A, p}}$ and $n_{A, p}$ is a diagonal matrix whose entries depend on the order $p$
of the  charge, $(n_{A, p})_{2j-1, 2j-1}=-j^p$ and $(n_{A, p})_{2j, 2j}=j^p$. For the squeezed states~\eqref{eq:squeezed_state}, the two-point correlation
matrix~\eqref{eq:two-point_corr} has the specific form in the thermodynamic limit 
\begin{equation}
\Gamma_{jj'}=\int_{-\pi}^\pi\frac{dk}{2\pi} e^{ik(j-j')}\left(\begin{array}{cc} 1-2\vartheta(k) & -2p(k) \\ 2p(k) & 2\vartheta(k)-1\end{array}\right),
\end{equation}
where $\vartheta(k)=\bra{\psi} \eta_k^\dagger \eta_k\ket{\psi}=\mathcal{M}(k)^2/(1+\mathcal{M}(k)^2)$ is the density of occupied modes in the state and $p(k)=\bra{\psi} \eta_k^\dagger \eta_{-k}^\dagger\ket{\psi}=i\mathcal{M}(k)/(1+\mathcal{M}(k)^2)$. When $p=0$, both $\Gamma$ and $e^{i\alpha n_{A,0}}$ are block Toeplitz matrices, i.e., the non-zero entries in the main diagonal and the parallels to it are periodic with period $2$, due to the translational invariance of both the squeezed states~\eqref{eq:squeezed_state} and the charge $\hat{Q}_0$. In that case, Eq.~\eqref{eq:asymm_var_gauss}  can be analytically derived from Eq.~\eqref{eq:charged_mom_gauss} by applying the results on the asymptotic behavior of the determinants of block Toeplitz matrices, as shown in Ref.~\cite{murciano2024entanglement}. However, for $p\neq 0$, the charge $\hat{Q}_p$ is no longer translational invariant, and the corresponding matrix $n_{A, p}$ is not block Toeplitz, preventing us from applying those techniques here. 

\begin{figure}[t]
\centering
\includegraphics[width=0.45\textwidth]{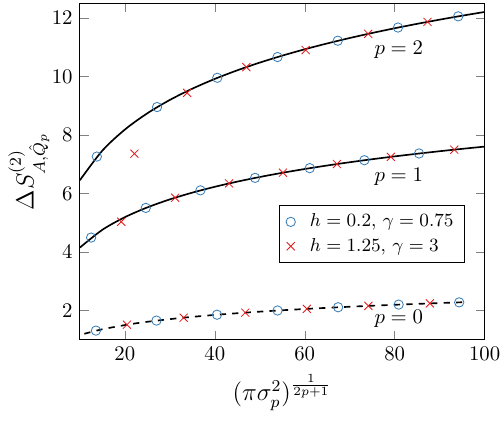}
\caption{R\'enyi-2 entanglement asymmetry as a function of the variance of the charge $\hat{Q}_p$ for different values of $p$ in the ground state of the Kitaev chain (see main text). Since this state is Gaussian, the symbols have been calculated numerically using Eqs.~\eqref{eq:charged_mom_gauss} and~\eqref{eq:var_gauss}  for different subsystem sizes and values of the  parameters of the Kitaev chain $h$ and $\gamma$. Dashed and solid curves correspond to Eqs.~\eqref{eq:asymm_var_gauss} and~\eqref{eq:asymm_var_gauss_p}, respectively. }
\label{fig:asymm_xy}
\end{figure}

Nevertheless, we can still efficiently compute the entanglement asymmetry numerically for $p\neq 0$
and large subsystems using Eq.~\eqref{eq:charged_mom_gauss}, and in this way check the validity of Eq.~\eqref{eq:asymm_var_gauss} in those cases. This is done in Fig.~\ref{fig:asymm_xy}, 
where we represent the R\'enyi-2 asymmetry as a function of the variance $\sigma_p^2=\langle \hat{Q}_p^2\rangle-\langle \hat{Q}_p\rangle^2$.
Applying Wick theorem, the variance $\sigma_p^2$ can be calculated
from the two-point correlators with the formula
\begin{equation}\label{eq:var_gauss}
\sigma_p^2=\Tr[N_{A, p}^2C]-\Tr[(CN_{A, p})^2]+\Tr[F^\dagger N_{A, p}FN_{A, p}],
\end{equation}
where $C_{jj'}=\langle c_j^\dagger c_{j'}\rangle$, $F_{jj'}=\langle c_j c_{j'}\rangle$, and $(N_{A, p})_{jj'}=\delta_{jj'}j^p$, with $j,j'\in A$. Symbols 
in Fig.~\ref{fig:asymm_xy} have been calculated numerically using Eqs.~\eqref{eq:charged_mom_gauss} and~\eqref{eq:asymm_var_gauss} for different subsytem sizes in the ground state of the Kitaev chain, which corresponds to taking $A_0=h/2$, $A_1=1/2$, $B_1=\gamma/2$, $A_{l>1}$, and $B_{l>1}=0$ in the Hamiltonian~\eqref{eq:quadratic_Ham}. For these states,
\begin{equation}
\vartheta(k)=\frac{1}{2}+\frac{1}{2}\frac{h-\cos k}{\sqrt{(h-\cos k)^2+\gamma^2\sin^2k}}
\end{equation}
and
\begin{equation}
p(k)=\frac{i}{2}\frac{\gamma\sin k}{\sqrt{(h-\cos k)^2+\gamma^2\sin^2k}}.
\end{equation}
Solid curves in Fig.~\ref{fig:asymm_xy} represent the function
\begin{equation}\label{eq:asymm_var_gauss_p}
\Delta S_{A, \hat{Q}_p}^{(2)}=\frac{1}{2}\log(8n^{1/(1-n)}\pi\sigma_p^2),
\end{equation}
for $p\neq 0$. The dashed curve corresponds instead to Eq.~\eqref{eq:asymm_var_gauss}. 
We obtain an excellent agreement, except for the smallest values of $\sigma_p^2$, which correspond to subsystems of size $\ell_A\sim 20$. In this case, the subleading corrections in $\ell_A$ to Eq.~\eqref{eq:asymm_var_gauss_p} become more relevant, particularly for larger values of $p$. 
The extra term $\log 2$ between Eqs.~\eqref{eq:asymm_var_gauss} and~\eqref{eq:asymm_var_gauss_p} stems from the fact that the $\mathbb{Z}_2$ subgroup $\{I, e^{i\pi \hat{Q}_p}\}$ is a symmetry of the Gaussian squeezed state when $p=0$, whereas this symmetry is broken for $p>0$. In fact, notice that, in terms of fermionic operators, $e^{i\pi \hat{Q}_0}=e^{i\pi\sum_j c_j^\dagger c_j}$
is the particle-number parity, which is well-defined in the squeezed states~\eqref{eq:squeezed_state}.  

\section{Entanglement asymmetry of inhomogeneous discrete subgroups}

In this Appendix, we study the asymmetry for discrete subgroups 
of the $U(1)$ group generated by the multipole moments. More specifically, we consider the $\mathbb{Z}_N$ 
cyclic subgroup with elements $\{e^{\frac{i2\pi m}{N}\hat Q_p} : m=0, 1\dots, N-1\}$. In this case, the symmetrized
density matrix $\hat{\rho}_{A, N}$ is obtained from
\begin{equation}
\hat\rho_{A,N}= \frac{1}{N}\sum_{m=0}^{N-1} e^{\frac{i2\pi m}{N}\hat Q_p}\hat \rho_A e^{-\frac{i2\pi m}{N}\hat Q_p}.
\end{equation} 
In general, for discrete groups, the entanglement asymmetry is bounded by the cardinality of the group, which in this case is $N$, as 
(see Refs.~\cite{gour2009measuring} and~\cite{ferro2024non})
\begin{equation}
\Delta S_{A, N}^{(n)}\leq \log N.
\end{equation}
Let us return to our initial state in Eq.~\eqref{Eq:prod_state} and use it as illustrative example. Its R\'enyi-2 entanglement asymmetry  with respect to 
the $\mathbb{Z}_N$ subgroup is
\begin{eqnarray}\label{eq:asymm_mult_discrete}
\Delta S_{A, N}^{(2)}&=&-\log \Tr[\hat \rho_{A,N}^2]\\
&=&\log N - \log\left[\sum_{m=0}^{N-1} \prod_{j\in \Lambda_e} \cos^{2}\left(\frac{\pi m j^p}{N}\right)\right]
\end{eqnarray}
for a subsystem of length $\ell_A=L/2$. In particular, if we take $p=0$, we have that  $\Delta S_{A, N}^{(2)}\to \log N$ in the limit $L\to\infty$, meaning that
the symmetry generated by any subgroup $\mathbb{Z}_N$ of the charge $\hat{Q}_0$ is maximally broken. On the other hand, in the dipole case $p=1$,
Eq.~\eqref{eq:asymm_mult_discrete}  gives
\begin{equation}\label{eq:asymm_mult_discrete_asymp}
\Delta S_{A, N}^{(2)}\to\left\{\begin{array}{l} \log N, \quad N\,\, \text{odd},\\
\log(N/2), \quad N\,\, \text{even},\end{array}\right.
\end{equation}
when $L\to\infty$. Observe that the asymmetry exactly vanishes for $N=2$, which corresponds to the $\mathbb{Z}_2$ subgroup $\{I, e^{i\pi \hat{Q}_1}\}$ and, thereby is a symmetry
of the state~\eqref{Eq:prod_state}. In fact, it is easy to check that~\eqref{Eq:prod_state} is an eigenstate  $e^{i\pi \hat Q_1}$ for any $L$. Notice that the elements of this subgroup correspond to 
the saddle points $\alpha=0$ and $\alpha=\pi$ in the integral~\eqref{Eq:renyi_2_prod}. Since this $\mathbb{Z}_2$ subgroup is contained in any other larger cyclic subgroup
$Z_N$ with $N$ even, then the associated asymmetry is not maximal, as we find in Eq.~\eqref{eq:asymm_mult_discrete_asymp}. On the other hand, the $\mathbb{Z}_2$ symmetric 
subgroup is not a subgoup of $\mathbb{Z}_N$ when $N$ is odd, and the asymmetry is maximal for them. 

An interesting case is $N=L$. Setting this in Eq.~\eqref{eq:asymm_mult_discrete}, we find that $\Delta S_{A, L}\sim 1/2\log L$  when $p=0$ and $L$ is large. 
For the dipole charge, $p=1$, we have instead that  $\Delta S_{A, L}\sim \log L$. 

\section{Explicit construction of the symmetrized state from a basis of the commutant algebra}\label{app:rho_S}

In this Appendix, we present an alternative derivation of Eq.~\eqref{eq:rhoS_proj} for the symmetrized density matrix $\hat{\rho}_S$ with respect to the commutant algebra $\mathcal{C}$, starting from its definition~\eqref{eq:rhoS1}.
To this end, we recall that  the commutant algebra $\mathcal{C}$ has the  structure, cf. Eq.~\eqref{Eq:decomp},
\begin{equation}
    \mathcal{C} \simeq \bigoplus_{\lambda}\left(\mathbb{I}_{D_{\lambda}}  \otimes M_{d_{\lambda}}(\mathbb{C})\right),
\end{equation}
where $\mathbb{I}_{D_\lambda}$ denotes the identity operator acting on the representation space of dimension $D_\lambda$ and 
$M_{d_\lambda}(\mathbb{C})$ stands for the algebra of complex $d_\lambda\times d_\lambda$ matrices. Therefore, the basis elements of the commutant, which is isomorphic to a direct sum of matrix algebras, read
\begin{equation}\label{eq:rep_basis_comm}
  \hat X_{\lambda,j,j'} = \sum_{i=1}^{D_{\lambda}} \ket{\lambda,i,j}\bra{\lambda,i,j'}=  \mathds{1}_{D_{\lambda}}\otimes \hat e^{(\lambda)}_{j,j'},
\end{equation}
where the last factor is the matrix of size $d_{\lambda}\times d_\lambda$ with entry $1$ in position $(j,j')$ and zeros elsewhere. 

From this result, we can explicitly 
construct the matrix $K$ that enters the definition~\eqref{eq:rhoS1} of
$\hat{\rho}_S$ in terms of the structure constants $T_{ab}^c$ of the algebra, given by the multiplication rule $\hat X_a \hat X_B=\sum_c T_{ab}^c \hat X_c$. Using Eq.~\eqref{eq:rep_basis_comm} in this formula and identifying the indices $a\equiv(\lambda,j,j')$, $b\equiv(\mu,l,l')$ and $c\equiv(\nu,m,m')$, we obtain  
\begin{equation}\label{eq:str_cons}
   T^{(\nu,m,m')}_{(\lambda,j,j')(\mu,l,l')}= \delta_{\lambda,\mu}\delta_{\lambda,\nu} \delta_{j',l}\delta_{j,m}\delta_{l',m'}.
\end{equation}
Taking into account that $K_{ab}={\Tr}(\hat X_a \hat X_b)= \sum_{m, n}T_{am}^nT_{bn}^m$ and applying Eq.~\eqref{eq:str_cons},
\begin{equation}
    \begin{split}
       & K_{(\lambda,j,j')(\mu,l,l')} = \sum_{\nu,m,m'}\sum_{\rho,s,s'}T^{(\nu,m,m')}_{(\lambda,j,j')(\rho,s,s')} T^{(\rho,s,s')}_{(\mu,l,l')(\nu,m,m')}=d_{\lambda} \delta_{j',l}\delta_{j,l'}\delta_{\lambda,\mu}.
    \end{split}
\end{equation}
Its inverse $\tilde K$ has matrix elements satisfying
\begin{equation}
    \sum_{\mu,l,l'} d_{\lambda} \delta_{j',l}\delta_{j,l'}\delta_{\lambda,\mu} \tilde K_{(\mu,l,l')(\nu,m,m')}= \delta_{\lambda,\nu}\delta_{j,m}\delta_{j',m'},
\end{equation}
which allows us to conclude that
\begin{equation}\label{eq:tilde_K}
    \tilde K_{(\mu,l,l')(\nu,m,m')}= \frac{1}{d_{\lambda}}\delta_{\mu,\nu} \delta_{l',m} \delta_{l,m'}.
\end{equation}

Let us now insert this expression in the definition~\eqref{eq:rhoS1} of the symmetrized state $\hat\rho_S$. We first rewrite the latter as
\begin{equation}
\hat\rho_S = \sum_{\lambda,j,j'}\sum_{\mu,l,l'} \tilde K_{(\lambda,j,j')(\mu,l,l')} \sum_{i=1}^{D_{\lambda}} \sum_{i'=1}^{D_{\mu}} \ket{\lambda,i,j}\bra{\lambda,i,j'}\hat\rho\ket{\mu,i',l}\bra{\mu,i',l'}.
\end{equation}
Plugging now Eq.~\eqref{eq:tilde_K} and simplifying, we finally arrive at
\begin{equation}    
    \hat\rho_S = \sum_{\lambda,j,j'} \frac{1}{d_{\lambda}}\sum_{i=1}^{D_{\lambda}} \sum_{i'=1}^{D_{\lambda}} \ket{\lambda,i,j}\bra{\lambda,i,j'}\hat\rho\ket{\lambda,i',j'}\bra{\lambda,i',j}= \sum_{\lambda} {\rm Tr}_{\mathcal{C}}\left(\hat\Pi_{\lambda}\hat \rho\hat\Pi_{\lambda}\right)\otimes\frac{\hat\Pi_{\mathcal{H}_{\lambda}^{(\mathcal{C})}}}{d_{\lambda}},
\end{equation}
which is consistent with the result found in Eq.~\eqref{eq:rhoS_proj} of the main text.

\end{document}